\documentclass[journal=jctcce,manuscript=article]{achemso}
\setkeys{acs}{chaptertitle=true}
\usepackage[T1]{fontenc}
\usepackage[version=4]{mhchem}
\usepackage{amsmath,amssymb}
\usepackage{bm}
\usepackage{booktabs}
\usepackage{array}
\newcolumntype{L}[1]{>{\raggedright\arraybackslash}p{#1}}
\usepackage{threeparttable}
\usepackage{graphicx}
\usepackage{subcaption}
\usepackage{xcolor}
\usepackage{url}
\usepackage{siunitx}
\usepackage{microtype}
\usepackage{placeins}
\usepackage{tikz}
\usetikzlibrary{arrows.meta,positioning,calc,fit,backgrounds,decorations.pathreplacing,decorations.pathmorphing,patterns,shapes.misc}
\usepackage[colorlinks=true,linkcolor=blue!60!black,citecolor=blue!60!black,urlcolor=blue!60!black]{hyperref}

\graphicspath{{figures/}}

\newif\ifdraftmarks
\draftmarkstrue

\newcommand{\J}{\mathbf{J}}
\newcommand{\K}{\mathbf{K}}
\newcommand{\F}{\mathbf{F}}
\newcommand{\D}{\mathbf{D}}
\newcommand{\vect}[1]{\mathbf{#1}}
\newcommand{\Eh}{\ensuremath{\,E_\mathrm{h}}}
\newcommand{\taustar}{\tau^{*}}
\newcommand{\taub}{\tau_\mathrm{Brc}}
\definecolor{fsKernel}{HTML}{EEF1F4}
\definecolor{fsKernelEdge}{HTML}{8E98A6}
\definecolor{fsBlock}{HTML}{FBEDE5}
\definecolor{fsBlockEdge}{HTML}{E8A688}
\definecolor{fsShared}{HTML}{DDD3E6}
\definecolor{fsSharedEdge}{HTML}{4B0D65}
\definecolor{fsGlobal}{HTML}{CFE0EE}
\definecolor{fsGlobalEdge}{HTML}{266798}
\definecolor{fsKetD}{HTML}{22043B}
\definecolor{fsKetL}{HTML}{ECE6F1}
\definecolor{fsBraD}{HTML}{D36135}
\definecolor{fsBraL}{HTML}{FBE6DC}
\definecolor{fsBack}{HTML}{F9F9F9}
\definecolor{fsBackEdge}{HTML}{BCC3CD}
\definecolor{fsOff}{HTML}{DADEE3}
\definecolor{fsInk}{HTML}{22043B}
\definecolor{fsExit}{HTML}{C0122F}
\definecolor{fsZoom}{HTML}{BCC3CD}
\definecolor{fsSteel}{HTML}{266798}
\definecolor{fsPurple}{HTML}{5A116E}
\definecolor{fsNavy}{HTML}{310754}
\definecolor{fsGray}{HTML}{5B5566}
\newcommand{\fstitle}{\fontsize{9}{11}\selectfont}
\newcommand{\fsnorm}{\fontsize{8}{9.6}\selectfont}
\newcommand{\fssmall}{\fontsize{7}{8.4}\selectfont}
\newcommand{\fstiny}{\fontsize{5.5}{6.6}\selectfont}
\tikzset{
  fs/.style={font=\sffamily\fsnorm, text=fsInk, >={Stealth[length=1.8mm]}, line join=round},
  ptitle/.style={font=\sffamily\fstitle\bfseries, text=fsInk, anchor=south west, inner sep=0pt},
  kernel/.style={draw=fsKernelEdge, fill=fsKernel, line width=0.6pt, rounded corners=2.5pt},
  gblock/.style={draw=fsBlockEdge, fill=fsBlock, line width=0.6pt, rounded corners=2.5pt},
  gexec/.style={draw=fsBlockEdge, fill=fsBlock, line width=0.8pt, rounded corners=5pt},
  back/.style={draw=fsBackEdge, fill=fsBack, line width=0.6pt, rounded corners=5pt},
  shared/.style={draw=fsSharedEdge, fill=fsShared, line width=0.6pt, inner sep=2.5pt, minimum height=4.5mm, rounded corners=2pt},
  global/.style={draw=fsGlobalEdge, fill=fsGlobal, line width=0.6pt, inner sep=1.6pt, rounded corners=1.5pt},
  zoom/.style={draw=fsZoom, line width=0.5pt},
  focus/.style={draw=fsKetD, line width=1.4pt, rounded corners=2.5pt},
  exit/.style={text=fsExit, font=\sffamily\fssmall, align=left},
  small/.style={font=\sffamily\fssmall},
  thread/.style={{Circle[length=1.1mm]}-{Stealth[length=1.6mm]}, draw=fsInk, line width=0.55pt},
  flow/.style={->, draw=fsInk, line width=0.6pt},
  fstep/.style={draw=fsSharedEdge, fill=white, rounded corners=2pt, align=center, inner sep=3pt, font=\sffamily\fssmall},
  hstep/.style={fstep, draw=fsBlockEdge, fill=fsBlock},
  note/.style={align=left, font=\sffamily\fssmall, text=fsInk},
}
\newcommand{\fslanes}[3]{\foreach \fsk in {1,...,#3} {\draw[thread] (#1+0.13*\fsk,#2) -- ++(0,-0.42);}}
\newcommand{\fslegend}[2]{%
  \begin{scope}[shift={(#1,#2)}, small]
    \node[shared, minimum width=5mm, minimum height=3mm, inner sep=0pt] at (0.25,0) {};
    \node[anchor=west] at (0.6,0) {Block-level data};
    \node[global, minimum width=5mm, minimum height=3mm, inner sep=0pt] at (0.25,-0.42) {};
    \node[anchor=west] at (0.6,-0.42) {Global memory};
    \shade[left color=fsKetD, right color=fsKetL] (0,-0.99) rectangle (0.5,-0.69);
    \draw[fsKetD, line width=0.4pt] (0,-0.99) rectangle (0.5,-0.69);
    \node[anchor=west] at (0.6,-0.84) {Sorted pair list};
    \node[kernel, minimum width=5mm, minimum height=3mm, inner sep=0pt] at (3.05,0) {};
    \node[anchor=west] at (3.4,0) {GPU kernel};
    \node[gblock, minimum width=5mm, minimum height=3mm, inner sep=0pt] at (3.05,-0.42) {};
    \node[anchor=west] at (3.4,-0.42) {Thread block};
    \draw[thread] (3.05,-0.66) -- ++(0,-0.36);
    \node[anchor=west] at (3.4,-0.84) {Thread};
  \end{scope}}

\newcommand{\GFourPRoute}{\texttt{get\_j} + \texttt{get\_k}}
\newcommand{\RefCrossMax}{\ensuremath{5.3\times10^{-8}}}

\newcommand{\RefCrossN}{7}

\newcommand{\tkRatioPa}{3.9}
\newcommand{\tkLooseEPa}{\ensuremath{5.2\times10^{-6}}}
\newcommand{\tkTightEPa}{\ensuremath{3.3\times10^{-6}}}
\newcommand{\swUMPaNine}{\ensuremath{2.2\times10^{-4}}}

\newcommand{\swUMPaTen}{\ensuremath{1.9\times10^{-5}}}

\newcommand{\swUMPaEleven}{\ensuremath{1.0\times10^{-5}}}

\newcommand{\swUMDZTen}{\ensuremath{9.1\times10^{-5}}}

\newcommand{\swUMDZEleven}{\ensuremath{9.4\times10^{-6}}}

\newcommand{\spbestPaGly}{1.74}

\newcommand{\spbestPaWat}{1.73}

\newcommand{\tplabDZGly}{$\mathrm{gly}_{110}$}

\newcommand{\spbestDZGly}{1.05}

\newcommand{\tplabDZWat}{$(\mathrm{H_2O})_{200}$}

\newcommand{\spbestDZWat}{0.98}
\newcommand{\tpUMTZGly}{67.3}

\newcommand{\tpMXTZGly}{26.3}

\newcommand{\tplabTZGly}{$\mathrm{gly}_{50}$}

\newcommand{\spbestTZGly}{1.56}

\newcommand{\tplabTZWat}{$(\mathrm{H_2O})_{140}$}

\newcommand{\spbestTZWat}{0.72}

\newcommand{\spAvgSTOGly}{3.1}
\newcommand{\spPeakSTOGly}{5.9}
\newcommand{\brcLastSTOGly}{$\mathrm{gly}_{30}$}

\newcommand{\spAvgSTOWat}{3.2}

\newcommand{\brcLastSTOWat}{$(\mathrm{H_2O})_{120}$}

\newcommand{\spAvgPaGly}{2.4}

\newcommand{\brcLastPaGly}{$\mathrm{gly}_{10}$}

\newcommand{\spAvgPaWat}{2.8}
\newcommand{\spPeakPaWat}{4.8}
\newcommand{\brcLastPaWat}{$(\mathrm{H_2O})_{40}$}

\newcommand{\spAvgPbGly}{2.4}

\newcommand{\spAvgPbWat}{2.6}

\newcommand{\spAvgDZGly}{1.2}
\newcommand{\spPeakDZGly}{1.2}

\newcommand{\spAvgDZWat}{1.4}

\newcommand{\spAvgTZGly}{1.6}

\newcommand{\spCrossTZWat}{$(\mathrm{H_2O})_{70}$}
\newcommand{\spAvgTZWat}{1.2}

\newcommand{\spAvgPopleMin}{2.4}
\newcommand{\spAvgPopleMax}{3.2}
\newcommand{\spAvgCcMin}{1.2}
\newcommand{\spAvgCcMax}{1.6}
\newcommand{\spQKMin}{15}
\newcommand{\spQKMax}{50}

\newcommand{\cvnfMXPaGly}{48}

\newcommand{\cvspbestPaGly}{1.51}

\newcommand{\cvnfMXPaWat}{22}

\newcommand{\cvspbestPaWat}{1.52}

\newcommand{\cvspbestDZGly}{1.05}

\newcommand{\cvslowDZWat}{1.06}
\newcommand{\cvCrossDZWat}{$(\mathrm{H_2O})_{140}$}

\newcommand{\cvnfMXTZGly}{16}

\newcommand{\cvspbestTZGly}{1.47}

\newcommand{\cvnfMXTZWat}{6}

\newcommand{\cvslowTZWat}{1.27}
\newcommand{\cvCrossTZWat}{$(\mathrm{H_2O})_{70}$}
\newcommand{\cvspMax}{1.62}

\newcommand{\pubMaxDE}{\ensuremath{2.4\times10^{-5}}}
\newcommand{\pubSys}{6-31G** $(\mathrm{H_2O})_{200}$}
\newcommand{\pubNOver}{11}
\newcommand{\pubN}{83}
\newcommand{\pubOthersMax}{\ensuremath{5.2\times10^{-6}}}
\newcommand{\itmedGP}{8.5--9}

\newcommand{\itmedMX}{9--10}

\newcommand{\itmedQK}{10--11}

\newcommand{\cvmaxdeGP}{\ensuremath{9.2\times10^{-6}}}
\newcommand{\cvmaxdeUM}{\ensuremath{1.2\times10^{-5}}}
\newcommand{\cvmaxdeMX}{\ensuremath{7.4\times10^{-6}}}
\newcommand{\cvmaxdeBRC}{\ensuremath{1.2\times10^{-5}}}
\newcommand{\cvmaxdeQK}{\ensuremath{3.1\times10^{-5}}}

\newcommand{\fdbworstUM}{\ensuremath{1.3\times10^{-5}}}

\newcommand{\fdbmissQK}{31}
\newcommand{\fdbnQK}{113}
\newcommand{\fdbworstQK}{\ensuremath{1.9\times10^{-4}}}
\newcommand{\fdbworstsysQK}{cc-pVDZ $\mathrm{gly}_{60}$}

\newcommand{\cvmaxdeNW}{\ensuremath{3.5\times10^{-5}}}
\newcommand{\cvmissBRC}{1}

\newcommand{\cvmissUM}{4}

\newcommand{\cvmissQK}{6}
\newcommand{\cvnQK}{114}

\newcommand{\cvmissOR}{8}
\newcommand{\cvnOR}{92}
\newcommand{\cvmaxdeOR}{\ensuremath{1.6\times10^{-5}}}

\newcommand{\cvmissNWDZ}{6}
\newcommand{\cvnNWDZ}{16}

\newcommand{\cvnNWF}{332}
\newcommand{\cvmaxdeNWF}{\ensuremath{5.9\times10^{-6}}}

\newcommand{\valNwDev}{2}
\newcommand{\valNwN}{15}
\newcommand{\NWSource}{the matched settings of Table~\ref{tab:thresholds}}

\newcommand{\nwspMin}{13}
\newcommand{\nwspMax}{35}
\newcommand{\nwCpuSpMin}{3.2}
\newcommand{\nwCpuSpMax}{8.3}
\newcommand{\nwBuildsLo}{10}
\newcommand{\nwBuildsHi}{21}

\newcommand{\cpuRatioMin}{8.0}
\newcommand{\cpuRatioMax}{19.3}
\newcommand{\cpuVsNwMin}{8.2}
\newcommand{\cpuVsNwMax}{17.4}
\newcommand{\cpuNwVsGpuMin}{66}
\newcommand{\cpuNwVsGpuMax}{207}
\newcommand{\cpuPyVsGpuMin}{22}
\newcommand{\cpuPyVsGpuMax}{175}
\newcommand{\cpuOrVsGpuMin}{43}
\newcommand{\cpuOrVsGpuMax}{114}
\newcommand{\tdpCPU}{350}
\newcommand{\tdpGPU}{300}
\newcommand{\tdpRatio}{1.17}

\newcommand{\orSpMin}{43}
\newcommand{\orSpMax}{124}
\newcommand{\orEffMin}{50}
\newcommand{\orEffMax}{144}
\newcommand{\orEffMaxSys}{$\mathrm{gly}_{40}$/cc-pVDZ}

\newcommand{\mgEffUM}{79}

\newcommand{\mgEffMX}{73}
\newcommand{\mgTMX}{0.551}

\newcommand{\mgEffBRC}{86}

\newcommand{\mgEffGP}{85}
\newcommand{\mgTGP}{0.564}

\newcommand{\mgEffQK}{91}
\newcommand{\mgTQK}{9.61}
\newcommand{\mgOneQKvsMX}{22}
\newcommand{\mgSys}{cc-pVDZ $\mathrm{gly}_{40}$}
\newcommand{\mgNbf}{3{,}025}

\newcommand{\jkKrPaGly}{2.36}

\newcommand{\jkKrTZWat}{0.59}

\newcommand{\jkcFirstSysPaGly}{$\mathrm{gly}_{10}$}
\newcommand{\jkcFirstJrPaGly}{3.80}

\newcommand{\jkcFirstJsharePaGly}{30}
\newcommand{\jkcLastSysPaGly}{$\mathrm{gly}_{120}$}
\newcommand{\jkcLastJrPaGly}{1.30}
\newcommand{\jkcLastKrPaGly}{2.36}
\newcommand{\jkcLastJsharePaGly}{69}

\newcommand{\jkcFirstJrDZGly}{1.97}
\newcommand{\jkcFirstKrDZGly}{0.91}
\newcommand{\jkcFirstJshareDZGly}{28}

\newcommand{\jkcLastJrDZGly}{0.66}
\newcommand{\jkcLastKrDZGly}{2.06}
\newcommand{\jkcLastJshareDZGly}{76}

\newcommand{\jkcFirstJrTZGly}{3.11}

\newcommand{\jkcFirstJshareTZGly}{16}

\newcommand{\jkcLastJrTZGly}{2.14}
\newcommand{\jkcLastKrTZGly}{1.17}
\newcommand{\jkcLastJshareTZGly}{38}

\newcommand{\jkcLastSysTZWat}{$(\mathrm{H_2O})_{140}$}

\newcommand{\jkcLastKrTZWat}{0.59}

\newcommand{\jkcKcrossTZWat}{$(\mathrm{H_2O})_{50}$}
\newcommand{\jkcxKMXTZWat}{2.61}
\newcommand{\jkcxKGPTZWat}{2.08}

\newcommand{\xKMXPa}{1.40}

\newcommand{\xKUMPa}{1.14}

\newcommand{\xJMXDZ}{2.04}
\newcommand{\xKMXDZ}{1.20}

\newcommand{\xJGPDZ}{1.81}
\newcommand{\xKGPDZ}{1.45}

\newcommand{\xKMXTZ}{1.25}

\newcommand{\xKGPTZ}{1.22}

\author{Miguel Alonso Mediavilla}
\email{miguel@mqs.dk}
\affiliation[UCPH]{Department of Chemistry, University of Copenhagen, Universitetsparken 5, DK-2100 Copenhagen \O, Denmark}
\alsoaffiliation[MQS]{Molecular Quantum Solutions ApS, Blegdamsvej 17, 2100 Copenhagen, Denmark}
\author{Mark Nicholas Jones}
\email{research@mqs.dk}
\affiliation[MQS]{Molecular Quantum Solutions ApS, Blegdamsvej 17, 2100 Copenhagen, Denmark}
\author{Stephan P. A. Sauer}
\affiliation[UCPH]{Department of Chemistry, University of Copenhagen, Universitetsparken 5, DK-2100 Copenhagen \O, Denmark}

\title{ForSCF: Portable Multicore and Multi-GPU Fock Builds for Electronic Structure Calculations}

\abbreviations{SCF, HF, ERI, MD, HGP, GPU, OpenACC}

\begin{document}

\begin{abstract}
The construction of the Fock matrix from electron repulsion integrals dominates the cost of Hartree--Fock calculations and limits the performance of the self-consistent field (SCF) step of density functional and post-Hartree--Fock methods.
This work discusses opt-UM and opt-Brc, two integral schemes of Palethorpe et al.\ for graphics processing units (GPUs), based on the McMurchie--Davidson and Head-Gordon--Pople algorithms and defined from recurrence graphs.
Whereas the original implementation uses CUDA, the schemes are expressed here in Fortran with compiler directives, so that one source builds with OpenACC for NVIDIA and AMD GPUs and with OpenMP for multicore CPUs.
They are adapted to the directive model and to GPUs with smaller caches, and a mixed scheme is introduced that builds the Coulomb matrix with opt-UM and the exchange matrix with opt-Brc.
The resulting code, ForSCF, can also serve as the Fock-build engine of other quantum chemistry programs.
At a matched per-build error below $10^{-5}\Eh$, for polyglycine chains and water clusters on one V100 GPU, ForSCF builds the Fock matrix on average \spAvgPopleMin--\spAvgPopleMax{} times as fast as GPU4PySCF with STO-3G and the Pople basis sets and \spAvgCcMin--\spAvgCcMax{} times as fast with cc-pVDZ and cc-pVTZ, although GPU4PySCF is faster for the larger water clusters with the latter two.
At the largest systems that NWChem completed, ForSCF on one GPU converges RHF calculations \cpuPyVsGpuMin--\cpuNwVsGpuMax{} times as fast as NWChem, PySCF and ORCA on all 32 CPU cores of the same node, and as the Fock-build engine of NWChem it makes direct SCF calculations \nwspMin--\nwspMax{} times faster.
\end{abstract}

\section{Introduction}
\label{sec:intro}
The self-consistent field (SCF) procedure lies at the heart of ab initio quantum chemical methods.
In Hartree--Fock (HF) theory and Kohn--Sham density functional theory (KS-DFT),\cite{Kohn1965} the electronic energy of a molecular system is minimized by solving the Roothaan--Hall equations\cite{Roothaan1951,Hall1951} iteratively, until the electronic density ceases to change beyond a defined threshold.
Post-HF methods, such as M{\o}ller--Plesset perturbation theory (MP2)\cite{Moller1934} and coupled-cluster (CC) theory,\cite{Bartlett2007} also begin with an SCF calculation, which makes the efficiency of the SCF procedure crucial for the overall performance of quantum chemical calculations.

The main computational bottleneck of the SCF procedure is the construction of the Fock matrix, which requires the evaluation of two-electron repulsion integrals (ERIs) over the atomic orbital basis.
The number of ERIs scales as $\mathcal{O}(N^4)$, where $N$ is the number of basis functions, leading to significant memory and computational demands for large molecular systems.
Conventional SCF implementations compute all ERIs once and store them, but this approach becomes impractical for large systems, such as biomolecules, or large basis sets.
For this reason, most modern quantum chemistry programs compute the ERIs on the fly in every Fock build, an approach known as the direct SCF method of Alml{\"o}f et al.\cite{Almlof1982}
Direct SCF trades storage for recomputation, whose cost is reduced by exploiting the permutational symmetry of the ERIs and by screening, commonly with Cauchy--Schwarz bounds\cite{Whitten1973} and the density matrix.\cite{Density_decay_1,Ochsenfeld1998}

The use of accelerators, such as graphics processing units (GPUs), has become increasingly popular in quantum chemistry due to their highly parallel architecture.
Since the first GPU evaluations of ERIs,\cite{Yasuda2008,Ufimtsev2008,Ufimtsev2009} the field has produced complete GPU Fock builds in TeraChem,\cite{Seritan2021,Wang2024} QUICK,\cite{Manathunga2021,Manathunga2023} GAMESS,\cite{Asadchev2012} EXESS,\cite{Barca2020,Barca2021} GPU4PySCF\cite{Li2025,Wu2025} and GANSU,\cite{GANSU} and integral libraries such as LibintX.\cite{Asadchev2023,Asadchev2024}
These codes differ in the integral scheme and in how the work of a Fock build is mapped onto threads.

Most recently, Palethorpe et al.\ described two optimized schemes, opt-Brc and opt-UM, based on the Head-Gordon--Pople\cite{HeadGordon1988} and McMurchie--Davidson\cite{McMurchie1978} schemes, respectively, and showed that their implementation outperformed TeraChem, QUICK and GPU4PySCF on single A100 GPUs by large factors.\cite{Palethorpe2024}

How these algorithms are expressed in code depends on the programming model, and consequently on the hardware.
NVIDIA and AMD GPUs share the same execution hierarchy of threads, blocks and on-chip shared memory, but differ in details such as the width of the SIMD group (32 threads on NVIDIA GPUs and 32 or 64 on AMD GPUs, depending on the architecture), the size of the register file and caches, and the available warp-level primitives.
CUDA\cite{CUDA} exposes these features directly for NVIDIA GPUs. HIP\cite{HIP} offers a near-identical syntax for AMD GPUs, so existing CUDA codes are commonly ported to it by source-to-source translation, which results in a second implementation to maintain.
Directive-based models such as OpenACC\cite{OpenACC33} and OpenMP\cite{OpenMP52} take a different route: loops in standard Fortran, C or C++ are annotated, and the compiler generates code for an NVIDIA or AMD GPU or for a multicore CPU from the same source.
This leaves less control over shared memory, warp-level primitives and per-thread early exits, but allows existing code to be reused and ported to new hardware.
It also makes the same algorithm available on CPUs and GPUs, so that the two can be benchmarked against each other without differences in the integral scheme or the screening.

This work presents ForSCF, a Fock-build engine and SCF program that implements both schemes of Palethorpe et al.\cite{Palethorpe2024}
Its contributions are:
\begin{enumerate}
\item Implementations of opt-Brc and opt-UM, with kernels generated from recurrence graphs, that follow the published algorithms and adapt or extend them where this is required by the cache and register budget of the GPU, by the programming model, or by details that the original description leaves open (Sections~\ref{sec:brc} and \ref{sec:um}).
\item A mixed integral scheme that combines the opt-UM Coulomb build with opt-Brc exchange (Section~\ref{sec:mixed}).
\item A comparison with GPU4PySCF, QUICK, PySCF, ORCA and NWChem at matched accuracy, with the screening threshold of every program calibrated on fixed densities of the largest test systems to the accuracy reported in Ref.~\citenum{Palethorpe2024}, $\Delta E \leq 10^{-5}\Eh$ per Fock build (Section~\ref{sec:results}).
\item An implementation in Fortran with OpenACC directives that builds from one source for NVIDIA GPUs and multicore CPUs and is prepared for AMD GPUs, and that runs on one or several GPUs or on the CPU (Section~\ref{sec:implementation}).
\item The use of ForSCF as the integral and Fock-build engine of other quantum chemistry programs, which pass a density and receive the Coulomb and exchange matrices, so that their SCF calculations are accelerated while their own algorithms are kept, with NWChem\cite{Apra2020} as the example (Section~\ref{sec:hosts}).
\end{enumerate}
This article covers restricted Hartree--Fock, although the current version also supports unrestricted calculations.

\section{Theoretical Background}
\label{sec:theory}
\subsection{Notation}
\label{sec:notation}
The notation follows that of Barca et al.\ and Palethorpe et al.\cite{Barca2021,Palethorpe2024,Palethorpe2025}
A contracted Gaussian function (CGF) with basis-set index $\alpha$ is a sum of $K$ primitive Gaussian functions (PGFs),
\begin{align}
|\alpha) &\equiv \psi_\alpha(\vect{r}) = \sum_{i=1}^{K} \phi_i^\alpha(\vect{r}), \label{eq:cgf}\\
|\alpha]_i &\equiv \phi_i^\alpha(\vect{r}) = D_i\,(r_x-A_x)^{a_x}(r_y-A_y)^{a_y}(r_z-A_z)^{a_z}\,e^{-\lambda_i|\vect{r}-\vect{A}|^2}, \label{eq:pgf}
\end{align}
where $\vect{a}=(a_x,a_y,a_z)$ is the angular momentum vector, $a=a_x+a_y+a_z$ the total angular momentum, $\vect{A}$ the center, $D_i$ the contraction coefficient and $\lambda_i$ the exponent.
The coefficient is absorbed into the primitive, so that $|\alpha]_i$ is the contribution of one Gaussian to one contracted function, and the primitive index $i$ is omitted where it is not needed.
Round brackets denote contracted quantities and square brackets primitive ones.
A function is also labeled by its angular momentum vector, $|\vect{a})\equiv|\alpha)$ and $|\vect{a}]\equiv|\alpha]$, and $\vect{1}_i$ denotes the unit vector in the Cartesian direction $i\in\{x,y,z\}$, so that $\vect{a}\pm\vect{1}_i$ raises or lowers one component of $\vect{a}$.

Functions are grouped into shells.
A primitive shell $|a]$ (non-bold) is the set of PGFs $|\vect{a}]$ with the same total angular momentum $a$, exponent and center. On the other hand, a contracted shell $|a)$ is the set of CGFs with the same $a$ and center that are contracted from the same list of exponents $\{\lambda_i\}$ and coefficients $\{D_i\}$. The $K$ PGFs within one CGF therefore differ in exponent and coefficient, whereas the CGFs of a shell differ only in $\vect{a}$.
A Cartesian shell of angular momentum $a$ has $(a+1)(a+2)/2$ functions; for example, a $p$ shell, $|p)\equiv|1)$, holds three functions: $p_x$, $p_y$ and $p_z$.
Pairs and quartets are formed as tensor products,
\begin{equation}
|ab) = |a)\otimes|b), \qquad |ab] = |a]\otimes|b], \qquad |abcd) = |ab)\otimes|cd),
\label{eq:pairs}
\end{equation}
and $|ab\}$ denotes a batch of contracted shell pairs.
In an integral $(ab|cd)$ the pair $(ab|$ is the bra and $|cd)$ the ket.
The electron repulsion integrals (ERIs) of a contracted quartet are sums over primitive quartets,
\begin{align}
(\vect{ab}|\vect{cd}) &= \sum_{i}^{K_A}\sum_{j}^{K_B}\sum_{k}^{K_C}\sum_{l}^{K_D}[\vect{ab}|\vect{cd}]_{ijkl}, \label{eq:eri}\\
[\vect{ab}|\vect{cd}]_{ijkl} &= \iint \phi_i^{\vect{a}}(\vect{r}_1)\,\phi_j^{\vect{b}}(\vect{r}_1)\,\frac{1}{|\vect{r}_1-\vect{r}_2|}\,\phi_k^{\vect{c}}(\vect{r}_2)\,\phi_l^{\vect{d}}(\vect{r}_2)\,d\vect{r}_1\,d\vect{r}_2 . \nonumber
\end{align}
Without bold indices, $(ab|cd)$ denotes the set of all ERIs of the shell quartet: $(pp|pp)\equiv(11|11)$ holds 81 integrals, from $(p_xp_x|p_xp_x)$ to $(p_zp_z|p_zp_z)$.
A class $(l_al_b|l_cl_d)$ collects the quartets with the same four angular momenta, written compactly as $l_al_bl_cl_d$ (for instance, 3231 for $(fd|fp)$), with total angular momentum $L=l_a+l_b+l_c+l_d$, whereas a sub-class also shares the contraction degrees.

For a primitive quartet with exponents $\lambda_i,\lambda_j,\lambda_k,\lambda_l$, the Gaussian product theorem gives
\begin{align}
\zeta &= \lambda_i+\lambda_j, & \eta &= \lambda_k+\lambda_l, \nonumber\\
\vect{P} &= \frac{\lambda_i\vect{A}+\lambda_j\vect{B}}{\zeta}, & \vect{Q} &= \frac{\lambda_k\vect{C}+\lambda_l\vect{D}}{\eta}, \label{eq:gpt}\\
\vartheta^2 &= \frac{\zeta\eta}{\zeta+\eta}, & T &= \vartheta^2|\vect{PQ}|^2, \nonumber
\end{align}
with difference vectors written as $\vect{AB}=\vect{A}-\vect{B}$, $\vect{PA}=\vect{P}-\vect{A}$, $\vect{PQ}=\vect{P}-\vect{Q}$, and their components as $PA_i$.
With the pair prefactors
\begin{align}
U_P &= n_a n_b D_i D_j\left(\frac{\pi}{\zeta}\right)^{3/2} e^{-\lambda_i\lambda_j|\vect{AB}|^2/\zeta}, \label{eq:uprefactor}\\
U_Q &= n_c n_d D_k D_l\left(\frac{\pi}{\eta}\right)^{3/2} e^{-\lambda_k\lambda_l|\vect{CD}|^2/\eta}, \nonumber
\end{align}
where $n_a$ is a normalization factor, the fundamental integrals are
\begin{equation}
[00|00]^{(m)} = U_P U_Q\,\frac{2\vartheta}{\sqrt{\pi}}\,F_m(T),
\label{eq:fund}
\end{equation}
with the Boys function
\begin{equation}
F_m(T)=\int_0^1 t^{2m}e^{-Tt^2}\,dt .
\label{eq:boys}
\end{equation}
Both schemes (Sections~\ref{sec:brc} and \ref{sec:um}) build all other integrals of a class from these by recurrence relations.

\subsection{Hartree--Fock and the SCF Procedure}
\label{sec:hf}
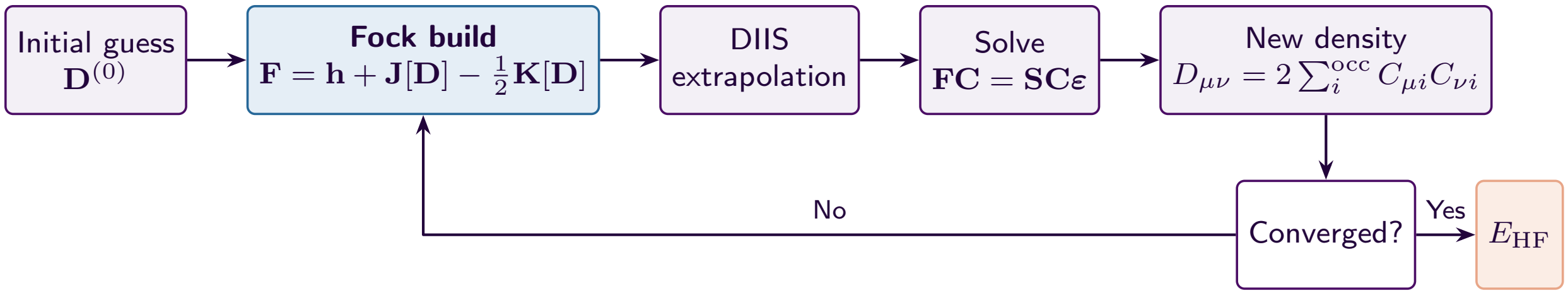
\begin{figure}[tbp]
\centering
\resizebox{\textwidth}{!}{\begin{tikzpicture}[fs, x=1cm, y=1cm, node distance=0.55cm,
  box/.style={rounded corners=2pt, align=center, inner sep=3pt, line width=0.6pt, draw=fsSharedEdge, fill=fsShared!35,
              minimum height=1.0cm, font=\sffamily\fsnorm},
  link/.style={->, draw=fsInk, line width=0.7pt}]
\node[box] (g) {Initial guess\\$\D^{(0)}$};
\node[box, draw=fsGlobalEdge, fill=fsGlobal!55, right=of g] (f) {\textbf{Fock build}\\$\F=\mathbf{h}+\J[\D]-\tfrac12\K[\D]$};
\node[box, right=of f] (d) {DIIS\\extrapolation};
\node[box, right=of d] (s) {Solve\\$\F\mathbf{C}=\mathbf{S}\mathbf{C}\bm{\varepsilon}$};
\node[box, right=of s] (n) {New density\\$D_{\mu\nu}=2\sum_i^{\mathrm{occ}}C_{\mu i}C_{\nu i}$};
\node[box, below=0.6cm of n, fill=white] (c) {Converged?};
\node[box, right=0.55cm of c, fill=fsBlock, draw=fsBlockEdge] (e) {$E_\mathrm{HF}$};
\draw[link] (g) -- (f);
\draw[link] (f) -- (d);
\draw[link] (d) -- (s);
\draw[link] (s) -- (n);
\draw[link] (n) -- (c);
\draw[link] (c) -- node[above, font=\sffamily\fssmall] {Yes} (e);
\draw[link] (c.west) -- node[above, font=\sffamily\fssmall] {No} (c.west -| f.south) -- (f.south);
\end{tikzpicture}}
\caption{The SCF procedure. In a direct SCF calculation the Fock build recomputes the two-electron integrals in every iteration and dominates the cost.}
\label{fig:scf}
\end{figure}

In restricted closed-shell HF, the molecular orbital coefficients $\mathbf{C}$ solve the Roothaan--Hall equations\cite{Roothaan1951,Hall1951,Szabo1996}
\begin{equation}
\F\mathbf{C}=\mathbf{S}\mathbf{C}\bm{\varepsilon},
\label{eq:rh}
\end{equation}
where $\mathbf{S}$ is the overlap matrix and $\bm{\varepsilon}$ the diagonal matrix of orbital energies.
With $\mu,\nu,\lambda,\sigma$ indexing basis functions, the Fock matrix is
\begin{align}
F_{\mu\nu} &= h_{\mu\nu} + J_{\mu\nu} - \tfrac{1}{2}K_{\mu\nu}, \label{eq:fock}\\
J_{\mu\nu} &= \sum_{\lambda\sigma}(\mu\nu|\lambda\sigma)\,D_{\lambda\sigma}, \nonumber\\
K_{\mu\nu} &= \sum_{\lambda\sigma}(\mu\lambda|\nu\sigma)\,D_{\lambda\sigma}, \nonumber
\end{align}
where $\mathbf{h}$ is the one-electron Hamiltonian, $\J$ and $\K$ the Coulomb and exchange matrices, and
\begin{equation}
D_{\mu\nu}=2\sum_i^{\mathrm{occ}}C_{\mu i}C_{\nu i}
\label{eq:density}
\end{equation}
the density matrix.
$J_{ab}$, $K_{ab}$ and $D_{ab}$ denote the blocks whose rows belong to shell $a$ and columns to shell $b$.
Since $\F$ depends on $\D$, the equations are solved iteratively (Figure~\ref{fig:scf}): from an initial density, $\F$ is built, extrapolated by direct inversion in the iterative subspace (DIIS),\cite{Pulay1980,Pulay1982} and diagonalized, and a new density is formed, until the density and the energy
\begin{equation}
E_\mathrm{HF}=\tfrac12\sum_{\mu\nu}D_{\mu\nu}(h_{\mu\nu}+F_{\mu\nu})+E_\mathrm{nuc}
\label{eq:ehf}
\end{equation}
no longer change.
The initial guess affects the number of iterations, and a superposition of atomic densities (SAD) is the common default.\cite{VanLenthe2006,Lehtola2019} ForSCF implements the MINAO guess of PySCF, a SAD guess that projects atomic densities from a minimal basis of atomic natural orbitals onto the basis of the calculation.\cite{Sun2020}

All steps other than the Fock build cost $\mathcal{O}(N^3)$ or less.
The two-electron part, $\J$ and $\K$, formally costs $\mathcal{O}(N^4)$. It dominates the iteration time for large systems and, in direct SCF, is repeated in every iteration.\cite{Almlof1982}
A Fock build can be full, from $\D$, or incremental, from the change $\Delta\D$ since the previous iteration, which is sparser and can be screened more aggressively.\cite{Almlof1982,Haser1989}
The benchmarks of this work use full builds, as Ref.~\citenum{Palethorpe2024} does.

\subsection{Integral Screening}
\label{sec:screening}
The main integral screening is based on the Cauchy--Schwarz inequalities\cite{Whitten1973,Haser1989}
\begin{align}
|(ab|cd)| &\le G_{ab}\,G_{cd}, & G_{ab} &= \max_{|\vect{ab})\in|ab)}\sqrt{(\vect{ab}|\vect{ab})}, \label{eq:schwarz}\\
|[ab|cd]| &\le G_{[ab]}\,G_{[cd]}, & G_{[ab]} &= \max_{|\vect{ab}]\in|ab]}\sqrt{[\vect{ab}|\vect{ab}]}. \nonumber
\end{align}
A contribution to the Fock matrix is skipped when the product of its bound and the density element that it multiplies falls below a threshold.
Each scheme applies the bounds differently (Sections~\ref{sec:brc} and \ref{sec:um}).
ForSCF has three thresholds: $\tau_J$ for the opt-UM Coulomb build and pair lists, $\tau_K$ for the target accuracy of the opt-UM exchange build, and $\taub$ for the opt-Brc quartets.

\section{The opt-Brc Scheme}
\label{sec:brc}
The opt-Brc scheme was first described by Barca et al.\cite{Barca2021} and later optimized by Palethorpe et al.\cite{Palethorpe2024} for GPU evaluation of ERIs.
\subsection{HGP Recursive Scheme}
\label{sec:hgp}
opt-Brc evaluates contracted quartets with the Head-Gordon--Pople (HGP) refinement of the Obara--Saika recurrences.\cite{HeadGordon1988,Obara1986,Barca2021,Palethorpe2024}
For each primitive quartet, vertical recurrence relations (VRRs) build angular momentum on the first center of the bra and of the ket, starting from the fundamental integrals of Eq.~\ref{eq:fund}:
\begin{align}
[\vect{e}{+}\vect{1}_i\,0|\vect{f}0]^{(m)} ={}& PA_i\,[\vect{e}0|\vect{f}0]^{(m)} - \frac{\eta}{\zeta+\eta}\,PQ_i\,[\vect{e}0|\vect{f}0]^{(m+1)} \nonumber\\
 &+ \frac{e_i}{2\zeta}\left([\vect{e}{-}\vect{1}_i\,0|\vect{f}0]^{(m)}-\frac{\eta}{\zeta+\eta}[\vect{e}{-}\vect{1}_i\,0|\vect{f}0]^{(m+1)}\right) \nonumber\\
 &+ \frac{f_i}{2(\zeta+\eta)}\,[\vect{e}0|\vect{f}{-}\vect{1}_i\,0]^{(m+1)}, \label{eq:vrr}\\[0.5em]
[\vect{e}0|\vect{f}{+}\vect{1}_i\,0]^{(m)} ={}& QC_i\,[\vect{e}0|\vect{f}0]^{(m)} + \frac{\zeta}{\zeta+\eta}\,PQ_i\,[\vect{e}0|\vect{f}0]^{(m+1)} \nonumber\\
 &+ \frac{f_i}{2\eta}\left([\vect{e}0|\vect{f}{-}\vect{1}_i\,0]^{(m)}-\frac{\zeta}{\zeta+\eta}[\vect{e}0|\vect{f}{-}\vect{1}_i\,0]^{(m+1)}\right) \nonumber\\
 &+ \frac{e_i}{2(\zeta+\eta)}\,[\vect{e}{-}\vect{1}_i\,0|\vect{f}0]^{(m+1)}, \nonumber
\end{align}
for $e\in[a,a+b]$ and $f\in[c,c+d]$.
The resulting $[e0|f0]$ are contracted inside the primitive loops,
\begin{equation}
(e0|f0)=\sum_{ijkl}[e0|f0]^{(0)}_{ijkl},
\label{eq:midcontract}
\end{equation}
and horizontal recurrence relations (HRRs), which do not depend on the exponents, transfer angular momentum to the second centers outside the loops:
\begin{align}
(\vect{e}0|\vect{c}\,\vect{d}{+}\vect{1}_i) &= (\vect{e}0|\vect{c}{+}\vect{1}_i\,\vect{d}) + CD_i\,(\vect{e}0|\vect{c}\vect{d}), \label{eq:hrr}\\
(\vect{a}\,\vect{b}{+}\vect{1}_i|\vect{c}\vect{d}) &= (\vect{a}{+}\vect{1}_i\,\vect{b}|\vect{c}\vect{d}) + AB_i\,(\vect{a}\vect{b}|\vect{c}\vect{d}). \nonumber
\end{align}
This mid-contraction makes HGP efficient for contracted shells: the VRRs run once per primitive quartet, and the HRRs once per contracted quartet.
Eqs.~\ref{eq:vrr} and \ref{eq:hrr} leave open which component is reduced at each step.
The choices define a directed acyclic graph (DAG) of intermediates, which Ref.~\citenum{Palethorpe2024} builds by heuristic tree searches and turns into unrolled code with a generator (Section~\ref{sec:codegen}).

\subsection{opt-Brc Screening}
\label{sec:brcscreen}
opt-Brc screens whole contracted quartets.
By the eightfold permutational symmetry of the ERIs, one quartet $(ab|cd)$ contributes to six blocks of the Fock matrix,
\begin{align}
J_{ab} &\mathrel{+}= (ab|cd)\,D_{cd}, & J_{cd} &\mathrel{+}= (ab|cd)\,D_{ab}, \nonumber\\
K_{ac} &\mathrel{+}= (ab|cd)\,D_{bd}, & K_{bd} &\mathrel{+}= (ab|cd)\,D_{ac}, \label{eq:digest}\\
K_{ad} &\mathrel{+}= (ab|cd)\,D_{bc}, & K_{bc} &\mathrel{+}= (ab|cd)\,D_{ad}, \nonumber
\end{align}
where the sums over the functions of each shell and the symmetry factors are implied.
The quartet is skipped when the product of its Schwarz bound (Eq.~\ref{eq:schwarz}) and the largest density element that it multiplies falls below the threshold $\taub$:
\begin{equation}
G_{ab}\,G_{cd}\,D^{\max}_{abcd} < \taub, \qquad
D^{\max}_{abcd} = \max\bigl(\|D_{ab}\|,\|D_{cd}\|,\|D_{ac}\|,\|D_{ad}\|,\|D_{bc}\|,\|D_{bd}\|\bigr),
\label{eq:brcskip}
\end{equation}
with $\|D_{ab}\|$ the largest absolute element of the block $D_{ab}$.
$\taub$ thus bounds each skipped contribution, not the error of the Fock matrix, which sums all of them.

Two cheaper tests act before Eq.~\ref{eq:brcskip}.
Primitive pairs with a negligible Schwarz factor are removed from their contracted pairs when the pair lists are formed.
Shell pairs are grouped into batches of equal class and contraction degree and split by magnitude, with batch $i$ holding the pairs with
\begin{equation}
\taub^{\,i/n} < G_{ab} \le \taub^{\,(i-1)/n}
\label{eq:brcbatch}
\end{equation}
($n=2$ in ForSCF), so that a pair of batches whose largest Schwarz product cannot reach $\taub$ is not evaluated at all.

\subsection{GPU Implementation of opt-Brc}
\label{sec:brcgpu}
\begin{figure}[tbp]
\centering
\resizebox{\textwidth}{!}{\begin{tikzpicture}[fs, x=1cm, y=1cm]
% ================================================================ kernel launches
\draw[back] (0,0) rectangle (3.85,-3.95);
\node[small, align=center] at (0.45, -0.38) {Class\\[-1pt]magn.};
\foreach \l [count=\i from 0] in {ss, ps, pp} {
  \node at (1.33 + 0.96*\i, -0.22) {\l};
  \node at (0.22, -1.0 - 0.96*\i) {\l};
  \node[small] at (1.09 + 0.96*\i, -0.55) {$1$};
  \node[small] at (1.57 + 0.96*\i, -0.55) {$\sqrt{\tau}$};
  \node[small, anchor=east] at (0.88, -0.98 + 0.12 - 0.96*\i) {$1$};
  \node[small, anchor=east] at (0.88, -1.46 + 0.12 - 0.96*\i) {$\sqrt{\tau}$};
}
\foreach \i in {0,...,5} {\foreach \j in {0,...,5} {
  \pgfmathtruncatemacro{\ci}{int(\i/2)}
  \pgfmathtruncatemacro{\cj}{int(\j/2)}
  \pgfmathtruncatemacro{\mi}{mod(\i,2)}
  \pgfmathtruncatemacro{\mj}{mod(\j,2)}
  \pgfmathtruncatemacro{\on}{(\cj > \ci || (\cj == \ci && \j >= \i)) && (\mi + \mj < 2) ? 1 : 0}
  \ifnum\on=1
    \draw[kernel] ({0.87 + 0.48*\j}, {-0.75 - 0.48*\i}) rectangle ++(0.42, -0.42);
  \else
    \fill[fsOff] ({0.87 + 0.48*\j}, {-0.75 - 0.48*\i}) rectangle ++(0.42, -0.42);
  \fi
}}
\node[ptitle] at (0, 0.47) {(a) Kernel launches};
\draw[focus] (3.27, -0.75) rectangle ++(0.42, -0.42);

% ================================================================ gang launches
\draw[zoom] (3.69, -0.75) -- (4.6, 0.25);
\draw[zoom] (3.69, -1.17) -- (4.6, -3.95);
\draw[kernel] (4.6, 0.25) rectangle (8.75, -3.95);
\fill[fsKetD!70!fsKetL] (5.3, 0.12) rectangle (8.6, -0.27);
\foreach \k in {0,...,7} {\draw[white, line width=0.6pt] (5.3 + 0.4125*\k, 0.12) rectangle ++(0.4125, -0.39);}
\node[text=white, font=\sffamily\fsnorm, fill=fsKetD!70!fsKetL, inner sep=1pt] at (6.95, -0.07) {$|cd)$};
\foreach \r [count=\i from 0] in {5, 4, 3, 3, 2, 2, 1} {
  \fill[fsBraD!70!fsBraL] (4.72, -0.42 - 0.5*\i) rectangle ++(0.45, -0.5);
  \draw[white, line width=0.6pt] (4.72, -0.42 - 0.5*\i) rectangle ++(0.45, -0.5);
  \draw[gblock] (5.3, -0.46 - 0.5*\i) rectangle (8.6, -0.88 - 0.5*\i);
  \foreach \m in {1,...,\r} {
    \fill[fsKetD!70!fsKetL] (5.36 + 0.4125*\m - 0.4125, -0.51 - 0.5*\i) rectangle ++(0.33, -0.32);
  }
}
\node[text=white, rotate=90, font=\sffamily\fsnorm] at (4.945, -1.15) {$|ab)$};
\node[ptitle] at (4.6, 0.47) {(b) Block launches};
\draw[focus] (5.3, -0.46) rectangle (8.6, -0.88);

% ================================================================ gang execution
\draw[zoom] (8.6, -0.46) -- (9.4, 0.3);
\draw[zoom] (8.6, -0.88) -- (9.4, -4.7);
\draw[gexec] (9.4, 0.3) rectangle (17.95, -4.7);
\begin{scope}[font=\sffamily\fssmall]
\node[shared, minimum width=2.8cm, font=\sffamily\fsnorm] at (11.2, -0.05) {$|ab)$ data};
\foreach \k in {0,...,3} {
  \draw[thread] (10.28 + 0.64*\k, -0.42) -- ++(0, -0.45);
  \fill[fsKetD!70!fsKetL] (9.98 + 0.64*\k, -0.98) rectangle ++(0.62, -0.38);
  \draw[white, line width=0.6pt] (9.98 + 0.64*\k, -0.98) rectangle ++(0.62, -0.38);
  \node[text=white] at (10.29 + 0.64*\k, -1.17) {$|cd)_{\k}$};
}
\node at (12.8, -1.17) {$\cdots$};
\node[anchor=west] at (13.0, -1.17) {Survivors of $|ab)$};
% HGP in one lane
\node[anchor=west] at (9.55, -1.72) {For $[ab|cd]\in(ab|cd)$:};
\node[anchor=west] at (9.75, -2.07) {VRRs, $\ (e0|f0)\mathrel{+}=[e0|f0]^{(0)}$};
\node[anchor=west] at (9.55, -2.42) {HRRs};
\draw[decorate, decoration={brace, amplitude=4pt}, line width=0.6pt] (13.45, -1.57) -- (13.45, -2.57);
\node[anchor=west] at (13.6, -2.07) {Compute $(ab|cd)$ in one thread};
% digest
\node[anchor=west] at (9.55, -3.0) {$k_{ac}\mathrel{+}=(ab|cd)D_{bd}$};
\node[anchor=west] at (12.2, -3.0) {$k_{ad}\mathrel{+}=(ab|cd)D_{bc}$};
\node[anchor=west] at (9.55, -3.38) {$k_{bc}\mathrel{+}=(ab|cd)D_{ad}$};
\node[anchor=west] at (12.2, -3.38) {$k_{bd}\mathrel{+}=(ab|cd)D_{ac}$};
\node[anchor=west] at (9.55, -3.76) {$j_{cd}\mathrel{+}=(ab|cd)D_{ab}$};
\node[shared, minimum height=3.6mm, inner sep=1.2pt, anchor=west] at (12.25, -3.76) {$j_{ab}$};
\node[anchor=west] at (12.62, -3.76) {$\mathrel{+}=(ab|cd)D_{cd}$};
\node[global, anchor=west] at (9.6, -4.27) {$K_{ac}$};
\node[anchor=west] at (10.12, -4.27) {$\mathrel{+}=k_{ac},\ \ldots,$};
\node[global, anchor=west] at (11.65, -4.27) {$J_{cd}$};
\node[anchor=west] at (12.14, -4.27) {$\mathrel{+}=j_{cd}$};
\draw[decorate, decoration={brace, amplitude=4pt}, line width=0.6pt] (14.85, -2.85) -- (14.85, -4.45);
\node[anchor=west, align=left] at (15.0, -3.3) {$\forall\,a\in a,\ b\in b,$\\$\phantom{\forall\,}c\in c,\ d\in d$};
\node[anchor=west, align=left] at (15.0, -4.12) {One atomic update\\per element and thread};
% J_ab after the kets
\node[anchor=north west, align=left] at (14.75, 0.22) {After the kets,\\summed over the threads,\\once per block:};
\node[global, anchor=west] at (15.0, -1.17) {$J_{ab}$};
\node[anchor=west] at (15.45, -1.17) {$\mathrel{+}=$};
\node[shared, minimum height=3.6mm, inner sep=1.2pt, anchor=west] at (15.9, -1.17) {$j_{ab}$};
\end{scope}
\node[ptitle] at (9.4, 0.47) {(c) Block execution};

\fslegend{0.05}{-4.95}
\end{tikzpicture}}
\caption{opt-Brc in ForSCF. (a) One kernel per pair of shell-pair batches with $ab\le cd$, and batch pairs whose Schwarz products cannot reach $\taub$ are not launched (gray). (b) For every bra pair, the ket pairs whose quartet passes the density test of Eq.~\ref{eq:brcskip} are listed. (c) A thread block owns a bra pair, and each thread computes all integrals of one contracted quartet by HGP with mid-contraction and adds them to the Fock blocks. Drawn after the scheme of Ref.~\citenum{Palethorpe2024}.}
\label{fig:brcexec}
\end{figure}
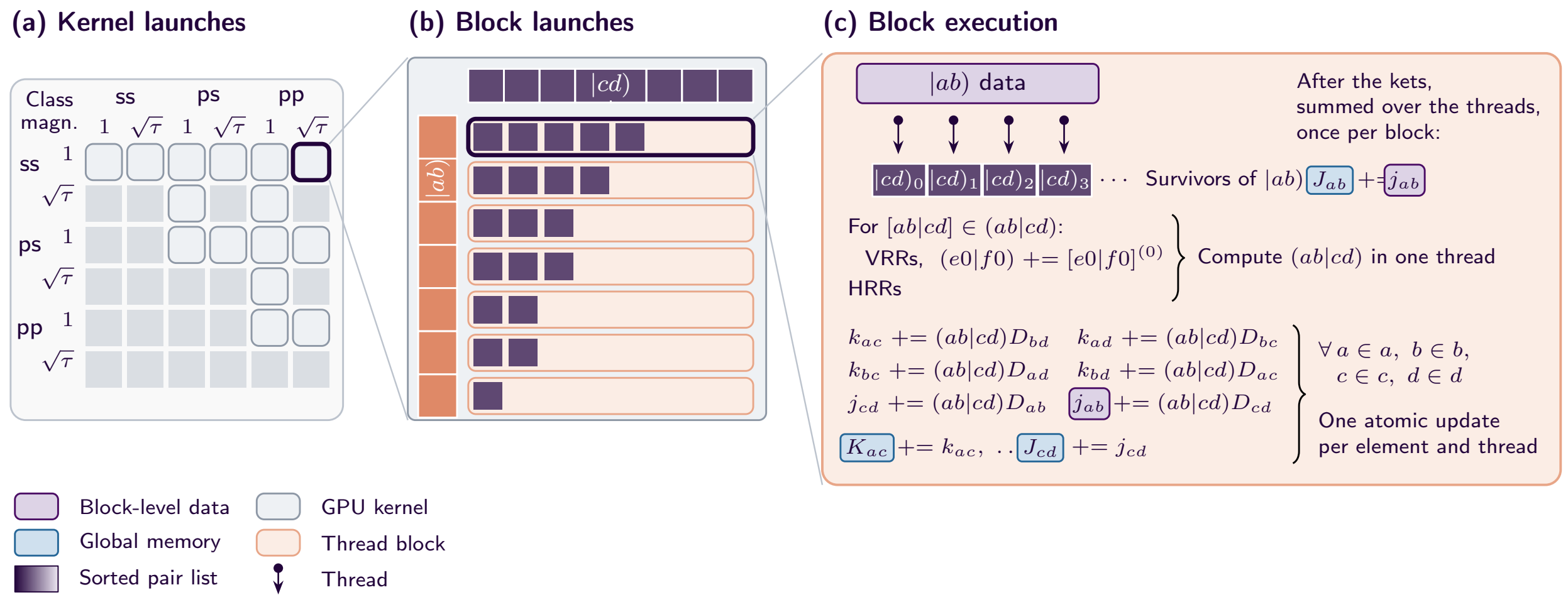

The GPU implementations are described in CUDA terms:\cite{CUDA} threads run in warps of 32 that follow one instruction stream, and warps are grouped into thread blocks that share a fast on-chip memory. HIP and OpenACC have equivalent concepts.\cite{HIP,OpenACC33}
\\
In the implementation of Ref.~\citenum{Palethorpe2024}, significant shell pairs are grouped into batches of equal angular momentum and contraction degree, so that all threads of a kernel evaluate the same recurrences at a similar cost.
The batches are further split by magnitude, so that the Schwarz part of Eq.~\ref{eq:brcskip} is applied to whole pairs of batches before launch, and kernels that could only produce negligible integrals are never launched.
Only pairs of batches with $ab\le cd$ are considered, following the bra--ket symmetry.
Each thread computes all integrals of one contracted quartet, and the threads of a block share one bra pair.
The full test of Eq.~\ref{eq:brcskip} is applied to each quartet inside the kernel, and the ket pairs are sorted by $G_{cd}\|D_{cd}\|$, so that neighboring threads tend to pass or fail it together.
Each thread adds its integrals to private copies of the six blocks of Eq.~\ref{eq:digest}, which are then added to the Fock matrix atomically.
\\ \\
ForSCF follows this design (Figure~\ref{fig:brcexec}) and changes it in three places, partly because OpenACC leaves the use of on-chip memory to the compiler and does not let a single thread leave a loop on its own.
First, the density test is separated from the integral evaluation. For every bra pair, the ket pairs that pass Eq.~\ref{eq:brcskip} are listed before any integral is evaluated (Figure~\ref{fig:brcexec}b).
All threads of a warp then work on significant quartets, which removes the divergence that sorting only reduces, and the ket pairs can keep the order of their shells, which makes the reads of the density local.
Since $J_{ab}$ is common to all threads of a block, it is summed over the threads and added once. \\
Second, the largest classes are split into parts that are evaluated separately, so that fewer intermediates are live at once.
On the A100 used in Ref.~\citenum{Palethorpe2024}, the large L2 cache absorbs the register spills of these classes, but on GPUs with smaller caches, such as the V100 used in this work, the spills dominate the cost.
The split is chosen from the recurrence graphs (Section~\ref{sec:codegen}). \\
Third, in the exchange build, shells on one atom that share their exponents, such as 1s and 2s in cc-pV$X$Z, are treated as one family, whose quartets are all computed from a single pass over the primitives.
The Supporting Information lists every difference from the published design, with its reason.

\section{The opt-UM Scheme}
\label{sec:um}
The opt-UM scheme was first described by Ufimtsev and Martinez\cite{Ufimtsev2009} and later optimized by Palethorpe et al.\cite{Palethorpe2024}
\subsection{McMurchie--Davidson Recursive Scheme}
\label{sec:md}
opt-UM evaluates integrals with the McMurchie--Davidson (MD) scheme.\cite{McMurchie1978,Palethorpe2024}
The product of the Cartesian factors of a primitive pair is expanded in Hermite Gaussians centered at $\vect{P}$.
They are indexed by a Hermite vector $\vect{p}=(p_x,p_y,p_z)$, with $\vect{0}\le\vect{p}\le\vect{a}+\vect{b}$ componentwise, and by $\vect{q}$ for the ket:
\begin{equation}
[\vect{ab}|\vect{cd}] = \sum_{\vect{p}}^{\vect{a}+\vect{b}}\sum_{\vect{q}}^{\vect{c}+\vect{d}} E^{\vect{ab}}_{\vect{p}}\,[\vect{p}|\vect{q}]\,E^{\vect{cd}}_{\vect{q}} .
\label{eq:md}
\end{equation}
The expansion coefficients factor over the Cartesian directions and follow, for $x$,
\begin{align}
E^{\vect{a}+\vect{1}_x,\vect{b}}_{\vect{p}} &= \frac{1}{2\zeta}\,E^{\vect{ab}}_{\vect{p}-\vect{1}_x} + PA_x\,E^{\vect{ab}}_{\vect{p}} + (p_x+1)\,E^{\vect{ab}}_{\vect{p}+\vect{1}_x}, \label{eq:erec}\\
E^{\vect{00}}_{\vect{p}} &= \delta_{\vect{p}\vect{0}}, \nonumber
\end{align}
with the analogous relation for $\vect{b}$ ($PB_x$ in place of $PA_x$).
They depend only on the primitive pair and are computed before the SCF.\cite{Palethorpe2024}
The two-center Hermite integrals reduce to one-center ones $[\vect{r}]^{(m)}$, with $\vect{r}=\vect{p}+\vect{q}$ and $q=q_x+q_y+q_z$,
\begin{align}
[\vect{p}|\vect{q}] &= (-1)^{q}\,[\vect{p}+\vect{q}]^{(0)}, \label{eq:rrec}\\
[\vect{r}+\vect{1}_x]^{(m)} &= PQ_x\,[\vect{r}]^{(m+1)} + r_x\,[\vect{r}-\vect{1}_x]^{(m+1)}, \nonumber
\end{align}
and likewise for $y$ and $z$, starting from
\begin{equation}
[\vect{0}]^{(m)} = U_P U_Q\,\frac{2\vartheta}{\sqrt{\pi}}\,(-2\vartheta^2)^m F_m(T).
\label{eq:r0}
\end{equation}
Eq.~\ref{eq:rrec} leaves open the direction in which each $[\vect{r}]^{(m)}$ is reduced.
As for HGP, the choice defines a DAG, whose size sets the FLOP count and the number of live intermediates.\cite{Johnson1991,Ryu1993}
Palethorpe and Barca reduce along the smallest non-zero component of $\vect{r}$ (the minimum-$L$ direction, MILD, heuristic) and prune the DAG with the sieve of Gill et~al.\cite{Gill1989}
For high $L$, they split the targets of a class by direction and magnitude into batches evaluated by separate kernels.\cite{Palethorpe2025}

\subsubsection{opt-UM Coulomb Matrix}
\label{sec:mdj}
The density enters the Coulomb matrix only on the ket, so the MD expansion allows pre- and post-contraction at the pair level:\cite{Ufimtsev2009,Palethorpe2024}
\begin{align}
D_{\vect{q}} &= \sum_{\vect{cd}}E^{\vect{cd}}_{\vect{q}}\,D_{\vect{cd}}, \label{eq:jmd}\\
J_{\vect{p}} &= \sum_{\vect{q}}[\vect{p}|\vect{q}]\,D_{\vect{q}}, \nonumber\\
J_{[\vect{ab}]} &= \sum_{\vect{p}} E^{\vect{ab}}_{\vect{p}}\,J_{\vect{p}}. \nonumber
\end{align}
Only $J_{\vect{p}}$ is formed at the quartet level, directly from Hermite integrals, while $D_{\vect{q}}$ and $J_{[ab]}$ are formed per pair, by batched BLAS calls, one per class.
The Coulomb build screens primitive quartets: a quartet is skipped when
\begin{equation}
G_{[ab]}\,G_{[cd]}\,\|D_{cd}\| < \tau_J .
\label{eq:jskip}
\end{equation}

\subsubsection{opt-UM Exchange Matrix}
\label{sec:mdk}
For exchange, Ref.~\citenum{Palethorpe2024} uses only the bra--ket symmetry $[ac|bd]=[bd|ac]$, so that each contribution to $K_{ab}$ is screened against the single density block $D_{cd}$.
Since $D_{cd}$ couples bra and ket, the test cannot be factored into a bra and a ket part.
It relies instead on the decay of the density matrix with distance,\cite{Kohn1959,Maslen1998,IsmailBeigi1999}
\begin{equation}
|D_{cd}| \le H\,e^{-\Lambda|\vect{CD}|},
\label{eq:decay}
\end{equation}
with $H$ and $\Lambda$ fitted in every iteration to the largest $|D|$ in bins of $|\vect{CD}|$.
For each primitive shell $a$, the extent $R_a$ and the largest Schwarz factor $G_a$ of its significant partners are
\begin{align}
R_a &= \max_c\left\{|\vect{AC}| : G_{[ac]} \ge \tau_J\right\}, \label{eq:ra}\\
G_a &= \max_c G_{[ac]} . \nonumber
\end{align}

Every significant contribution has $|\vect{CD}|\ge|\vect{AB}|-R_a-R_b$, so that
\begin{equation}
|K_{[ab]}| \le N^{ab}_{cd}\,H\,e^{-\Lambda(|\vect{AB}|-R_a-R_b)}\,G_a G_b ,
\label{eq:kbound}
\end{equation}
where $N^{ab}_{cd}$, the number of significant ket pairs, is bounded by a constant $N^*$.
A block is skipped before any integral is evaluated when
\begin{equation}
H\,e^{-\Lambda(|\vect{AB}|-R_a-R_b)}\,G_a G_b < \taustar, \qquad \taustar = \tau_K/N^* .
\label{eq:kskip}
\end{equation}
$\tau_K$ is therefore a target for the error of each block of $\K$, unlike $\tau_J$ and $\taub$, which bound single contributions.

Ref.~\citenum{Palethorpe2024} specifies neither the bins of the fit of Eq.~\ref{eq:decay} nor how $N^*$ is obtained.
ForSCF uses bins of fixed width, fits only the exponential regime of the binned maxima, and evaluates $N^*$ from the fitted model on the first build.
The first two choices matter for long chains rather than for large molecules in general, because they depend on the largest distance $|\vect{CD}|$ and not on the number of atoms (e.g., about \SI{44}{\angstrom} for gly$_{120}$ against \SI{24}{\angstrom} for (H$_2$O)$_{200}$). With a fixed number of bins, the bins widen with the length of the molecule, so that ever fewer of them describe the exponential decay (7 of 50 for gly$_{120}$), and the flat tail of the binned maxima at the level of numerical noise, if it is included in the fit, lowers the fitted $\Lambda$. In a first version that did both, the bound of Eq.~\ref{eq:kbound} decayed so slowly for the longer chains that Eq.~\ref{eq:kskip} no longer removed any block.
ForSCF also adds two tests.
The block bound of Luehr et al., $|K_{ab}|\le(\mathbf{G}\mathbf{D}\mathbf{G})_{ab}$,\cite{Luehr2016} which Ref.~\citenum{Palethorpe2024} discusses but does not use, removes further blocks.
In addition, the last groups of pairs of a block, whose summed bound fits a per-element error budget $\tau_e$, are skipped.
On compact water clusters the decay bound removes no blocks, since its cut-off distance exceeds the diameter of the cluster, and these two tests are what remove work there.

\subsection{GPU Implementation of opt-UM}
\label{sec:umgpu}
\subsubsection{opt-UM Coulomb Matrix on the GPU}
\label{sec:umgpuj}
In Ref.~\citenum{Palethorpe2024}, each warp computes $J_{\vect{p}}$ for one primitive bra pair $[ab|$ and loops over the ket pairs $|cd]$ of a class.
The bra pairs are sorted by $G_{[ab]}$ and the ket pairs by $G_{[cd]}\|D_{cd}\|$, so that the loop stops at the first ket that fails Eq.~\ref{eq:jskip} and all threads of a warp stop together.
The partial $J_{\vect{p}}$ of the threads are summed over the warp and written without atomic operations.
$D_{\vect{q}}$ and $J_{[ab]}$ are formed per class by batched matrix products.

\begin{figure}[tbp]
\centering
\resizebox{\textwidth}{!}{\begin{tikzpicture}[fs, x=1cm, y=1cm]
% ================================================================ kernel launches
\draw[back] (0,0) rectangle (3.55,-3.25);
\foreach \l [count=\i] in {ss, ps, pp} {
  \node at (0.55 + 0.82*\i, -0.3) {\l};
  \node at (0.4, -0.25 - 0.82*\i) {\l};
}
\foreach \i in {1,2,3} {\foreach \j in {1,2,3} {
  \draw[kernel] ({0.2 + 0.82*\j}, {-0.6 - 0.82*(\i-1)}) rectangle ++(0.68, -0.68);
}}
\node[ptitle] at (0, 0.47) {(a) Kernel launches};
\draw[focus] (2.66, -0.6) rectangle ++(0.68, -0.68);

% ================================================================ gang launches
\draw[zoom] (3.34, -0.60) -- (4.35, 0.25);
\draw[zoom] (3.34, -1.28) -- (4.35, -3.75);
\draw[kernel] (4.35, 0.25) rectangle (8.85, -3.75);
% ket list: factor-2 bins, decreasing
\foreach \c [count=\b from 0] in {100, 80, 62, 45, 28, 12} {
  \fill[fsKetD!\c!fsKetL] (5.15 + 0.6*\b, 0.12) rectangle ++(0.6, -0.38);
  \draw[white, line width=0.6pt] (5.15 + 0.6*\b, 0.12) rectangle ++(0.6, -0.38);
}
% bra list: decreasing
\foreach \c [count=\r from 0] in {100, 86, 72, 58, 44, 30, 18, 8} {
  \fill[fsBraD!\c!fsBraL] (4.5, -0.42 - 0.405*\r) rectangle ++(0.5, -0.405);
  \draw[white, line width=0.6pt] (4.5, -0.42 - 0.405*\r) rectangle ++(0.5, -0.405);
}
% gangs of two workers, with the cut-off of each worker
\foreach \g in {0,...,3} {
  \draw[gblock] (5.15, -0.42 - 0.81*\g) rectangle (8.75, -1.23 - 0.81*\g);
  \draw[fsBlockEdge, line width=0.3pt] (5.15, -0.825 - 0.81*\g) -- (8.75, -0.825 - 0.81*\g);
}
\foreach \cut [count=\r from 0] in {8.6, 8.35, 8.0, 7.7, 7.15, 6.85, 6.2, 5.75} {
  \draw[fsExit, line width=1.1pt] (\cut, -0.47 - 0.405*\r) -- ++(0, -0.31);
}
\node[text=white, font=\sffamily\fsnorm] at (5.5, -0.07) {$|cd]$};
\node[text=white, rotate=90, font=\sffamily\fsnorm] at (4.75, -0.95) {$[ab|$};
\node[ptitle] at (4.35, 0.47) {(b) Block launches};
\draw[focus] (5.15, -0.42) rectangle (8.75, -1.23);

% ================================================================ gang execution
\draw[zoom] (8.75, -0.42) -- (9.5, 0.3);
\draw[zoom] (8.75, -1.23) -- (9.5, -4.0);
\draw[gexec] (9.5, 0.3) rectangle (17.5, -4.0);
\node[shared, minimum width=2.0cm] (jp) at (11.3, -0.05) {$J_{\vect{p}}$};
\node[shared, minimum width=2.0cm] (jq) at (14.6, -0.05) {$J_{\vect{p}'}$};
\fslanes{10.75}{-0.42}{5}
\fslanes{14.05}{-0.42}{5}
\node[small, anchor=west] at (14.75, -0.65) {Warp 2:};
\node[small, anchor=west] at (14.75, -0.95) {Next bra};
% ket list of the worker
\foreach \c [count=\k from 0] in {100, 90, 80, 70, 60, 50, 40, 30, 20} {
  \fill[fsKetD!\c!fsKetL] (9.85 + 0.62*\k, -1.45) rectangle ++(0.62, -0.42);
  \draw[white, line width=0.6pt] (9.85 + 0.62*\k, -1.45) rectangle ++(0.62, -0.42);
}
\foreach \k/\t in {0/{[cd]_0}, 1/{[cd]_1}, 2/{[cd]_2}} {
  \node[text=white, font=\sffamily\fssmall] at (10.16 + 0.62*\k, -1.66) {$\t$};
}
\node[font=\sffamily\fssmall] at (12.02, -1.66) {$\cdots$};
\foreach \k in {0,...,3} {
  \draw[->, fsInk, line width=0.4pt] (11.1, -0.95) .. controls (10.9 + 0.3*\k, -1.15) .. (10.16 + 0.62*\k, -1.43);
}
% cut-off
\draw[fsExit, line width=1.3pt] (14.85, -1.38) -- ++(0, -0.56);
\fill[white, opacity=0.75] (14.86, -1.45) rectangle (15.43, -1.87);
\draw[->, fsExit, line width=0.8pt] (15.2, -2.28) -- (14.93, -1.97);
\node[exit, anchor=north west] at (13.98, -2.3) {Cut-off, found before the loop:\\last ket with\\$G_{[ab]}G_{[cd]}\max|D_{cd}|$\\$\ge\tau_J$};
% accumulation and reduction
\node[shared, minimum height=4mm, inner sep=1.5pt] (acc) at (10.25, -2.45) {$J_{\vect{p}}$};
\node[anchor=west] at (10.62, -2.45) {$\mathrel{+}=\,[\vect{p}|\vect{q}]$};
\node[global, anchor=west] at (11.9, -2.45) {$D_{\vect{q}}$};
\node[small, anchor=west] at (9.62, -3.1) {Reduction};
\foreach \k in {0,...,3} {
  \draw[thread] (11.15 + 0.13*\k, -2.85) -- ++(0, -0.32);
  \draw[fsSharedEdge, line width=0.5pt] (11.2 + 0.13*\k, -3.18) -- (11.4, -3.35);
}
\node[global, inner sep=1.5pt] (out) at (10.15, -3.62) {$J_{\vect{p}}$};
\node[anchor=west] at (10.45, -3.62) {$=$};
\node[shared, minimum height=4mm, inner sep=1.5pt, anchor=west] at (10.8, -3.62) {$J_{\vect{p}}$};
\draw[decorate, decoration={brace, amplitude=4pt}, line width=0.6pt] (12.5, -2.25) -- (12.5, -3.8);
\node[anchor=west] at (12.62, -3.02) {$\forall\vect{p}\in[ab|$};
\node[ptitle] at (9.5, 0.47) {(c) Block execution};

\fslegend{0.05}{-4.35}
\end{tikzpicture}}
\caption{opt-UM Coulomb build in ForSCF. (a) One kernel per bra and ket class. (b) Bras are sorted by $G_{[ab]}$, kets grouped by magnitude and ordered spatially within a group, and the cut-off of each bra is found before its loop. (c) Each warp owns one primitive bra, and its threads accumulate $J_{\vect{p}}$, which is summed over the warp. Drawn after the scheme of Refs.~\citenum{Palethorpe2024} and~\citenum{Ufimtsev2009}.}
\label{fig:umj}
\end{figure}
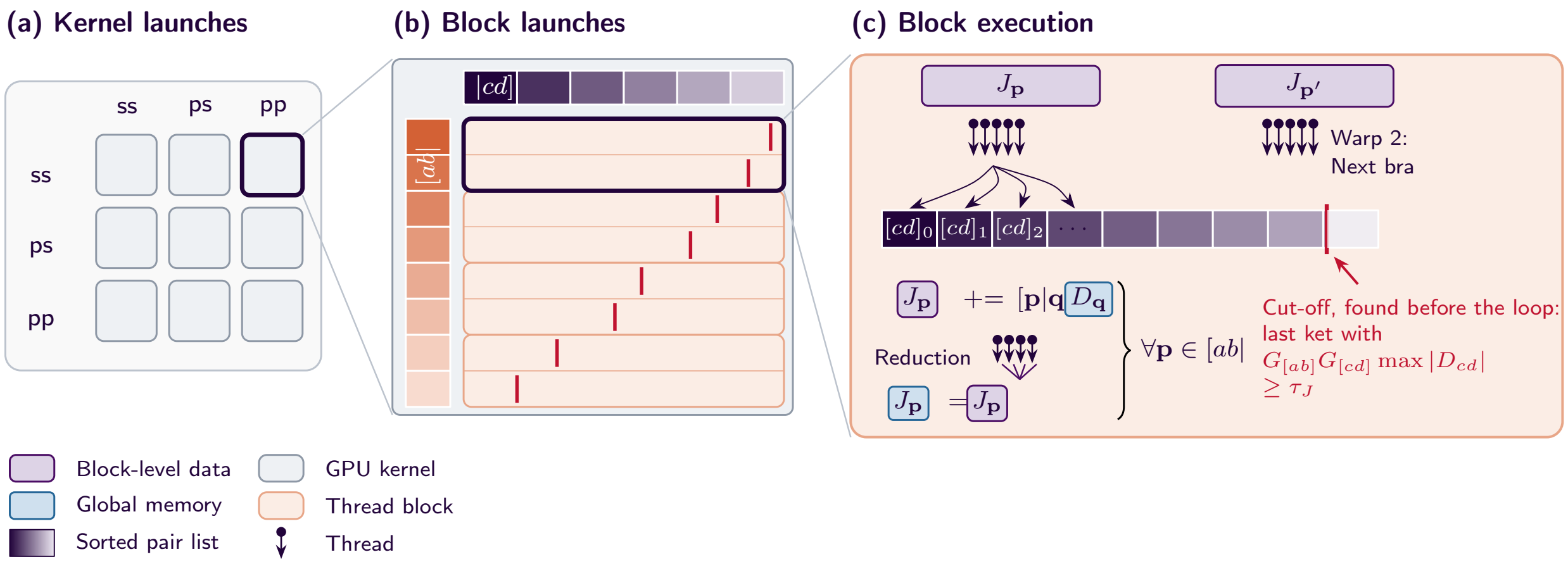

ForSCF follows this design (Figure~\ref{fig:umj}) but is limited by the lack of per-thread exit in OpenACC. Since a thread cannot leave the loop on its own, the last ket that can pass Eq.~\ref{eq:jskip} is found before the loop, and the threads of the warp share the kets up to it.
Also, the ket pairs are grouped by magnitude and ordered spatially within a group, so that neighboring threads read neighboring data and the few near-field kets, for which the Boys function is more expensive, are processed together.

\subsubsection{opt-UM Exchange Matrix on the GPU}
\label{sec:umgpuk}
In Ref.~\citenum{Palethorpe2024}, each thread block computes one block $K_{ab}$ of contracted shells.
Its threads share a bra pair $[ac|$, and each takes a different ket pair $|bd]$, and their private copies of $K_{ab}$ are summed at the end.
The block test of Eq.~\ref{eq:kskip} is applied before any integral is evaluated.
Inside the block, the bound is tightened, with $G_a$ replaced by $G_{[ac]}$ in the loop over $[ac|$ and $G_b$ by $G_{[bd]}$ in the loop over $|bd]$.
Since the pairs are sorted by their Schwarz factors, all threads leave these loops together, and the cost of exchange grows linearly with system size.

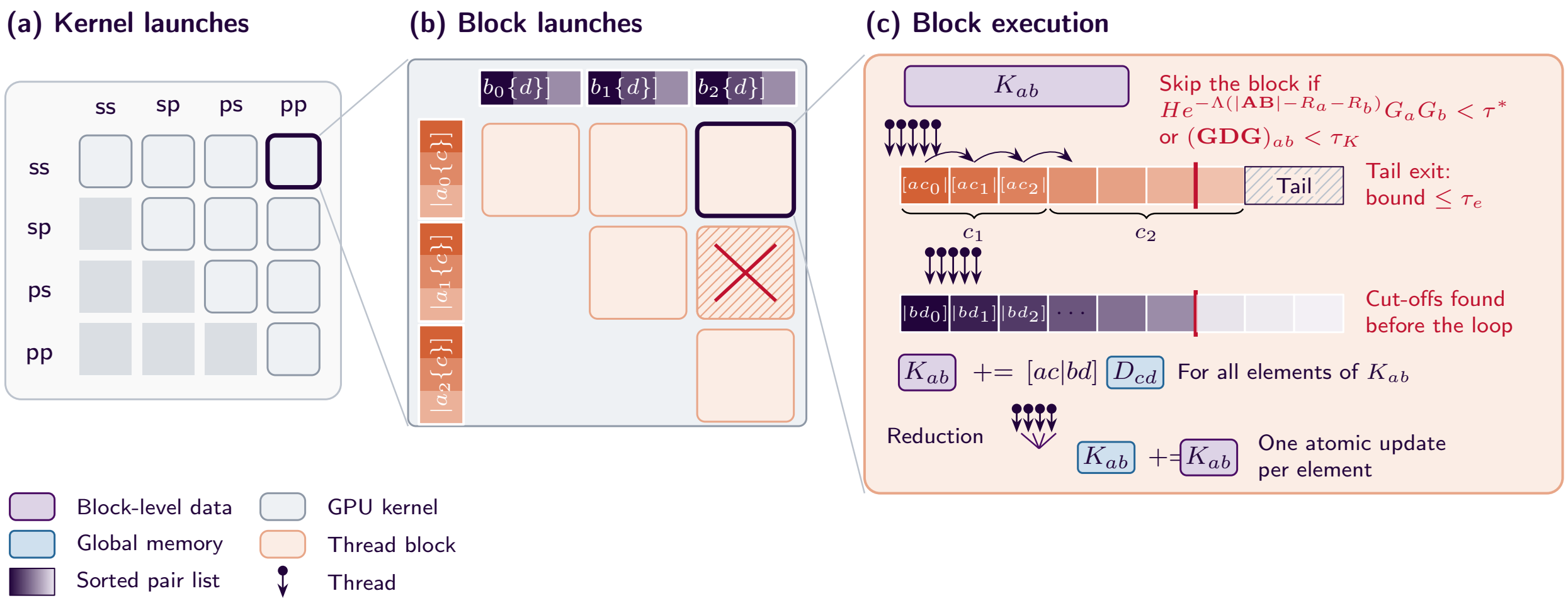
\begin{figure}[tbp]
\centering
\resizebox{\textwidth}{!}{\begin{tikzpicture}[fs, x=1cm, y=1cm]
% ================================================================ kernel launches
\draw[back] (0,0) rectangle (3.75,-3.55);
\foreach \l [count=\i] in {ss, sp, ps, pp} {
  \node at (0.42 + 0.7*\i, -0.3) {\l};
  \node at (0.38, -0.27 - 0.7*\i) {\l};
}
\foreach \i in {1,...,4} {\foreach \j in {1,...,4} {
  \pgfmathtruncatemacro{\up}{\j >= \i ? 1 : 0}
  \ifnum\up=1
    \draw[kernel] ({0.13 + 0.7*\j}, {-0.6 - 0.7*(\i-1)}) rectangle ++(0.58, -0.58);
  \else
    \fill[fsOff] ({0.13 + 0.7*\j}, {-0.6 - 0.7*(\i-1)}) rectangle ++(0.58, -0.58);
  \fi
}}
\node[ptitle] at (0, 0.47) {(a) Kernel launches};
\draw[focus] (2.93, -0.6) rectangle ++(0.58, -0.58);

% ================================================================ gang launches
\draw[zoom] (3.51, -0.6) -- (4.5, 0.25);
\draw[zoom] (3.51, -1.18) -- (4.5, -3.85);
\draw[kernel] (4.5, 0.25) rectangle (8.95, -3.85);
\foreach \j in {0,1,2} {
  \foreach \c [count=\k from 0] in {100, 70, 40} {
    \fill[fsKetD!\c!fsKetL] (5.3 + 1.2*\j + 0.38*\k, 0.13) rectangle ++(0.38, -0.4);
  }
  \draw[white, line width=0.8pt] (5.3 + 1.2*\j, 0.13) rectangle ++(1.14, -0.4);
  \node[text=white, font=\sffamily\fssmall] at (5.68 + 1.2*\j, -0.07) {$|b_{\j}\{d\}]$};
}
\foreach \i in {0,1,2} {
  \foreach \c [count=\k from 0] in {100, 70, 40} {
    \fill[fsBraD!\c!fsBraL] (4.62, -0.42 - 1.15*\i - 0.37*\k) rectangle ++(0.5, -0.37);
  }
  \draw[white, line width=0.8pt] (4.62, -0.42 - 1.15*\i) rectangle ++(0.5, -1.11);
  \node[text=white, rotate=90, font=\sffamily\fssmall] at (4.87, -0.98 - 1.15*\i) {$|a_{\i}\{c\}]$};
}
\foreach \i in {0,1,2} {\foreach \j in {0,1,2} {
  \pgfmathtruncatemacro{\up}{\j >= \i ? 1 : 0}
  \ifnum\up=1
    \draw[gblock] (5.33 + 1.2*\j, -0.47 - 1.15*\i) rectangle ++(1.08, -1.03);
  \fi
}}
% one block skipped at gang entry
\draw[gblock, pattern=north east lines, pattern color=fsBlockEdge] (7.73, -1.62) rectangle ++(1.08, -1.03);
\draw[fsExit, line width=1.1pt] (7.93, -1.82) -- ++(0.68, -0.63) (8.61, -1.82) -- ++(-0.68, -0.63);
\node[ptitle] at (4.5, 0.47) {(b) Block launches};
\draw[focus] (7.73, -0.47) rectangle ++(1.08, -1.03);

% ================================================================ gang execution
\draw[zoom] (8.81, -0.47) -- (9.6, 0.3);
\draw[zoom] (8.81, -1.5) -- (9.6, -4.6);
\draw[gexec] (9.6, 0.3) rectangle (17.4, -4.6);
\node[shared, minimum width=2.5cm] (kab) at (11.3, -0.05) {$K_{ab}$};
\node[exit, anchor=north west] at (12.75, 0.22) {Skip the block if\\$He^{-\Lambda(|\vect{AB}|-R_a-R_b)}G_aG_b<\tau^*$\\or $(\mathbf{GDG})_{ab}<\tau_K$};
% [ac| walk, groups by shell c
\fslanes{9.75}{-0.42}{5}
\foreach \c [count=\k from 0] in {100, 88, 76, 64, 52, 40, 28} {
  \fill[fsBraD!\c!fsBraL] (10.0 + 0.55*\k, -0.95) rectangle ++(0.55, -0.42);
  \draw[white, line width=0.6pt] (10.0 + 0.55*\k, -0.95) rectangle ++(0.55, -0.42);
}
\foreach \k/\t in {0/{[ac_0|}, 1/{[ac_1|}, 2/{[ac_2|}} {\node[text=white, font=\sffamily\fstiny] at (10.275 + 0.55*\k, -1.16) {$\t$};}
\draw[fsInk, line width=0.4pt, pattern=north east lines, pattern color=fsBackEdge] (13.85, -0.95) rectangle ++(1.1, -0.42);
\node[small, fill=fsBlock, inner sep=1pt] at (14.4, -1.16) {Tail};
\foreach \k in {0,...,2} {\draw[->, fsInk, line width=0.4pt] (10.275 + 0.55*\k, -0.9) to[bend left=40] (10.825 + 0.55*\k, -0.9);}
\draw[decorate, decoration={brace, mirror, amplitude=3pt}, line width=0.5pt] (10.02, -1.42) -- (11.63, -1.42) node[midway, below=2pt, small] {$c_1$};
\draw[decorate, decoration={brace, mirror, amplitude=3pt}, line width=0.5pt] (11.67, -1.42) -- (13.83, -1.42) node[midway, below=2pt, small] {$c_2$};
\draw[fsExit, line width=1.3pt] (13.3, -0.9) -- ++(0, -0.52);
\node[exit, anchor=west] at (15.05, -1.16) {Tail exit:\\bound $\le\tau_e$};
% |bd] lanes
\fslanes{10.2}{-1.85}{5}
\foreach \c [count=\k from 0] in {100, 88, 76, 64, 52, 40, 28, 18, 10} {
  \fill[fsKetD!\c!fsKetL] (10.0 + 0.55*\k, -2.38) rectangle ++(0.55, -0.42);
  \draw[white, line width=0.6pt] (10.0 + 0.55*\k, -2.38) rectangle ++(0.55, -0.42);
}
\foreach \k/\t in {0/{|bd_0]}, 1/{|bd_1]}, 2/{|bd_2]}} {\node[text=white, font=\sffamily\fstiny] at (10.275 + 0.55*\k, -2.59) {$\t$};}
\node[font=\sffamily\fssmall] at (11.93, -2.59) {$\cdots$};
\draw[fsExit, line width=1.3pt] (13.3, -2.33) -- ++(0, -0.52);
\fill[white, opacity=0.7] (13.31, -2.38) rectangle (14.95, -2.8);
\node[exit, anchor=west] at (15.05, -2.59) {Cut-offs found\\before the loop};
% accumulation
\node[shared, minimum height=4mm, inner sep=1.5pt] at (10.3, -3.25) {$K_{ab}$};
\node[anchor=west] at (10.66, -3.25) {$\mathrel{+}=\,[ac|bd]$};
\node[global, anchor=west] at (12.3, -3.25) {$D_{cd}$};
\node[small, anchor=west] at (9.7, -3.95) {Reduction};
\foreach \k in {0,...,3} {
  \draw[thread] (11.3 + 0.13*\k, -3.6) -- ++(0, -0.32);
  \draw[fsSharedEdge, line width=0.5pt] (11.35 + 0.13*\k, -3.93) -- (11.55, -4.1);
}
\node[global, inner sep=1.5pt] at (12.3, -4.2) {$K_{ab}$};
\node[anchor=west] at (12.62, -4.2) {$\mathrel{+}=$};
\node[shared, minimum height=4mm, inner sep=1.5pt, anchor=west] at (13.12, -4.2) {$K_{ab}$};
\node[small, anchor=west, align=left] at (13.85, -4.2) {One atomic update\\per element};
\node[small, anchor=west] at (12.95, -3.25) {For all elements of $K_{ab}$};
\node[ptitle] at (9.6, 0.47) {(c) Block execution};

\fslegend{0.05}{-4.75}
\end{tikzpicture}}
\caption{opt-UM exchange build in ForSCF. (a) One kernel per class with $[ac|bd]=[bd|ac]$ in the upper triangle, and one thread block per contracted block $K_{ab}$. (b) The walk of one thread block: the block tests at entry, common cut-offs for the groups of $[ac|$ pairs that share the shell $c$ and for the $|bd]$ pairs of the threads, and the tail exit. (c) The work of one thread per visit. Drawn after the scheme of Refs.~\citenum{Palethorpe2024} and~\citenum{Ufimtsev2009}.}
\label{fig:umk}
\end{figure}

ForSCF follows this design (Figure~\ref{fig:umk}) with important refinements.
First, the block tests at entry include the bound of Luehr et al., as previously mentioned, and the $[ac|$ pairs are visited in groups that share the shell $c$.
As in the Coulomb build, the cut-offs of both loops are common to all threads and found before each loop, and, inside the block, the actual distances and density elements replace the bounds of Eq.~\ref{eq:kbound} wherever they are known.
The elements of $K_{ab}$ are added to $\K$ atomically, since blocks of different classes update the same elements.
For generally contracted basis sets, the primitives shared by a family of shells are visited once for all its members.

The exchange evaluation keeps the most intermediates live, and for classes with $f$ shells it is limited by register spills.
This is the regime in which Ref.~\citenum{Palethorpe2024} recommends opt-Brc, and it motivates the mixed scheme (Section~\ref{sec:mixed}).
The Supporting Information lists every difference from the published design, with its reason.

\section{Implementation}
\label{sec:implementation}
\subsection{Overview}
\label{sec:overview}
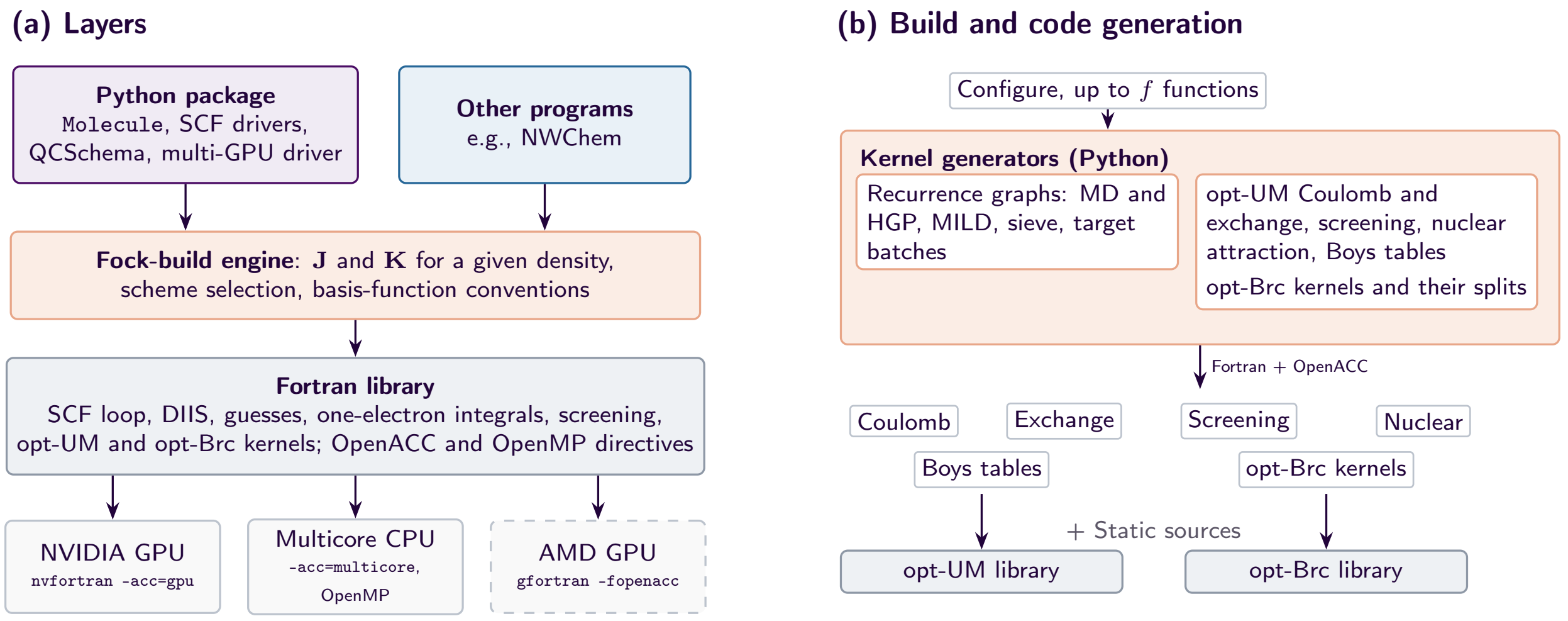
\begin{figure}[tbp]
\centering
\resizebox{\textwidth}{!}{\begin{tikzpicture}[fs, x=1cm, y=1cm,
  box/.style={rounded corners=2pt, align=center, inner sep=3pt, line width=0.6pt, font=\sffamily\fssmall},
  file/.style={draw=fsBackEdge, fill=white, rounded corners=1pt, inner sep=2pt, line width=0.5pt, font=\ttfamily\fssmall},
  link/.style={->, draw=fsInk, line width=0.6pt},
  lab/.style={font=\sffamily\fstiny, text=fsInk, fill=white, inner sep=1pt}]
% ================================================================ layers
\node[box, draw=fsSharedEdge, fill=fsShared!45, minimum width=3.55cm, minimum height=1.2cm] (py) at (1.85, -0.6)
  {\textbf{Python package}\\\texttt{Molecule}, SCF drivers,\\QCSchema, multi-GPU driver};
\node[box, draw=fsGlobalEdge, fill=fsGlobal!45, minimum width=3.0cm, minimum height=1.2cm] (nw) at (5.55, -0.6)
  {\textbf{Other programs}\\e.g., NWChem};
\node[box, draw=fsBlockEdge, fill=fsBlock, minimum width=7.1cm, minimum height=0.9cm] (abi) at (3.6, -2.15)
  {\textbf{Fock-build engine}: $\J$ and $\K$ for a given density,\\scheme selection, basis-function conventions};
\node[box, draw=fsKernelEdge, fill=fsKernel, minimum width=7.1cm, minimum height=1.2cm] (lib) at (3.6, -3.6)
  {\textbf{Fortran library}\\SCF loop, DIIS, guesses, one-electron integrals, screening,\\opt-UM and opt-Brc kernels; OpenACC and OpenMP directives};
\draw[link] (py) -- (py |- abi.north);
\draw[link] (nw) -- (nw |- abi.north);
\draw[link] (abi) -- (lib);
\foreach \x/\t/\c/\d in {1.1/{NVIDIA GPU}/{\texttt{nvfortran -acc=gpu}}/solid, 3.6/{Multicore CPU}/{\texttt{-acc=multicore}, OpenMP}/solid, 6.1/{AMD GPU}/{\texttt{gfortran -fopenacc}}/dashed} {
  \node[box, draw=fsBackEdge, fill=fsBack, minimum width=2.15cm, text width=2.0cm, minimum height=0.95cm, \d] (b) at (\x, -5.15) {\t\\[-1pt]{\fstiny\c}};
  \draw[link] (lib.south -| b) -- (b);
}
\node[ptitle] at (0.05, 0.25) {(a) Layers};

% ================================================================ build
\node[file, font=\sffamily\fssmall] (cm) at (11.35, -0.25) {Configure, up to $f$ functions};
\node[box, draw=fsBlockEdge, fill=fsBlock, minimum width=7.4cm, minimum height=2.2cm, anchor=north] (gk) at (12.3, -0.65) {};
\node[anchor=north west, font=\sffamily\fssmall\bfseries] at (8.65, -0.7) {Kernel generators (Python)};
\node[box, draw=fsBlockEdge, fill=white, text width=3.1cm, anchor=north west, align=flush left] at (8.75, -1.1)
  {Recurrence graphs: MD and HGP, MILD, sieve, target batches};
\node[box, draw=fsBlockEdge, fill=white, text width=3.3cm, anchor=north west, align=flush left] at (12.25, -1.1)
  {opt-UM Coulomb and exchange, screening, nuclear attraction, Boys tables\\[2pt]opt-Brc kernels and their splits};
\draw[link] (cm) -- (cm |- gk.north);
\draw[link] (gk.south) -- node[lab, right=2pt, fill=none] {Fortran + OpenACC} ++(0, -0.45);
\foreach \x/\t in {9.25/Coulomb, 10.9/Exchange, 12.7/Screening, 14.6/Nuclear} {\node[file, font=\sffamily\fssmall] at (\x, -3.65) {\t};}
\foreach \x/\t in {10.05/Boys tables, 13.6/opt-Brc kernels} {\node[file, font=\sffamily\fssmall] at (\x, -4.15) {\t};}
\node[box, draw=fsKernelEdge, fill=fsKernel, minimum width=2.9cm] (so1) at (10.05, -5.2) {opt-UM library};
\node[box, draw=fsKernelEdge, fill=fsKernel, minimum width=2.9cm] (so2) at (13.6, -5.2) {opt-Brc library};
\draw[link] (10.05, -4.4) -- (so1);
\draw[link] (13.6, -4.4) -- (so2);
\node[small, text=fsGray] at (11.82, -4.78) {+ Static sources};
\node[ptitle] at (8.55, 0.25) {(b) Build and code generation};
\end{tikzpicture}}
\caption{(a) Layers of ForSCF and the programs that can use it as their Fock-build engine, where one Fortran source builds for NVIDIA GPUs and multicore CPUs. (b) At configure time, the generators build the recurrence graphs and write the kernels as Fortran with OpenACC directives.}
\label{fig:stack}
\end{figure}

ForSCF consists of a Python package, which defines molecules and basis sets, drives the SCF and offers a QCSchema interface, and a Fortran library with OpenACC directives, which contains the integrals, the Fock build and the SCF loop (Figure~\ref{fig:stack}a).
The library can also serve other programs as the engine of their Fock builds (Section~\ref{sec:hosts}).
Kernels specialized by angular momentum are generated in Python at configure time (Figure~\ref{fig:stack}b, Section~\ref{sec:codegen}). This article presents results up to $f$ functions, but higher angular momentum can be used.

One source builds for NVIDIA and AMD GPUs and for multicore CPUs.
The kernels contain no CUDA. Vendor library and runtime calls sit behind a backend guard with a portable fallback, and no register limits or vendor tuning flags are used.
Device data are held either in managed memory or under explicit OpenACC data clauses, selected at build time.

Basis sets are put in Davidson's optimized form:\cite{Davidson1996} a primitive that also forms a single-primitive shell of the same angular momentum is removed from the contracted shells.
The span is unchanged and the contractions are shorter, which matters most for the correlation-consistent sets.
In the benchmarks, every program reads the same basis functions (Section~\ref{sec:design}).

The SCF driver supports restricted and unrestricted HF.
The default guess is MINAO in the atom-blocked form of GPU4PySCF,\cite{Li2025} in which each atom's density is projected onto that atom's functions only, while SAD, the full MINAO projection and the core Hamiltonian are also available.\cite{VanLenthe2006,Lehtola2019}
Convergence is accelerated by DIIS.\cite{Pulay1980,Pulay1982}
Fock builds are full or incremental, and incremental builds screen relative to $\Delta\D$.

\subsection{Code Generation}
\label{sec:codegen}
For the MD integrals, the generator builds the DAG of $[\vect{r}]^{(m)}$ with the MILD heuristic and the sieve, and splits the targets of each class into directional batches, as in Ref.~\citenum{Palethorpe2025}.
The batches are evaluated one after another inside one kernel rather than as separate kernels, which bounds the number of live intermediates without repeating the screening of the exchange kernels in every batch.
For opt-Brc, the generator builds an HGP DAG of vertical-recurrence, contracted and horizontal-recurrence nodes, pruned by the sieve.
The VRRs reduce the ket index first, which gave the fewest live values at nearly the lowest FLOP count, and the HRRs follow Ref.~\citenum{Palethorpe2024}, ket before bra.
Intermediates are scheduled depth-first in the primitive, contracted and horizontal phases instead of with the priority queue of Ref.~\citenum{Palethorpe2024}, since the contracted accumulators that the mid-contraction keeps alive dominate the live set.
From the FLOPs and live values of every class the generator chooses the kernel splits of Section~\ref{sec:brcgpu}.

\subsection{Integral Engines and the Mixed Scheme}
\label{sec:engines}
\label{sec:mixed}
Three schemes are available: \texttt{um} builds $\J$ and $\K$ with opt-UM, \texttt{brc} with opt-Brc, and \texttt{mixed} builds $\J$ with opt-UM and $\K$ with opt-Brc.
The scheme is chosen per calculation.
All three share the density preparation, the distribution over GPUs and the assembly of $\F$, so that the rest of the program is the same for all.

Ref.~\citenum{Palethorpe2024} chooses one scheme per system: opt-Brc for small systems and highly contracted triple-$\zeta$ basis sets, and opt-UM for large double-$\zeta$ ones.
The mixed scheme instead chooses per matrix, because the two halves of the Fock build favor different schemes.
The Hermite pre- and post-contraction of Eq.~\ref{eq:jmd} makes the opt-UM Coulomb build much cheaper than the integral-by-integral digest of opt-Brc.
For contracted shells, on the other hand, opt-Brc exchange does far less work than the primitive-pair walk of opt-UM, whose kernels are also the ones bound by spills (Section~\ref{sec:umgpuk}).
Each half keeps its own threshold, $\tau_J$ for Coulomb and $\taub$ for exchange, since the two need different values for the same accuracy (Section~\ref{sec:sweep}).

Both schemes evaluate the Boys function from the same tables, which Ref.~\citenum{Palethorpe2024} leaves open.
The highest order needed is interpolated from tabulated Taylor expansions, the lower orders follow by downward recursion, and large arguments use the asymptotic form, with a relative error below $10^{-14}$.
The prefactors of Eq.~\ref{eq:r0} are folded into the returned values, which removes the divisions.

\subsection{Multi-GPU Support}
\label{sec:multigpu}
A Fock build can be divided into disjoint slices of the work, whose results add up.
The opt-UM Coulomb build deals the bra pairs of each class, sorted by their Schwarz bound, to the GPUs in turn; opt-UM exchange deals the blocks $K_{ab}$ of each class, sorted by estimated cost, to the GPUs in turn; and opt-Brc assigns each kernel launch whole to the GPU with the least work so far, sharing only the largest launches among all GPUs.
A Python driver runs one process per GPU, each of which builds one slice; it exchanges the density and the result through shared memory, keeps the SCF loop in the parent process, and needs neither MPI nor peer-to-peer links between the GPUs.
Every GPU holds a complete copy of the screening and pair data, so more GPUs make a build faster but do not raise the largest system that fits. A multi-GPU version that uses sliced memory to run larger systems is currently in development.

\subsection{Use as a Fock-Build Engine}
\label{sec:hosts}
Any quantum chemistry program that runs its own SCF can use ForSCF to build the two-electron part of its Fock matrix: it supplies a density and receives the Coulomb and exchange matrices.
The interface does not depend on the host program: the host describes its molecule and basis set once, and ForSCF converts between the basis-function conventions of the host, Cartesian or spherical, and its own.
The host keeps its guess, its solver, its convergence control and its properties, so that its algorithms are unchanged while the most expensive step of its SCF runs on the GPU, or on the CPU build of ForSCF.
NWChem 7.3.1\cite{Apra2020} serves here as the example, with its direct SCF handing its two-electron builds to ForSCF.
The energies agree with those of NWChem's own builder to $10^{-9}$--$2\times10^{-8}\Eh$ for Cartesian and spherical basis sets up to $f$ functions.

\section{Results and Discussion}
\label{sec:results}
All calculations ran on one node with an AMD Ryzen Threadripper PRO 7975WX processor (32 cores, 64 hardware threads), \SI{30}{GB} of memory and four NVIDIA V100-SXM2 GPUs with \SI{16}{GB} each (driver 580.159.04), and every timed calculation ran alone on the node, so that no other process could influence the timings.
ForSCF was compiled with the NVIDIA HPC SDK 26.3 (\texttt{nvfortran}, CUDA 13, compute capability 7.0) with managed memory and kernels up to $f$ functions, and for the CPU with OpenMP.
On the GPU, it is compared with GPU4PySCF 1.8.1 (PySCF 2.14.0)\cite{Li2025,Wu2025} and QUICK (git f33543c, built with CUDA 12.4.1, as its developers recommend for GPUs older than Hopper),\cite{Manathunga2021,Manathunga2023} and on the CPU with NWChem 7.3.1,\cite{Apra2020} PySCF 2.14.0\cite{Sun2020} and ORCA 6.1.1,\cite{Neese2020} with the settings of Table~\ref{tab:programs}.
QUICK was modified only to read an input density (Supporting Information).
GANSU,\cite{GANSU} released recently, and its successor GANSU2, currently in development, were used in early tests but are not part of the comparison, because their direct Fock builds have no non-incremental mode, so that a build on a fixed density is not a full build.
The test systems are the polyglycine chains gly$_n$ = H(C$_2$H$_3$NO)$_n$OH ($n$ = 5--120) and water clusters (H$_2$O)$_n$ ($n$ = 10--200) of Barca et al.,\cite{Barca2021} which are also used in the benchmarks of opt-UM and opt-Brc\cite{Palethorpe2024} and of GPU4PySCF,\cite{Li2025,Wu2025} with 15 sizes each. They are run with the STO-3G,\cite{Hehre1969} 6-31G*, 6-31G**,\cite{Hehre1972,Hariharan1973} cc-pVDZ and cc-pVTZ\cite{Dunning1989} basis sets in Cartesian functions. The largest system run has \num{9315} basis functions (gly$_{50}$/cc-pVTZ).
All calculations are closed-shell RHF.

\begin{table}[tbp]
\caption{Programs, screening, guess and convergence settings}
\label{tab:programs}
\centering\footnotesize
\begin{threeparttable}
\setlength{\tabcolsep}{4pt}
\begin{tabular}{L{0.15\textwidth}L{0.2\textwidth}L{0.14\textwidth}L{0.2\textwidth}L{0.2\textwidth}}
\toprule
program & screening value(s) & published & initial guess & convergence \\
\midrule
ForSCF opt-UM & $\tau_J$, $\tau_K$ & $10^{-10}$, $10^{-5}$ & MINAO (atom-blocked) & $\Delta E<10^{-6}$, DIIS error $<10^{-3}$ \\
ForSCF mixed & $\tau_J$, $\taub$ & $10^{-10}$, $10^{-10}$ & as above & as above \\
ForSCF opt-Brc & $\taub$ & $10^{-10}$ & as above & as above \\
GPU4PySCF & \texttt{direct\_scf\_tol} & $10^{-10}$ & MINAO (atom-blocked) & \texttt{conv\_tol} $10^{-6}$ \\
QUICK & \texttt{CUTOFF} & $10^{-10}$ & SAD & density RMS $10^{-6}$ \\
PySCF & \texttt{direct\_scf\_tol} & $10^{-10}$ & MINAO & \texttt{conv\_tol} $10^{-6}$ \\
ORCA & \texttt{Thresh} (\texttt{TCut} a decade below) & $10^{-10}$ & atomic model densities (PModel) & $\Delta E<10^{-6}$ \\
NWChem & \texttt{tol2e} & $10^{-10}$ & atomic SCF densities & orbital gradient $10^{-4}$ \\
NWChem + ForSCF & $\tau_J$, $\taub$ & as ForSCF & as NWChem & as NWChem \\
\bottomrule
\end{tabular}
\begin{tablenotes}
\item ``Published'': the values Ref.~\citenum{Palethorpe2024} used for every program, whereas the values used in this work are those of Table~\ref{tab:thresholds}. Every program builds the full Fock matrix in every iteration and uses DIIS, except NWChem, which uses its Newton--Raphson solver. ORCA uses spherical functions only.
\end{tablenotes}
\end{threeparttable}
\end{table}

\subsection{Study Design}
\label{sec:design}\label{sec:designsweep}\label{sec:designfdb}\label{sec:designconv}\label{sec:designcpu}\label{sec:designnw}
Palethorpe et al.\ set all screening thresholds to $10^{-10}$, except $\tau_K=10^{-5}$, and reported that the programs then agreed on the final energy to within $10^{-5}\Eh$.\cite{Palethorpe2024}
With these settings, the converged energies of ForSCF deviate from the references by up to \pubMaxDE\Eh{} (\pubSys) on \pubNOver{} of \pubN{} systems, while those of the other programs stay within \pubOthersMax\Eh.
Equal nominal thresholds also skip different amounts of work in different programs (Sections~\ref{sec:brc} and \ref{sec:um}), so no common value gives a fair comparison.
Every program is therefore first calibrated to the same accuracy and only then timed.

\paragraph{}
The accuracy is measured on one Fock build instead of a converged energy, which separates the error of screening from the effects of SCF convergence.
For every system and basis set, the density $\D$ of a GPU4PySCF calculation converged to $10^{-9}\Eh$ is read by every program in its own ordering and normalization, and each program builds $\mathbf{G}=\J-\tfrac12\K$ on it.
The error of a build is
\begin{align}
|\Delta E| &= \left|\tfrac12\operatorname{tr}\bigl[\D\,(\mathbf{G}_{\mathrm{prog}}-\mathbf{G}_{\mathrm{ref}})\bigr]\right|, \label{eq:fdberr}
\end{align}
with $\mathbf{G}_{\mathrm{ref}}$ from GPU4PySCF at \texttt{direct\_scf\_tol} $10^{-13}$, which agrees with ForSCF at $10^{-13}$ to \RefCrossMax\Eh{} on the \RefCrossN{} largest systems.
On a converged density this is, to first order, the error of the final energy, so $|\Delta E|<10^{-5}\Eh$ is the per-build form of the criterion of Ref.~\citenum{Palethorpe2024} and serves as the target.

\paragraph{}
For each program, one screening value is swept while everything else is kept fixed; for ForSCF, $\tau_J$ is swept together with the exchange threshold ($\tau_K$ for opt-UM, $\taub$ for the mixed scheme), and opt-Brc has the single threshold $\taub=\tau_J$.
The sweep runs on the largest glycine chain and the largest water cluster that the program completes within one hour and the memory of the node.
The selected value is the loosest at which these builds, and those at all tighter values, meet the target, so that an isolated dip of the error below the target does not decide the setting.
PySCF shares the screening of GPU4PySCF and takes its value, and ORCA, which cannot start from a given density, is swept on complete calculations against references with spherical functions.
One setting per program and basis set is then used for all systems.

\paragraph{}
Two quantities are timed at the selected settings: the Fock-build time and the time to solution.
The Fock-build time is the shortest warm build on the fixed density, from at least three builds after a discarded first one (six for GPU4PySCF, which is called through \GFourPRoute, the route its SCF takes).
The time to solution is the wall time of a complete SCF calculation, from the setup of the molecule to the converged energy, with each program's own guess, convergence acceleration and convergence test, set to about $10^{-6}\Eh$, and a full Fock build in every iteration.
ForSCF and GPU4PySCF are timed after a calculation on a water molecule in the same process, which absorbs just-in-time compilation and device initialization.
Every program runs each series up to the first system that does not complete within one hour or the memory of the node.

\paragraph{}
The CPU programs use the whole node in the layouts that were fastest in a scaling test (Supporting Information): NWChem with 64 MPI ranks, ORCA with 32 MPI processes, and PySCF and the CPU build of ForSCF with 64 OpenMP threads.
The cc-pVTZ basis set is not run on the CPU.
ORCA uses spherical functions, 4--6\% fewer than the Cartesian functions of the other programs with 6-31G* and cc-pVDZ, which favors it, and its times are drawn at its own number of functions.
NWChem with ForSCF runs NWChem's direct SCF with ForSCF building $\J$ and $\K$ in the mixed scheme, on one V100 with 32 MPI ranks or one, or on the CPU with one rank on 64 threads.

\subsection{Threshold Sweep}
\label{sec:sweep}
\begin{figure}[!htb]
\centering
\begin{subfigure}[t]{\textwidth}\centering
\includegraphics[width=\textwidth]{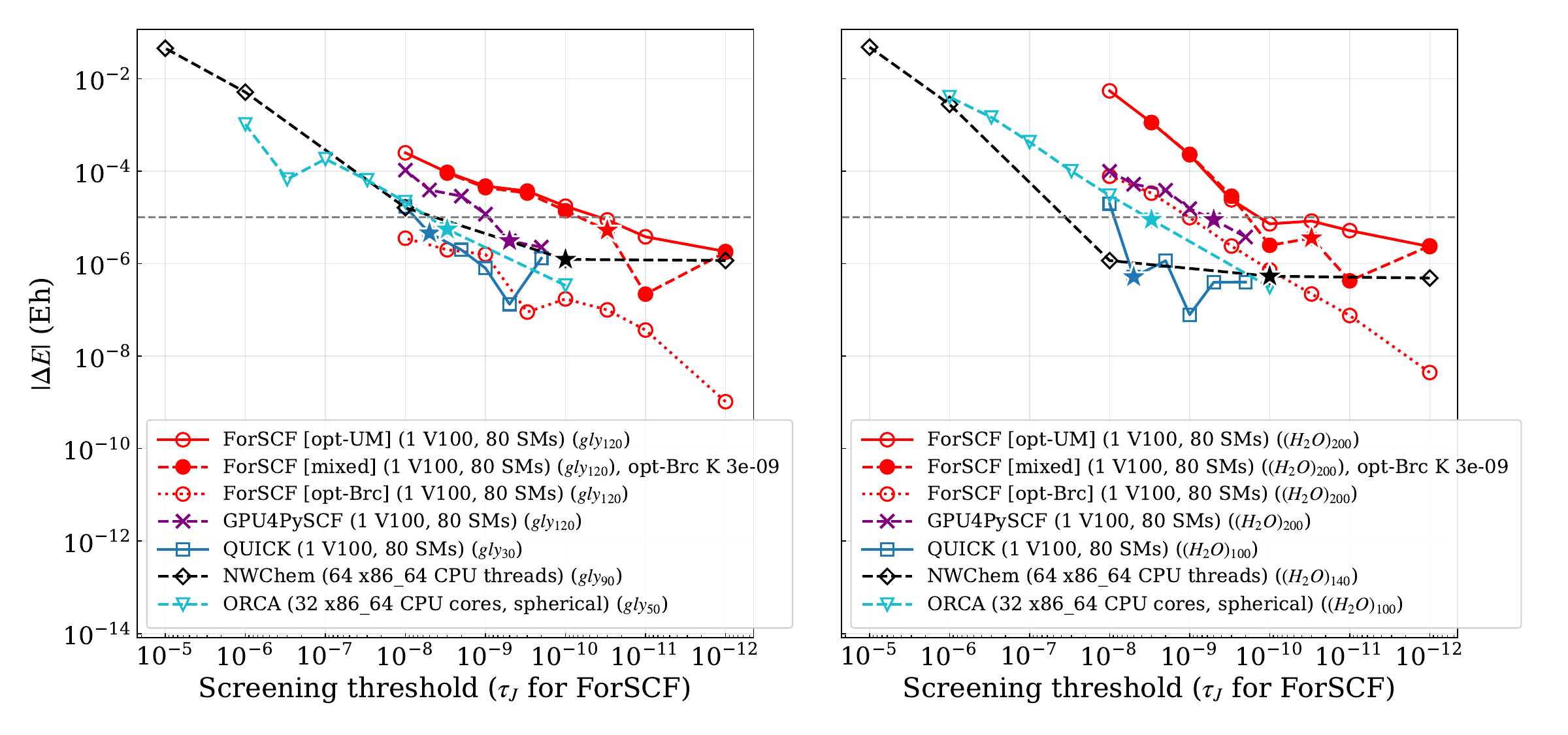}
\caption{6-31G*}\label{fig:sweep_631gs}
\end{subfigure}\\[0.5em]
\begin{subfigure}[t]{\textwidth}\centering
\includegraphics[width=\textwidth]{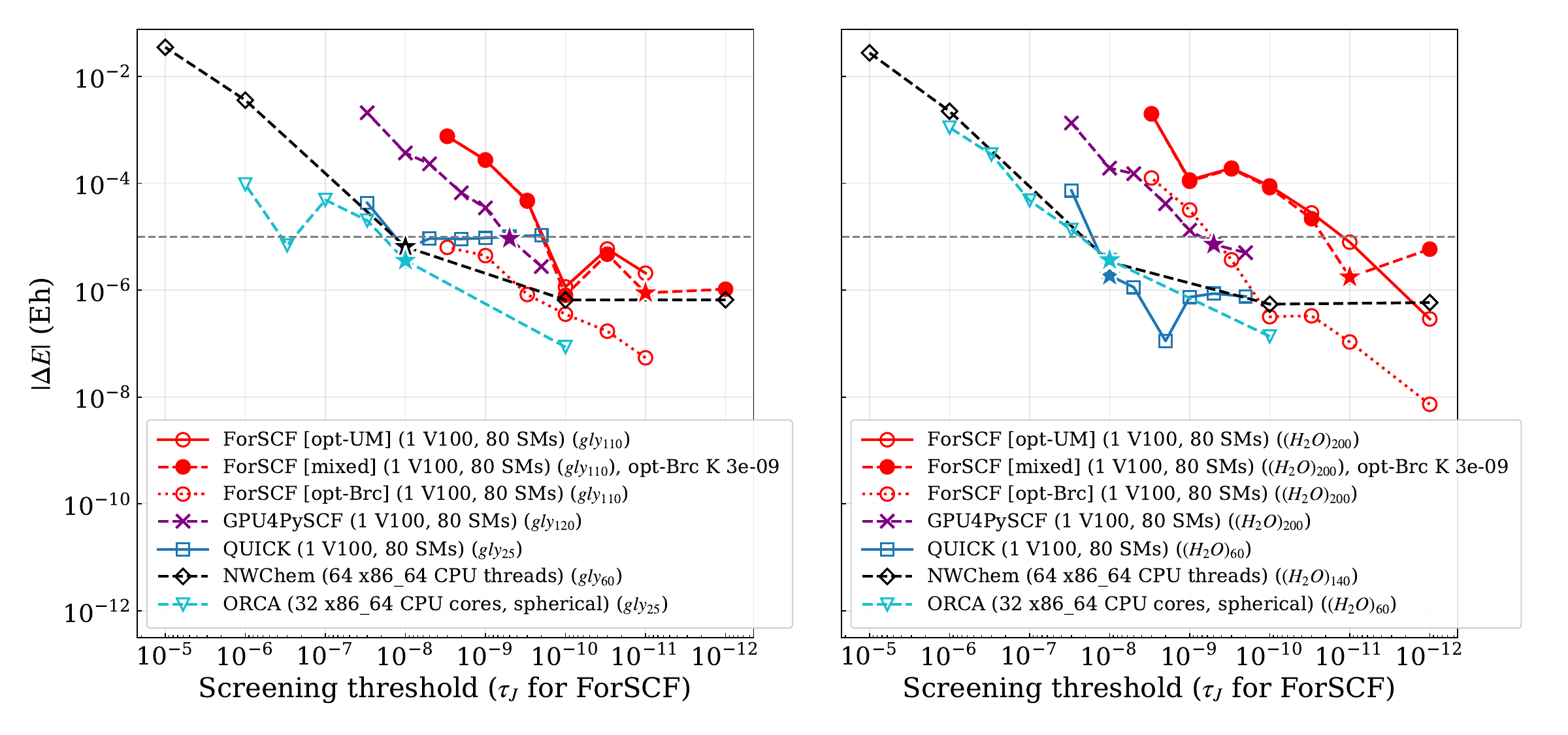}
\caption{cc-pVDZ}\label{fig:sweep_ccpvdz}
\end{subfigure}
\caption{Per-build error $|\Delta E|$ (Eq.~\ref{eq:fdberr}, larger of the errors on $\D$ and $\D-\D_{\mathrm{MINAO}}$) against the screening value, for polyglycine (left) and water clusters (right), each program on its own calibration system (in parentheses), with ForSCF against $\tau_J$ at the selected $\tau_K$ (opt-UM) or $\taub$ (mixed), and with $\taub=\tau_J$ (opt-Brc). Stars mark the selected settings and the dashed line the $10^{-5}\Eh$ target. The other basis sets are given in the Supporting Information.}
\label{fig:sweep}
\end{figure}

\begin{figure}[!htb]
\centering
\includegraphics[width=\textwidth]{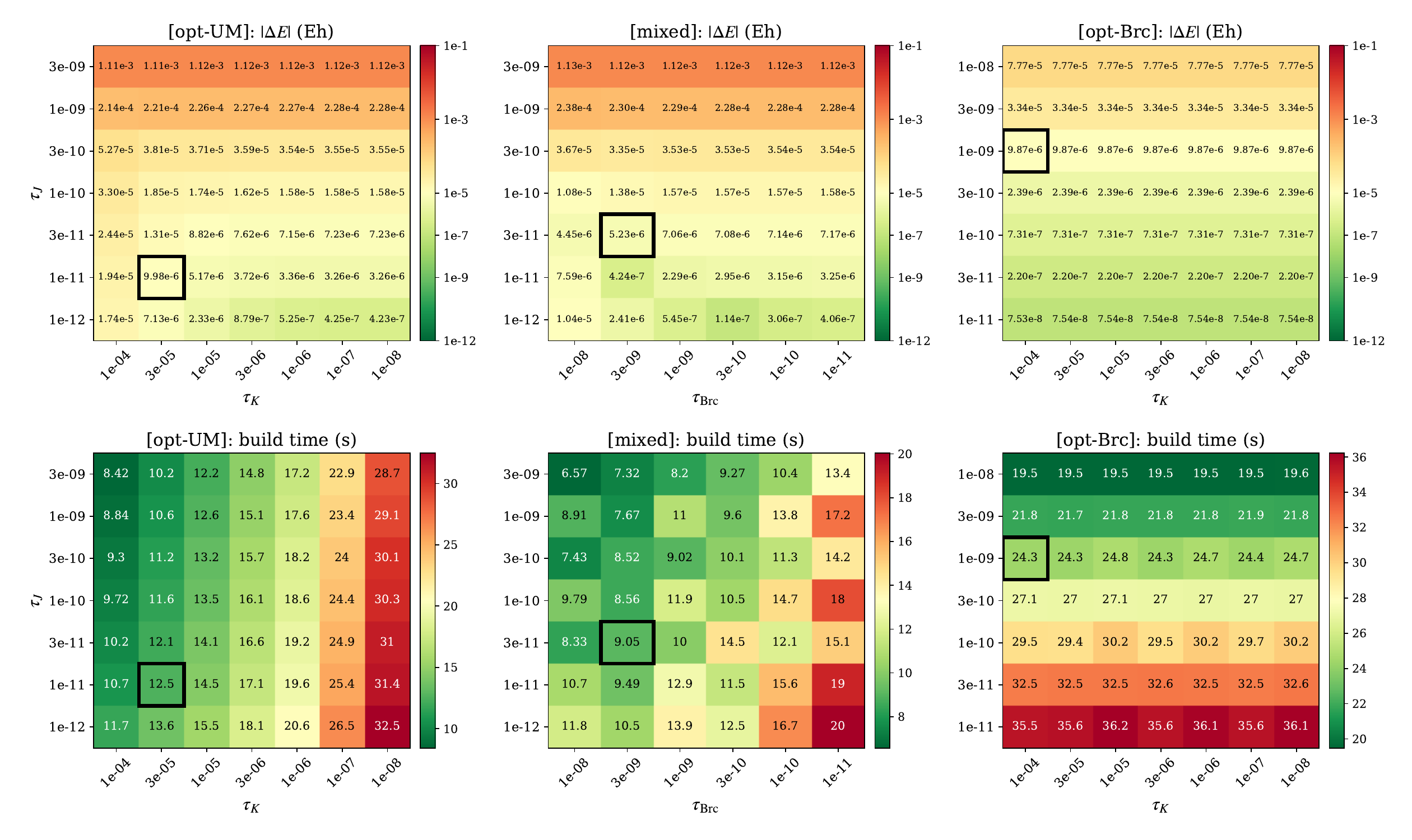}
\caption{ForSCF threshold grids for 6-31G* on gly$_{120}$ and (H$_2$O)$_{200}$: opt-UM against $\tau_K$, mixed against its exchange threshold $\taub$, and opt-Brc, which does not depend on $\tau_K$. Top: largest $|\Delta E|$ of the two systems (green below, red above $10^{-5}\Eh$); bottom: build time summed over both systems. The selected pair is outlined.}
\label{fig:grid}
\end{figure}

Figure~\ref{fig:sweep} shows the error of each program against its screening value, Figure~\ref{fig:grid} the ForSCF grids for 6-31G*, and Table~\ref{tab:thresholds} the selected settings.
In the opt-UM Coulomb build, $\tau_J$ sets the accuracy: for 6-31G* the error falls from \swUMPaNine\Eh{} at $\tau_J=10^{-9}$ to \swUMPaTen{} and \swUMPaEleven\Eh{} at $10^{-10}$ and $10^{-11}$, and for cc-pVDZ from \swUMDZTen{} to \swUMDZEleven\Eh{} between $10^{-10}$ and $10^{-11}$.
This follows from the primitive criterion of Eq.~\ref{eq:jskip}, the test of Ref.~\citenum{Palethorpe2024}, which an independent Python implementation reproduces: every skipped primitive quartet is below $\tau_J$, but their number grows with the system and with the contraction length, and their contributions do not cancel.
opt-Brc and GPU4PySCF, which screen contracted quartets, meet the target at values looser than the published $10^{-10}$, while every Coulomb build by opt-UM needs $\tau_J$ between $10^{-11}$ and $3\times10^{-11}$.
Once $\tau_J$ is fixed, $\tau_K$ mainly changes the cost of opt-UM exchange: for 6-31G* at $\tau_J=10^{-11}$, tightening it from $10^{-5}$ to $10^{-12}$ lowers the error from \tkLooseEPa{} to \tkTightEPa\Eh{} but makes the build \tkRatioPa{} times slower, and the sweep selects values looser than the published $10^{-5}$.
The mixed scheme selects an exchange threshold $\taub$ two orders of magnitude looser than its $\tau_J$, which a single threshold could not do.
\begin{table}[!htb]
\caption{Screening settings selected at matched accuracy}
\label{tab:thresholds}
\centering\small
\begin{threeparttable}
% Generated by scripts/make_results.py; do not edit.
\begin{tabular}{lccccc}
\toprule
basis & opt-UM & mixed & opt-Brc & GPU4PySCF & QUICK \\
\midrule
STO-3G & $10^{-11}$ / $3\times10^{-5}$ & $3\times10^{-11}$ / $3\times10^{-9}$ & $10^{-9}$ & $5\times10^{-10}$ & $2\times10^{-10}$ \\
6-31G* & $10^{-11}$ / $3\times10^{-5}$ & $3\times10^{-11}$ / $3\times10^{-9}$ & $10^{-9}$ & $5\times10^{-10}$ & $5\times10^{-9}$ \\
6-31G** & $10^{-11}$ / $3\times10^{-5}$ & $3\times10^{-11}$ / $3\times10^{-9}$ & $10^{-9}$ & $2\times10^{-10}$ & $5\times10^{-9}$ \\
cc-pVDZ & $10^{-11}$ / $3\times10^{-5}$ & $10^{-11}$ / $3\times10^{-9}$ & $3\times10^{-10}$ & $5\times10^{-10}$ & $10^{-8}$\textsuperscript{\emph{b}} \\
cc-pVTZ & $3\times10^{-11}$ / $10^{-4}$ & $10^{-11}$ / $10^{-9}$ & $3\times10^{-10}$ & $5\times10^{-10}$ & $3\times10^{-8}$ \\
\bottomrule
\end{tabular}

\begin{tablenotes}
\item ForSCF: $\tau_J/\tau_K$ (opt-UM), $\tau_J/\taub$ (mixed) and $\taub$ (opt-Brc), and for other programs their screening value (Table~\ref{tab:programs}). NWChem, PySCF and ORCA: Supporting Information.
\item[\emph{b}] No admissible setting, so the most accurate completed setting is used.
\end{tablenotes}
\end{threeparttable}
\end{table}

NWChem, whose SCF converges tightly, has the same error in its fixed-density builds as in its converged energies, within \valNwDev\% on the \valNwN{} systems where both exceed $2\times10^{-6}\Eh$, which supports Eq.~\ref{eq:fdberr} directly.
QUICK and NWChem, which read orbitals, keep a small error at the tightest values, and QUICK has no admissible setting for cc-pVDZ.
A setting selected on the calibration systems does not hold on every other system for three programs.
QUICK, calibrated on smaller systems than the others, misses the target in \fdbmissQK{} of its \fdbnQK{} fixed-density builds, by up to \fdbworstQK\Eh{} (\fdbworstsysQK), but in only \cvmissQK{} of \cvnQK{} converged energies, since it tightens its cutoff near convergence. NWChem misses in \cvmissNWDZ{} of its \cvnNWDZ{} converged cc-pVDZ energies (up to \cvmaxdeNW\Eh), whose error does not grow monotonically with the size of the system, and ORCA in \cvmissOR{} of \cvnOR{} (up to \cvmaxdeOR\Eh).
These results are kept: they favor the program concerned, so the speedups of ForSCF over it are lower bounds.
Of the ForSCF schemes, only opt-UM with cc-pVTZ, calibrated on gly$_{50}$, misses on fixed densities, for the two largest water clusters (up to \fdbworstUM\Eh).
\FloatBarrier

\subsection{Single-GPU Study}
\label{sec:gpures}
\subsubsection{Fock-Build Throughput}
\label{sec:throughput}
\begin{figure}[!htb]
\centering
\begin{subfigure}[t]{0.49\textwidth}\centering
\includegraphics[width=\textwidth]{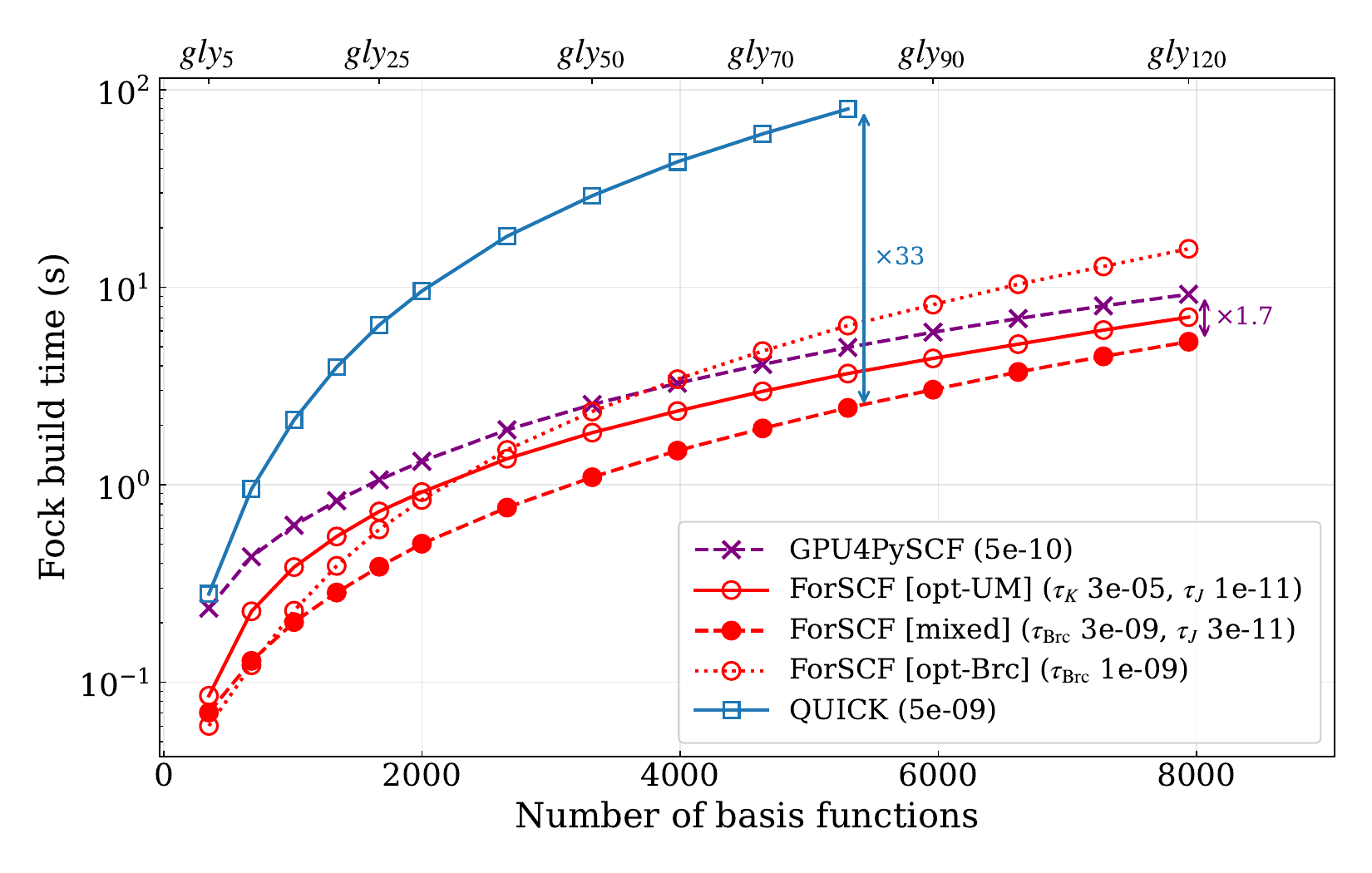}\caption{6-31G*}
\end{subfigure}\hfill
\begin{subfigure}[t]{0.49\textwidth}\centering
\includegraphics[width=\textwidth]{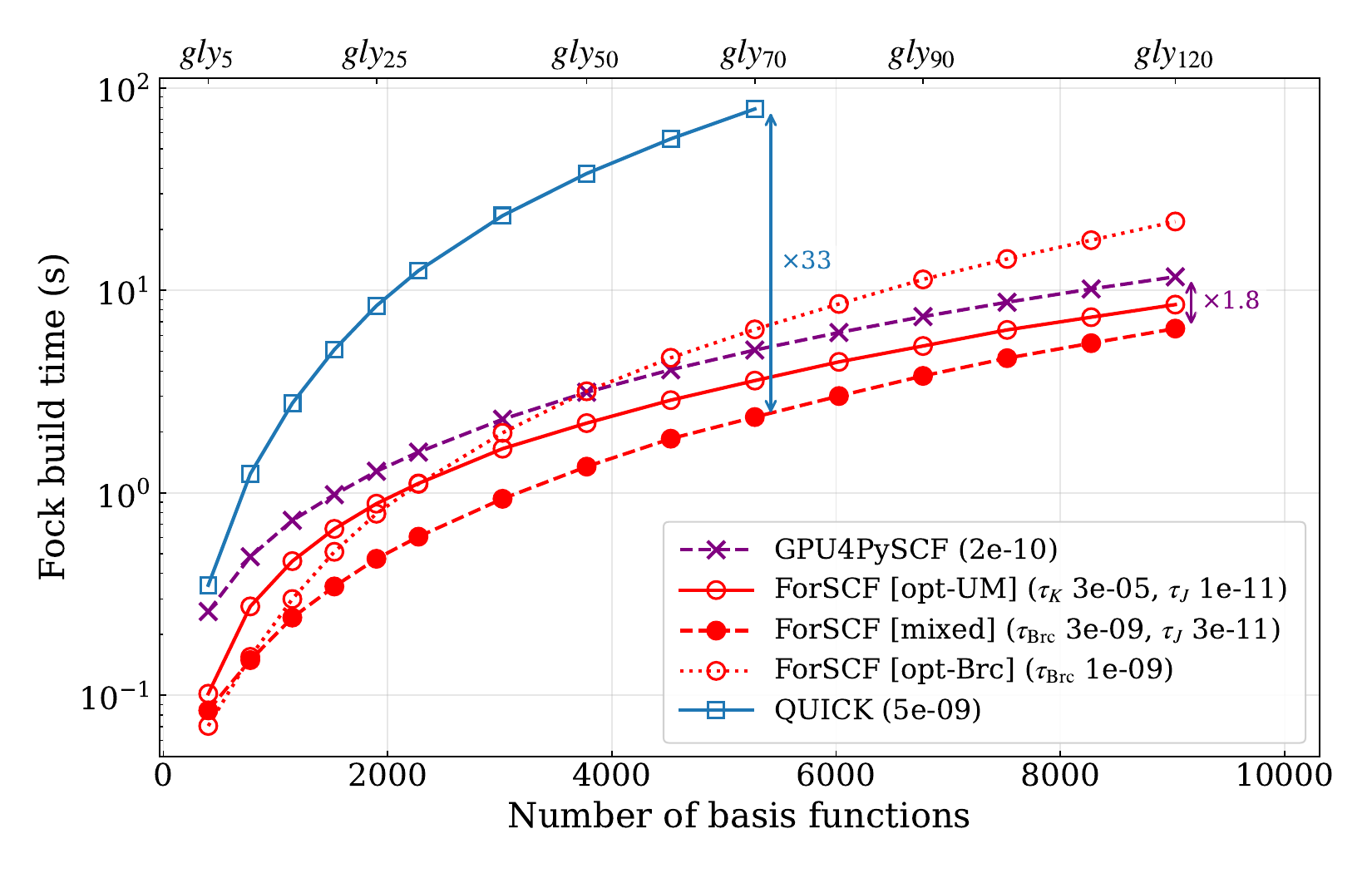}\caption{6-31G**}
\end{subfigure}\\[0.5em]
\begin{subfigure}[t]{0.49\textwidth}\centering
\includegraphics[width=\textwidth]{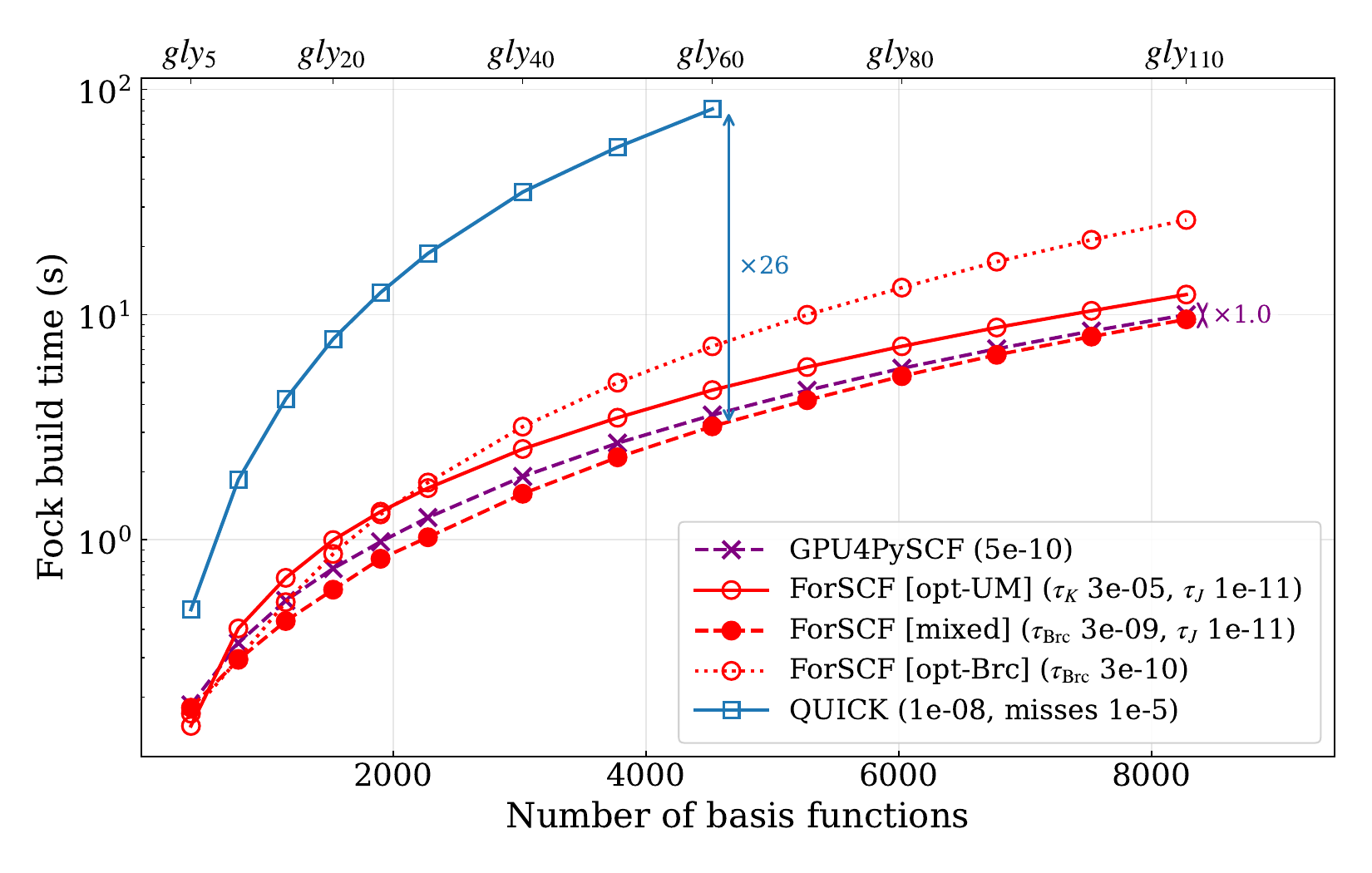}\caption{cc-pVDZ}
\end{subfigure}\hfill
\begin{subfigure}[t]{0.49\textwidth}\centering
\includegraphics[width=\textwidth]{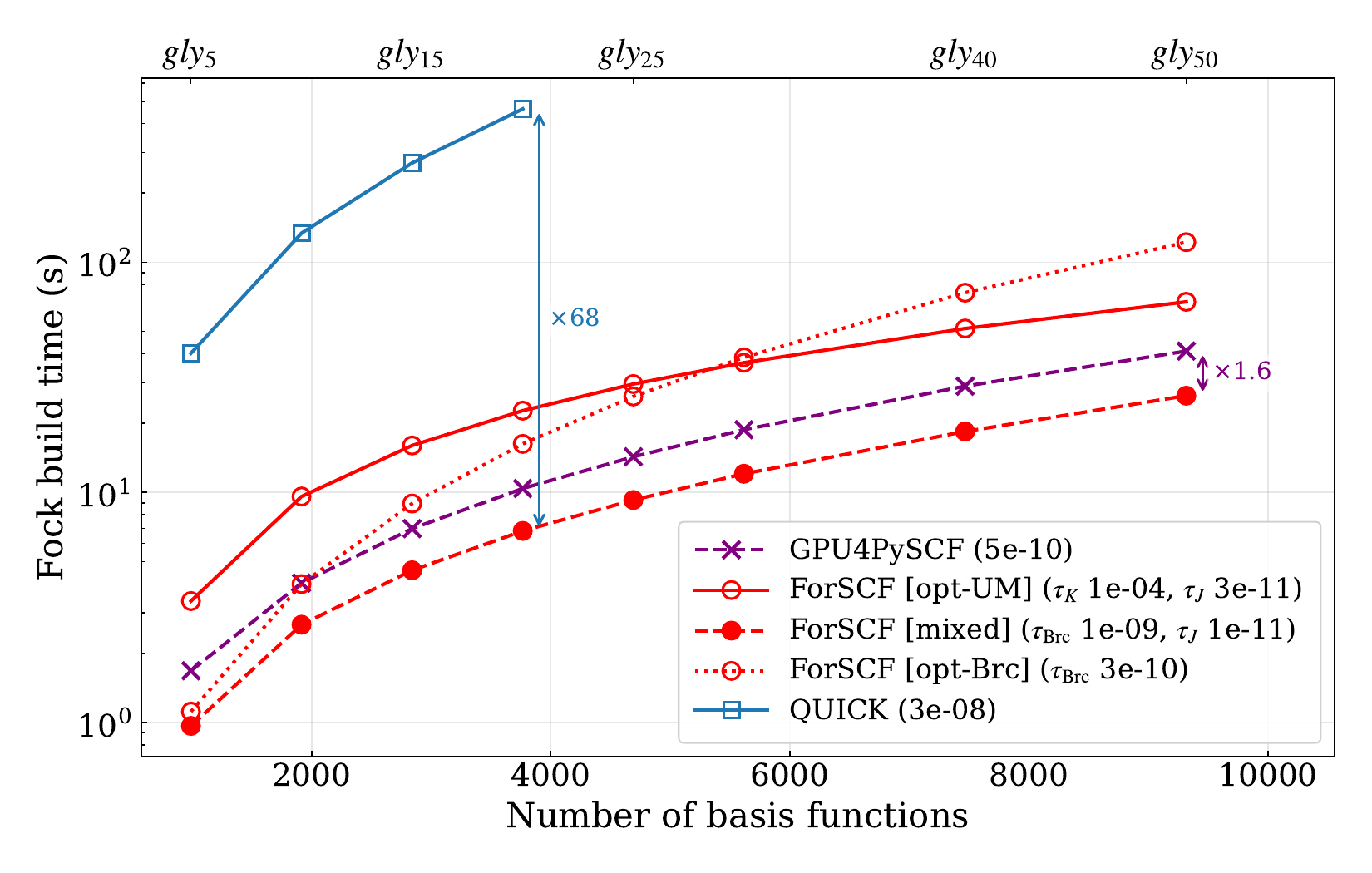}\caption{cc-pVTZ}
\end{subfigure}
\caption{Fock-build time of polyglycine chains at matched accuracy on one V100, against the number of basis functions (top axis: chain length), each program at its selected setting (legend, Table~\ref{tab:thresholds}). Arrows give the ratio of each other program to the fastest ForSCF scheme at the largest system both ran. STO-3G is given in the Supporting Information.}
\label{fig:fdbgly}
\end{figure}

\begin{figure}[!htb]
\centering
\begin{subfigure}[t]{0.49\textwidth}\centering
\includegraphics[width=\textwidth]{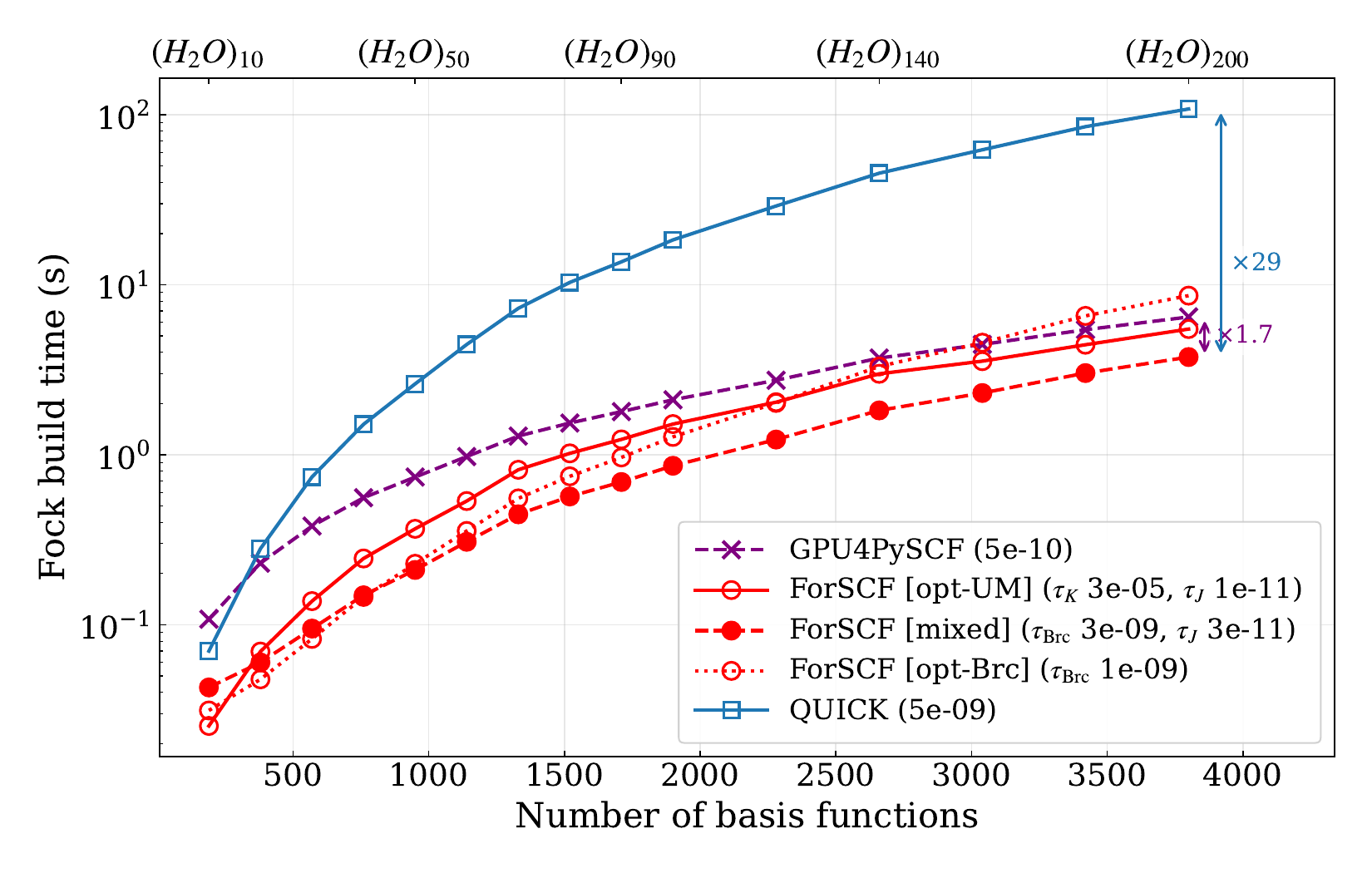}\caption{6-31G*}
\end{subfigure}\hfill
\begin{subfigure}[t]{0.49\textwidth}\centering
\includegraphics[width=\textwidth]{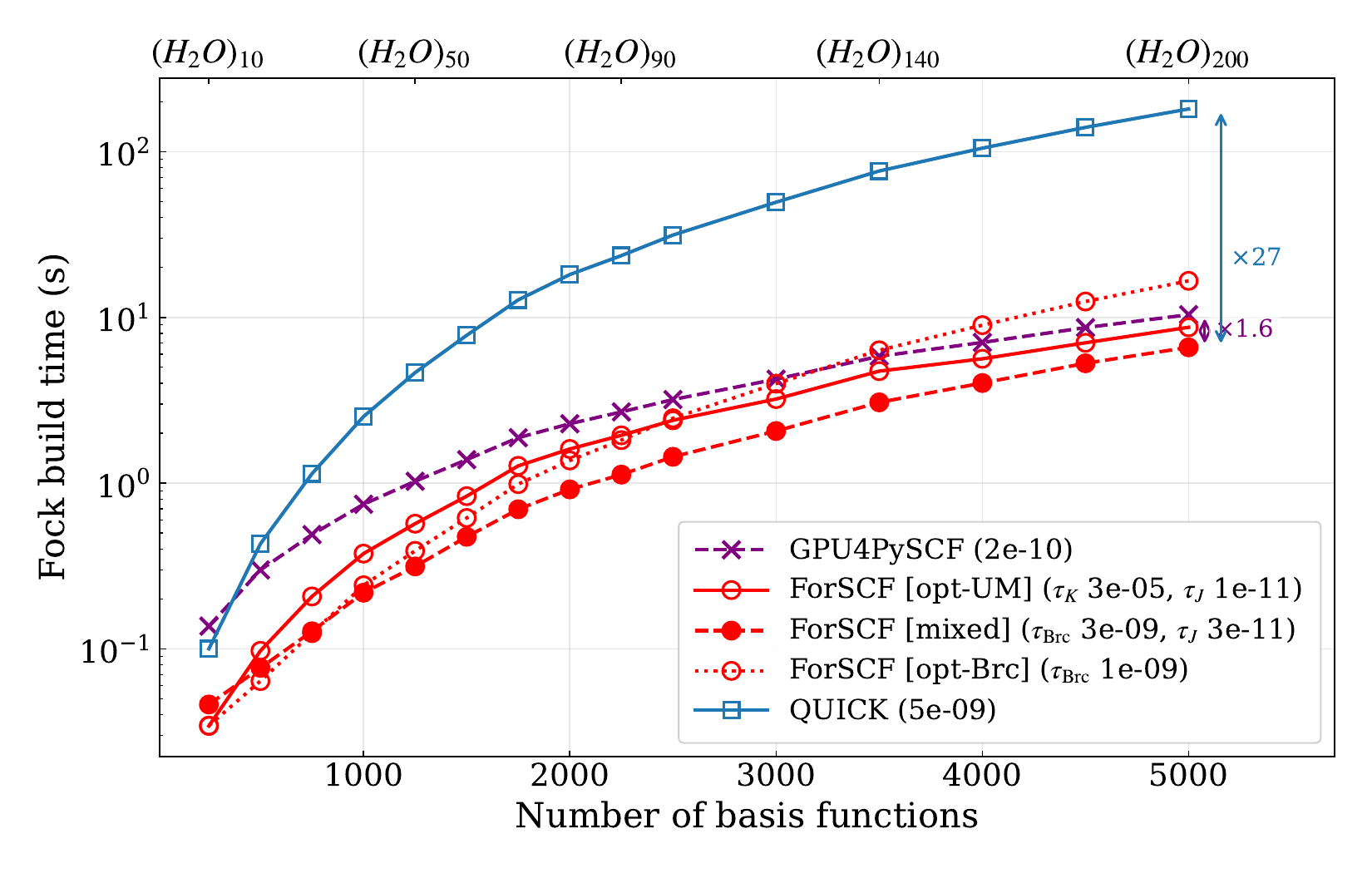}\caption{6-31G**}
\end{subfigure}\\[0.5em]
\begin{subfigure}[t]{0.49\textwidth}\centering
\includegraphics[width=\textwidth]{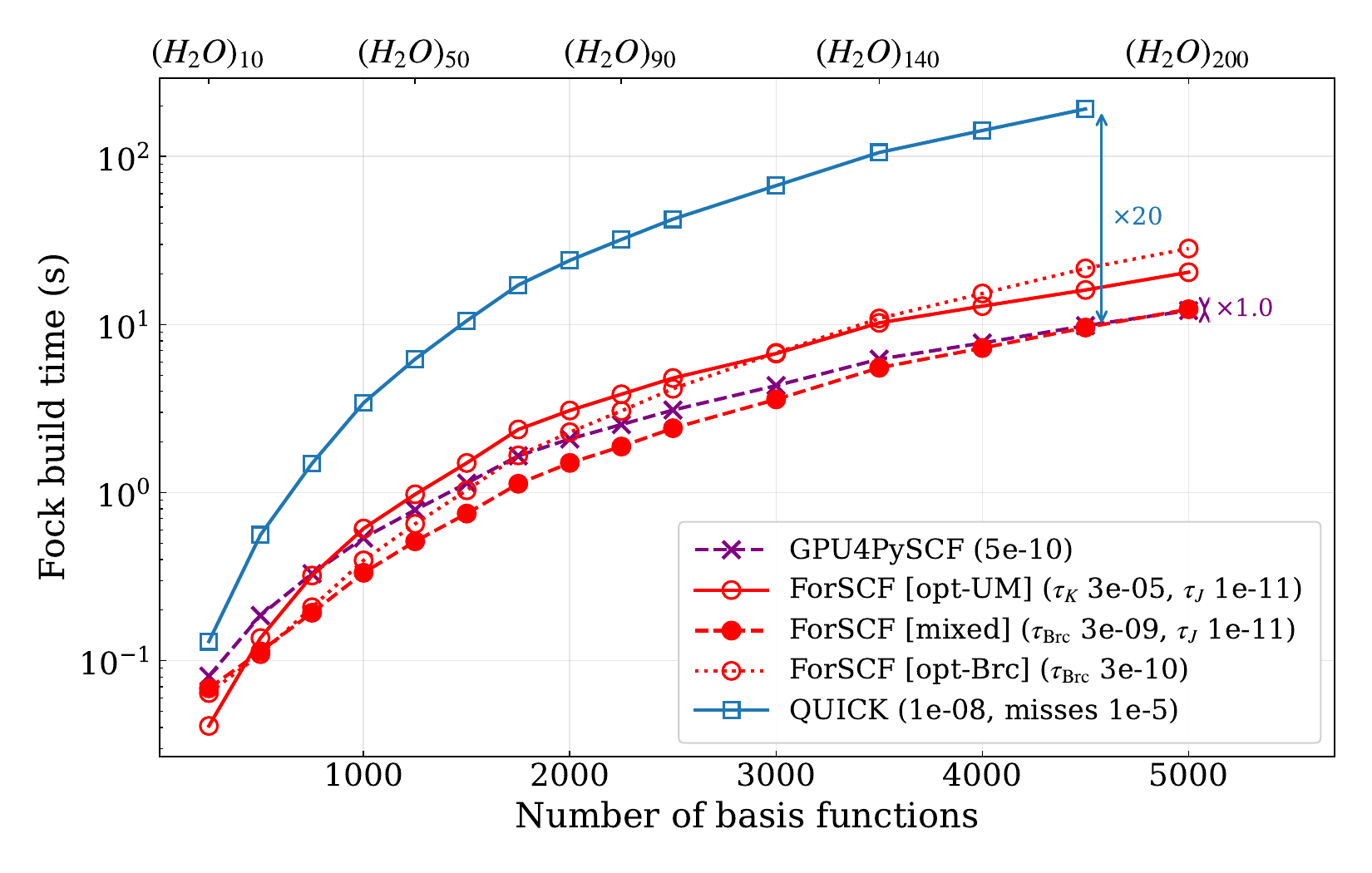}\caption{cc-pVDZ}
\end{subfigure}\hfill
\begin{subfigure}[t]{0.49\textwidth}\centering
\includegraphics[width=\textwidth]{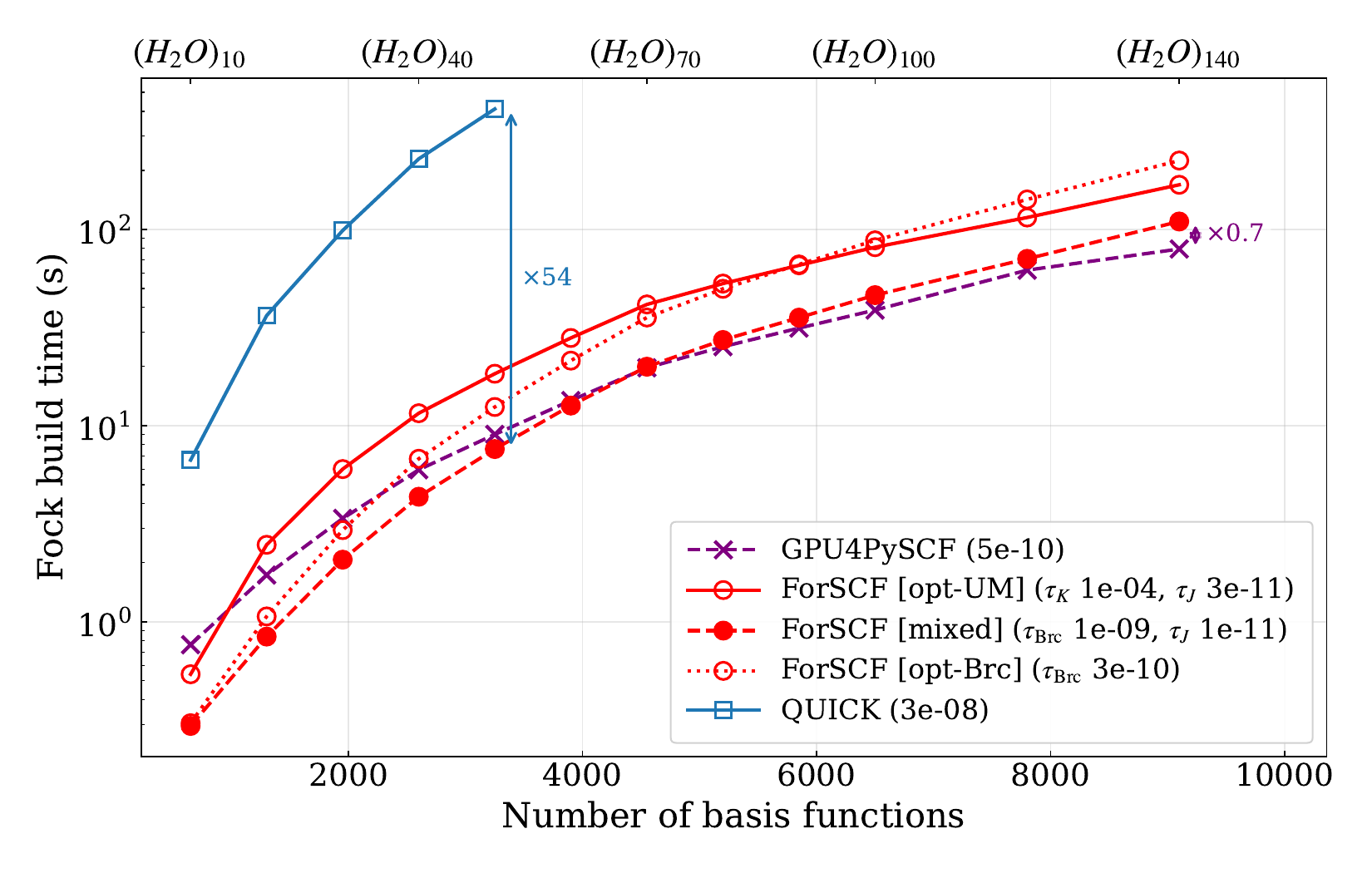}\caption{cc-pVTZ}
\end{subfigure}
\caption{Fock-build time of water clusters at matched accuracy on one V100, with details as in Figure~\ref{fig:fdbgly}.}
\label{fig:fdbwat}
\end{figure}

Figures~\ref{fig:fdbgly} and \ref{fig:fdbwat} show the Fock-build times at matched accuracy, and Table~\ref{tab:throughput} lists only the largest systems. To summarize each complete series of Figures~\ref{fig:fdbgly} and \ref{fig:fdbwat}, and the STO-3G series of the Supporting Information, the time of GPU4PySCF was divided by that of the fastest ForSCF scheme at every system size that both completed, and the geometric mean of these ratios was taken over the series. The resulting average speedups are
\spAvgSTOGly{} and \spAvgSTOWat{} for STO-3G polyglycine and water clusters, \spAvgPaGly{} and \spAvgPaWat{} for 6-31G*, \spAvgPbGly{} and \spAvgPbWat{} for 6-31G**, \spAvgDZGly{} and \spAvgDZWat{} for cc-pVDZ, and \spAvgTZGly{} and \spAvgTZWat{} for cc-pVTZ.
\begin{table}[!htb]
\caption{Fock-build time (s) at matched accuracy for the largest systems run by GPU4PySCF and ForSCF}
\label{tab:throughput}
\centering\small
\begin{threeparttable}
% Generated by scripts/make_results.py; do not edit.
\begin{tabular}{llrrrrrr}
\toprule
basis & system & $N_\mathrm{bf}$ & GPU4PySCF & opt-UM & mixed & opt-Brc & speedup\textsuperscript{\emph{b}} \\
\midrule
STO-3G & gly120 & 2767 & 1.68 & 0.883 & 0.783 & 0.973 & 2.15 \\
STO-3G & water200 & 1400 & 1.33 & 0.763 & 0.541 & 0.631 & 2.45 \\
6-31G* & gly120 & 7939 & 9.20 & 7.05 & 5.29 & 15.3 & 1.74 \\
6-31G* & water200 & 3800 & 6.46 & 5.48 & 3.74 & 8.63 & 1.73 \\
6-31G** & gly120 & 9025 & 11.7 & 8.51 & 6.48 & 21.1 & 1.80 \\
6-31G** & water200 & 5000 & 10.4 & 8.71 & 6.59 & 16.6 & 1.57 \\
cc-pVDZ & gly110 & 8275 & 9.98 & 12.3 & 9.52 & 26.2 & 1.05 \\
cc-pVDZ & water200 & 5000 & 12.1 & 20.5 & 12.3 & 28.2 & 0.98 \\
cc-pVTZ & gly50 & 9315 & 41.1 & 67.3 & 26.3 & 122 & 1.56 \\
cc-pVTZ & water140 & 9100 & 79.5 & -- & 110 & 224 & 0.72 \\
\bottomrule
\end{tabular}

\begin{tablenotes}
\item Fastest warm build at the settings of Table~\ref{tab:thresholds} on one V100. --: misses the target.
\item[\emph{b}] Time of GPU4PySCF divided by that of the fastest ForSCF scheme.
\end{tablenotes}
\end{threeparttable}
\end{table}
The ratio is largest for small and medium systems, up to \spPeakSTOGly{} for STO-3G and \spPeakPaWat{} for 6-31G*, and falls with system size for the STO-3G and Pople basis sets, to \spbestPaGly{} and \spbestPaWat{} for gly$_{120}$ and (H$_2$O)$_{200}$ with 6-31G*.
With the correlation-consistent basis sets it changes less over the polyglycine chains, from at most \spPeakDZGly{} to \spbestDZGly{} for \tplabDZGly{} with cc-pVDZ, and stays near \spbestTZGly{} with cc-pVTZ up to \tplabTZGly. For the water clusters it falls, below one from \spCrossTZWat{} with cc-pVTZ, where GPU4PySCF is faster (\spbestTZWat{} for \tplabTZWat), and to \spbestDZWat{} for \tplabDZWat{} with cc-pVDZ.
Section~\ref{sec:jkres} traces these trends to the Coulomb and exchange builds.
The mixed scheme is the fastest ForSCF scheme from medium sizes on, and with cc-pVTZ for all sizes.
opt-Brc is the fastest for most small systems, up to \brcLastSTOGly{} and \brcLastSTOWat{} with STO-3G and \brcLastPaGly{} and \brcLastPaWat{} with 6-31G*, as Ref.~\citenum{Palethorpe2024} recommends, but its Coulomb build makes it the slowest scheme for large systems, whereas opt-UM is held back by its exchange build wherever the contractions are long (\tpUMTZGly{} against \tpMXTZGly\,s for gly$_{50}$/cc-pVTZ).
At the largest systems that all programs ran, ForSCF builds the Fock matrix \spQKMin--\spQKMax{} times as fast as QUICK (Supporting Information).
GPU4PySCF also offers \texttt{get\_jk}, which builds both matrices by Rys quadrature and is 1.1--2.0 times slower than the route its SCF takes, so that comparisons through it overstate the lead of a competing program.

\FloatBarrier
\subsubsection{Time to Solution}
\label{sec:conv}
\begin{figure}[!htb]
\centering
\begin{subfigure}[t]{0.49\textwidth}\centering
\includegraphics[width=\textwidth]{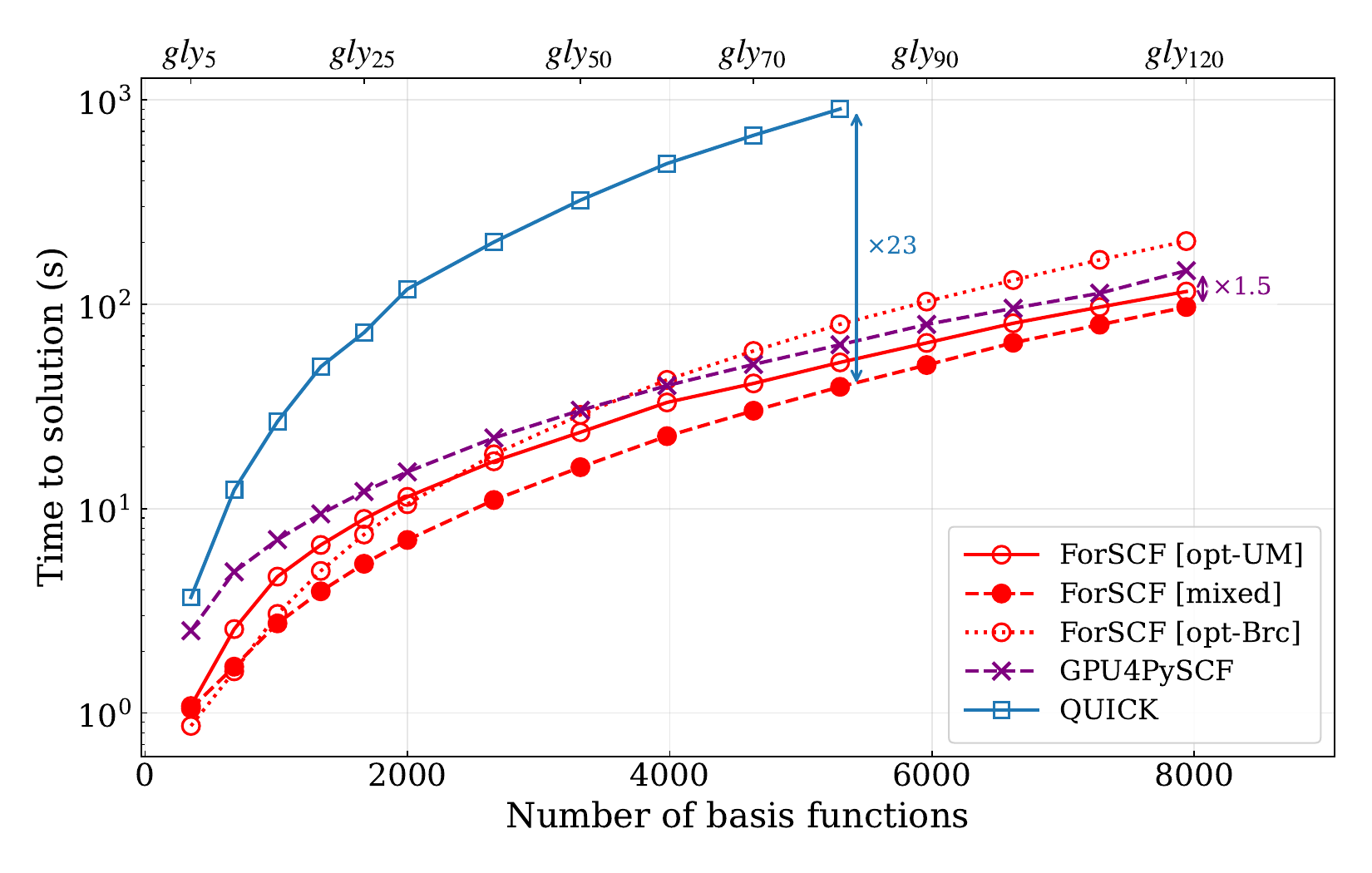}\caption{6-31G*}
\end{subfigure}\hfill
\begin{subfigure}[t]{0.49\textwidth}\centering
\includegraphics[width=\textwidth]{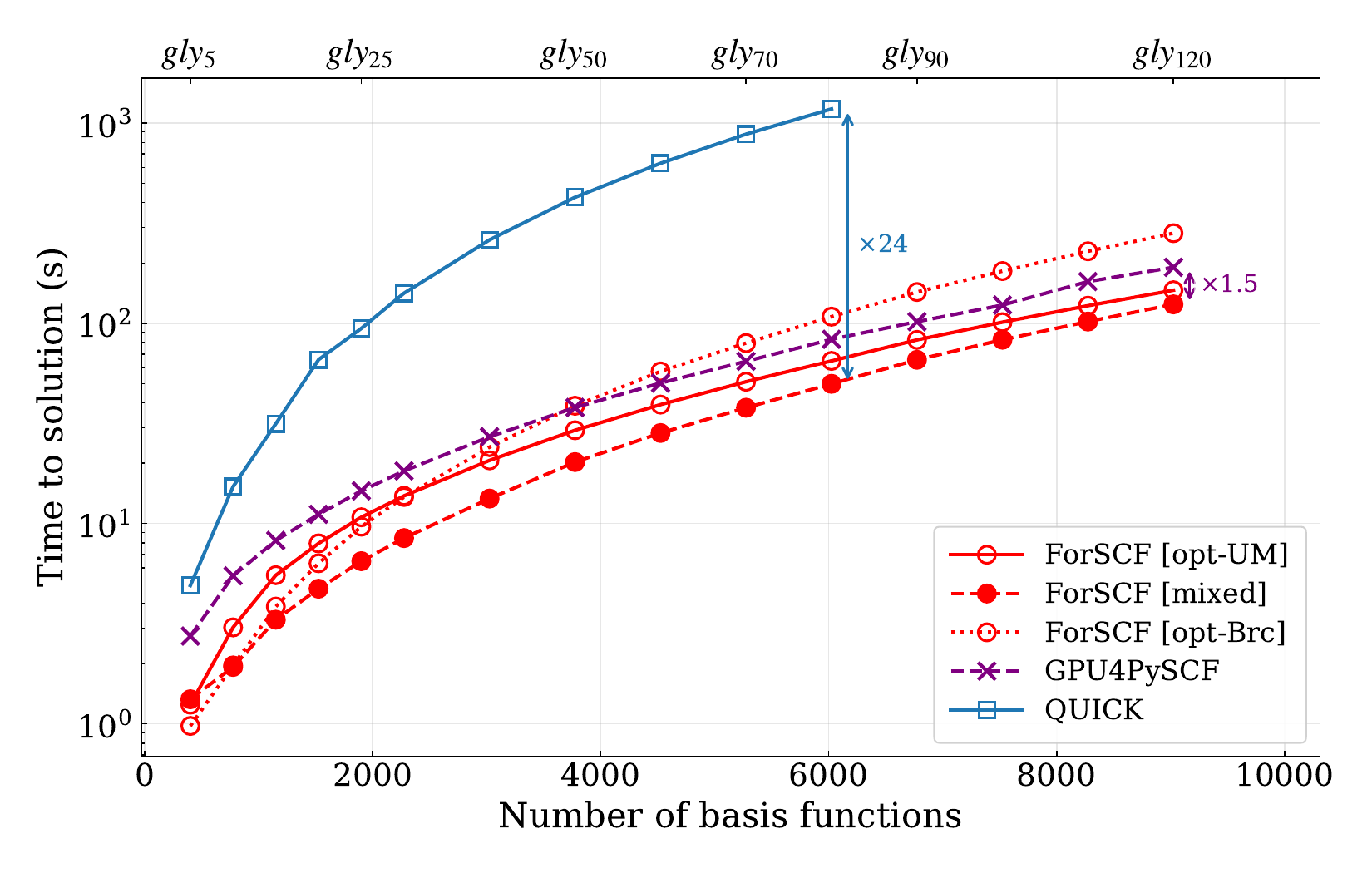}\caption{6-31G**}
\end{subfigure}\\[0.5em]
\begin{subfigure}[t]{0.49\textwidth}\centering
\includegraphics[width=\textwidth]{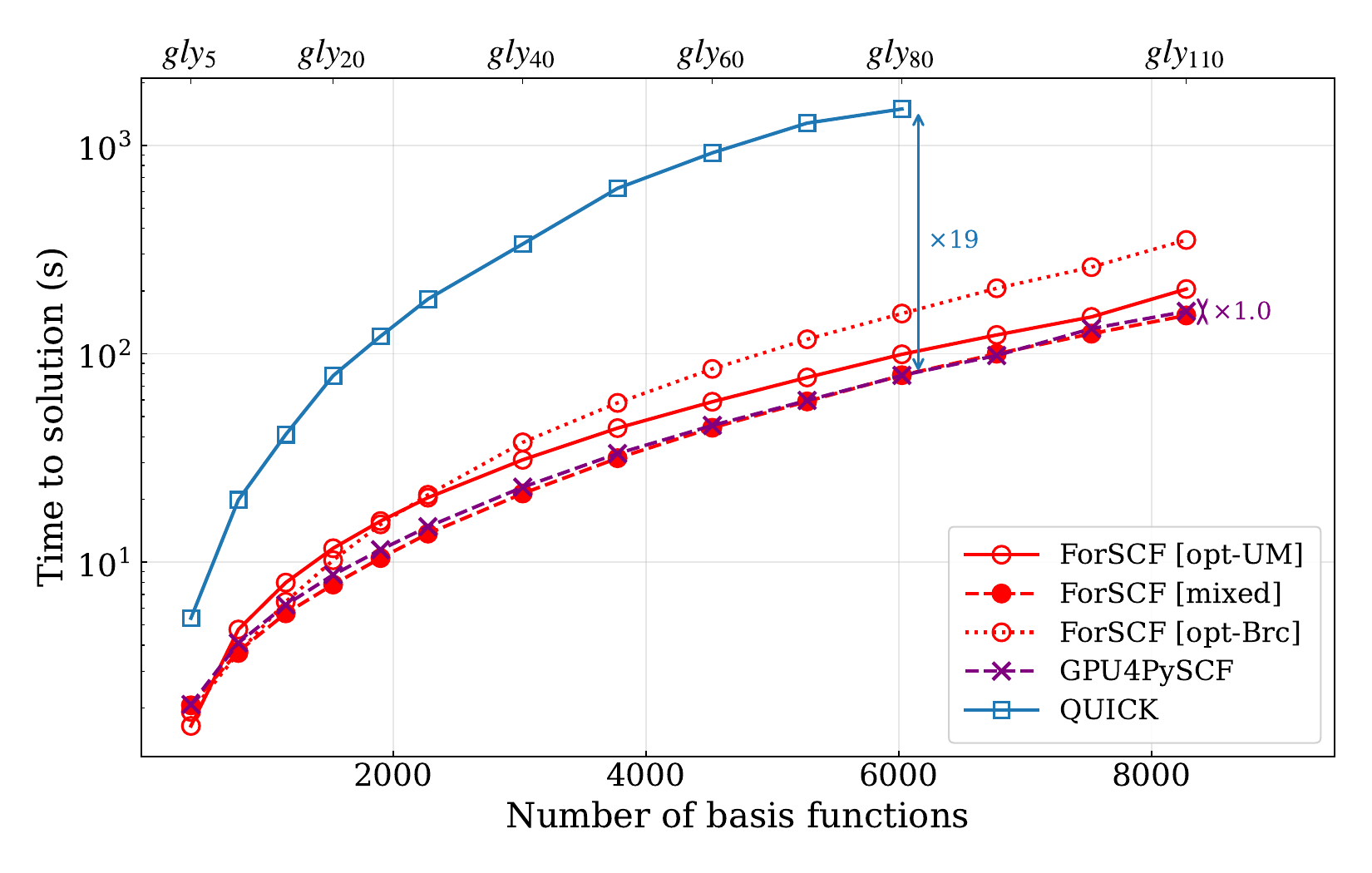}\caption{cc-pVDZ}
\end{subfigure}\hfill
\begin{subfigure}[t]{0.49\textwidth}\centering
\includegraphics[width=\textwidth]{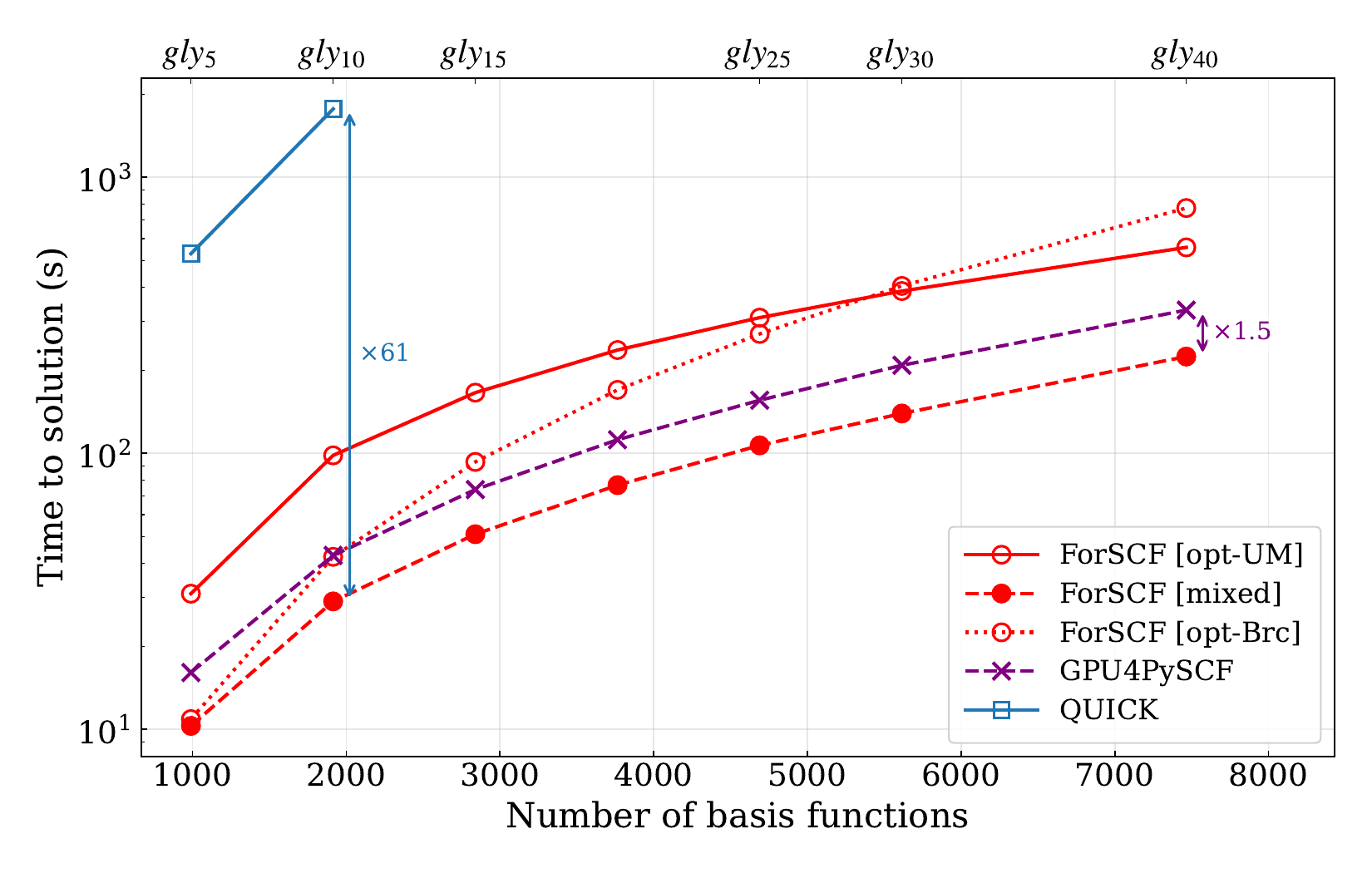}\caption{cc-pVTZ}
\end{subfigure}
\caption{Time to solution of complete RHF calculations on polyglycine chains at matched accuracy on one V100, at the settings of Table~\ref{tab:thresholds}, with arrows giving the ratio of each other program to the fastest ForSCF scheme at the largest system both ran.}
\label{fig:convgly}
\end{figure}

\begin{figure}[!htb]
\centering
\begin{subfigure}[t]{0.49\textwidth}\centering
\includegraphics[width=\textwidth]{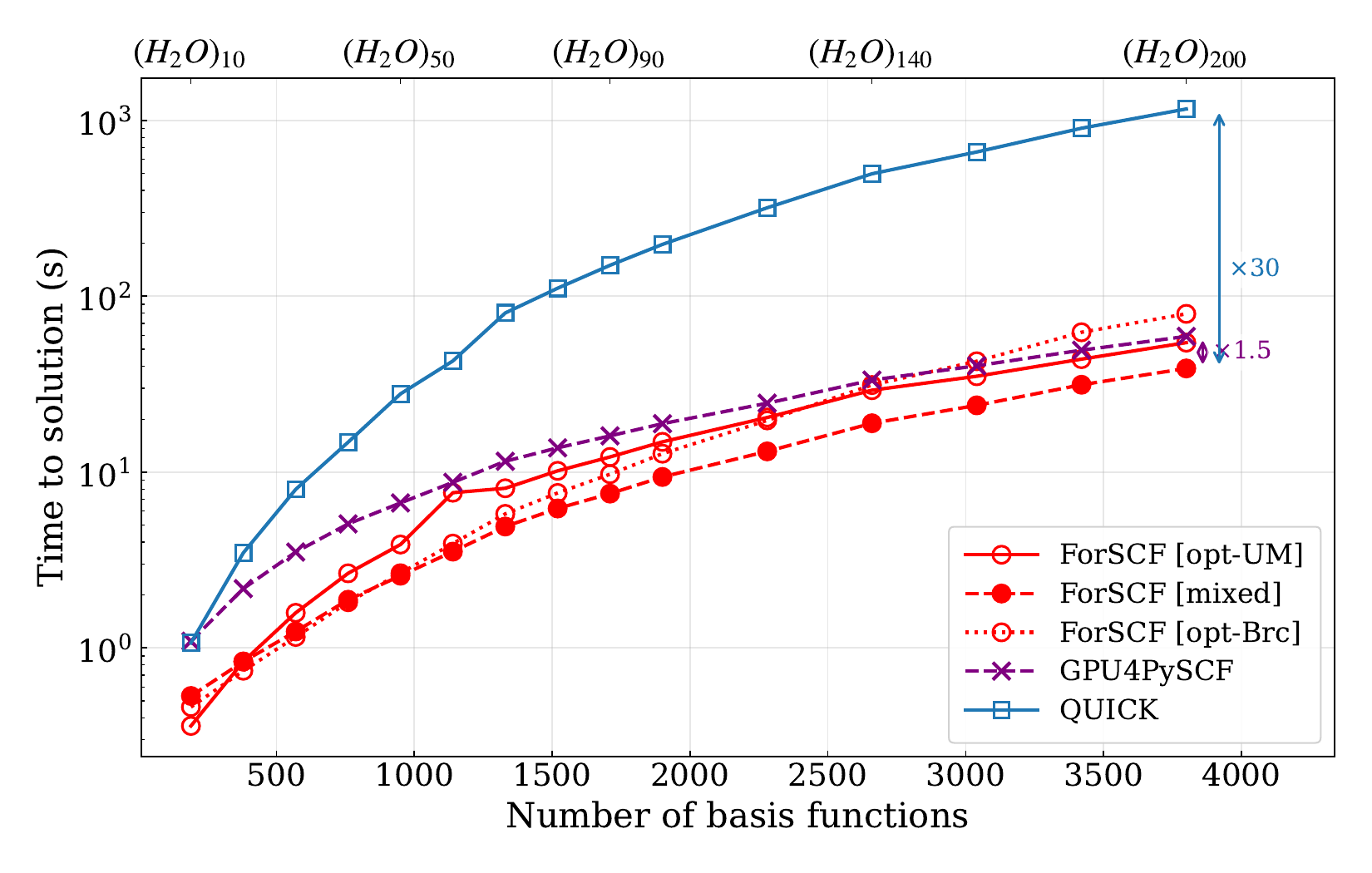}\caption{6-31G*}
\end{subfigure}\hfill
\begin{subfigure}[t]{0.49\textwidth}\centering
\includegraphics[width=\textwidth]{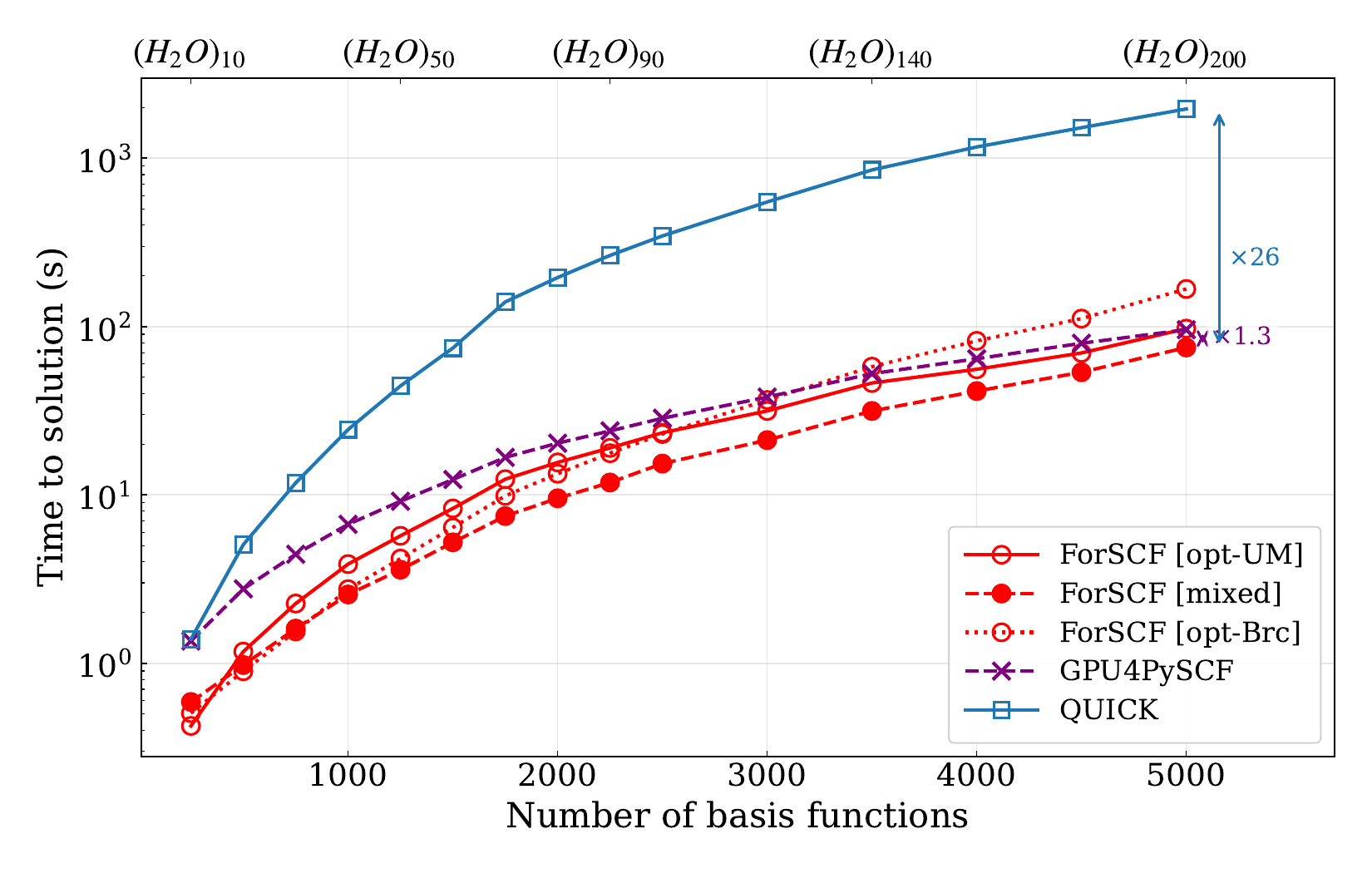}\caption{6-31G**}
\end{subfigure}\\[0.5em]
\begin{subfigure}[t]{0.49\textwidth}\centering
\includegraphics[width=\textwidth]{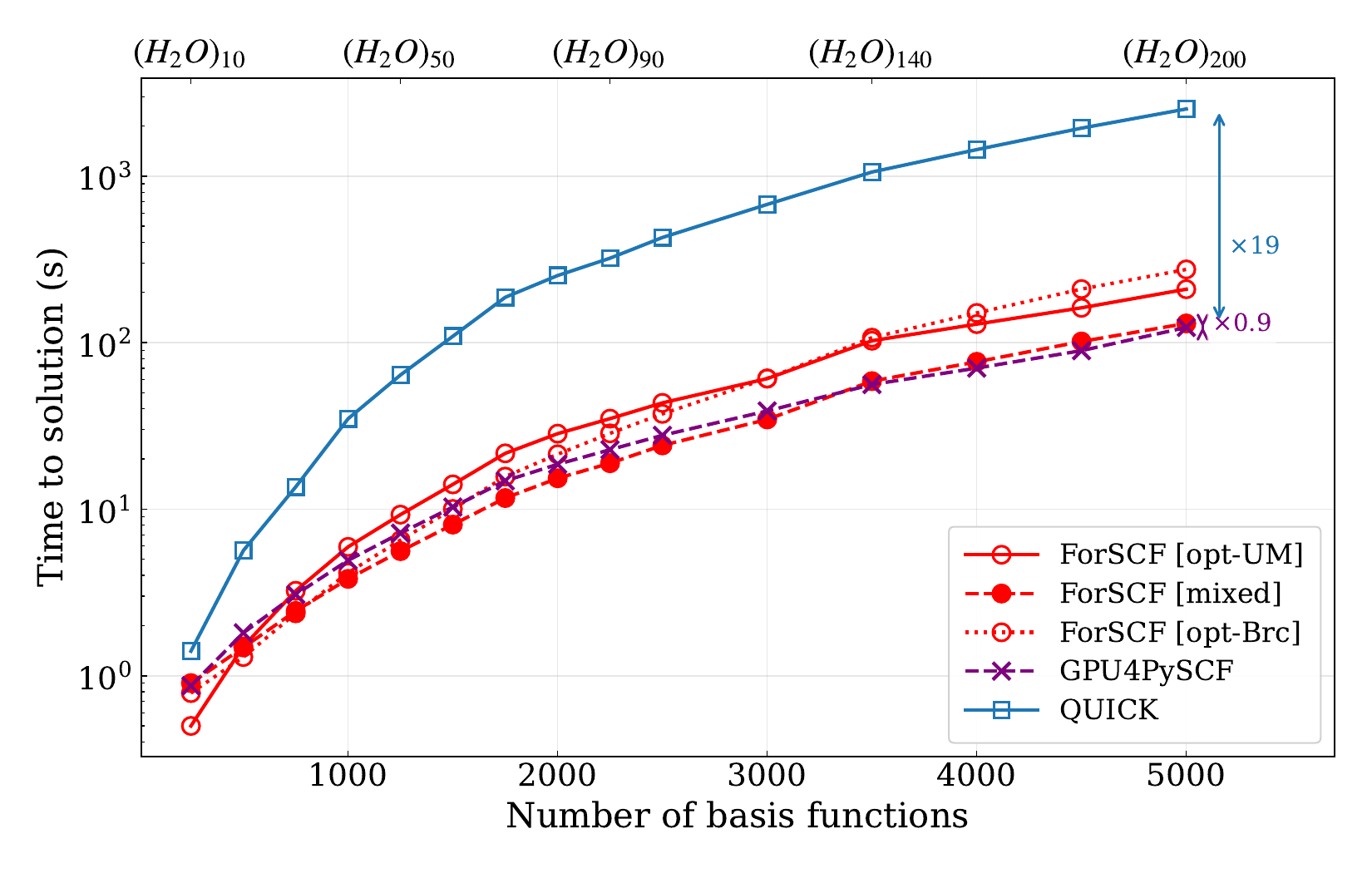}\caption{cc-pVDZ}
\end{subfigure}\hfill
\begin{subfigure}[t]{0.49\textwidth}\centering
\includegraphics[width=\textwidth]{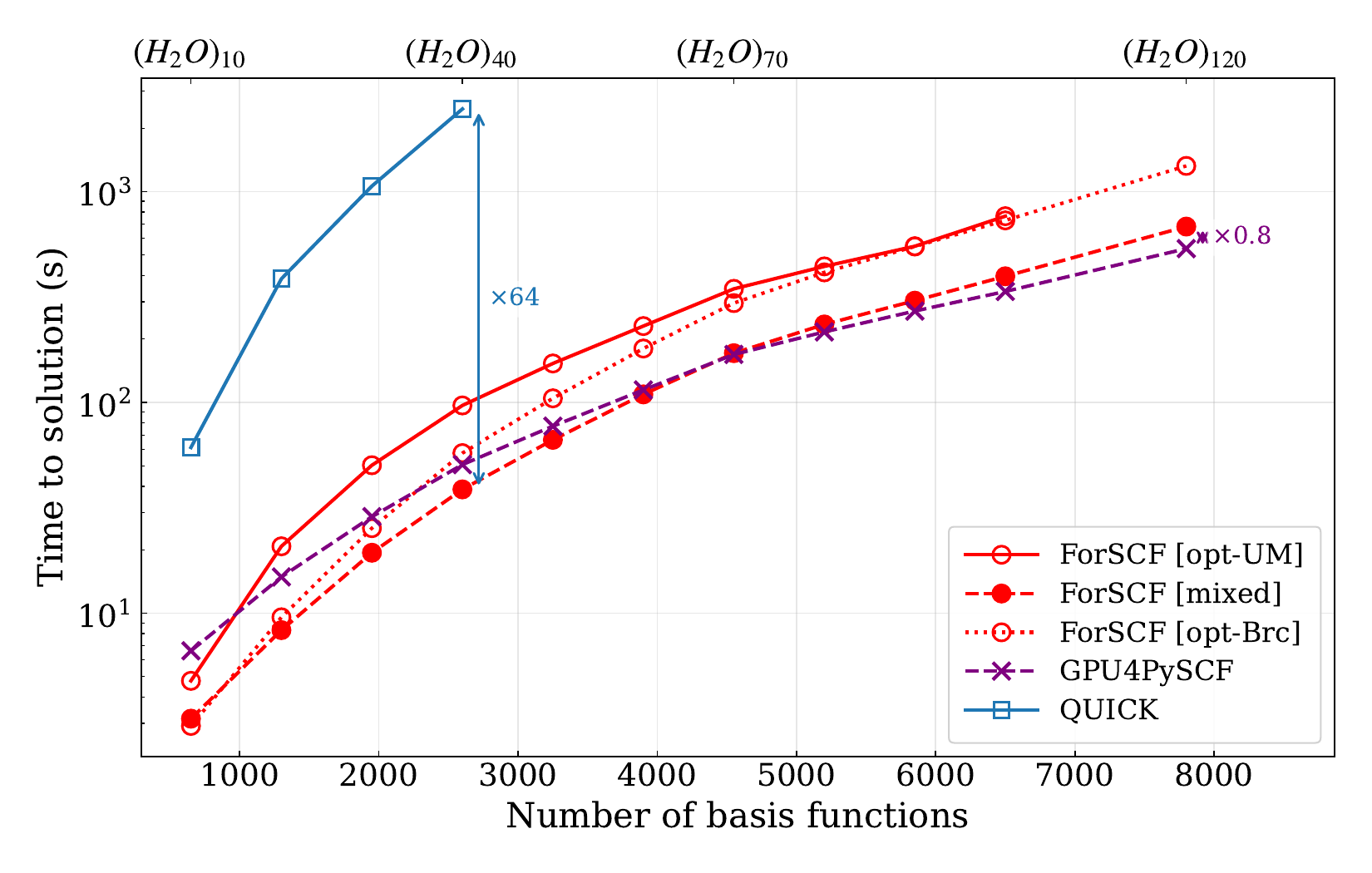}\caption{cc-pVTZ}
\end{subfigure}
\caption{Time to solution for water clusters, with details as in Figure~\ref{fig:convgly}.}
\label{fig:convwat}
\end{figure}

Figures~\ref{fig:convgly} and \ref{fig:convwat} and Table~\ref{tab:convergence} give the time to solution of complete RHF calculations at the same settings.
At the largest systems both ran, ForSCF (mixed) converges \cvspbestPaGly{} and \cvspbestPaWat{} times as fast as GPU4PySCF for 6-31G* polyglycine and water clusters, and \cvspbestDZGly{} and \cvspbestTZGly{} times as fast for polyglycine with cc-pVDZ and cc-pVTZ, while for the water clusters with these basis sets GPU4PySCF is faster, from \cvCrossDZWat{} and \cvCrossTZWat{} on, by factors of \cvslowDZWat{} and \cvslowTZWat{} at the largest clusters.
The ForSCF series end earlier than those of GPU4PySCF with cc-pVDZ and cc-pVTZ, where ForSCF needs more host memory than the node has.
\begin{table}[!htb]
\caption{Time to solution (s) and iterations at matched accuracy}
\label{tab:convergence}
\centering\small
\begin{threeparttable}
% Generated by scripts/make_results.py; do not edit.
\begin{tabular}{llrrrrrr}
\toprule
basis & system & $N_\mathrm{bf}$ & GPU4PySCF & opt-UM & mixed & opt-Brc & speedup\textsuperscript{\emph{b}} \\
\midrule
STO-3G & gly120 & 2767 & 20.3 (10) & 13.2 (11) & 12.5 (11) & 14.3 (11) & 1.62 \\
STO-3G & water200 & 1400 & 16.8 (11) & 15.8 (15) & 10.7 (15) & 11.3 (15) & 1.57 \\
6-31G* & gly120 & 7939 & 146 (11) & 115 (11) & 96.8 (11) & 204 (11) & 1.51 \\
6-31G* & water200 & 3800 & 59.1 (8) & 54.5 (9) & 38.9 (9) & 79.4 (9) & 1.52 \\
6-31G** & gly120 & 9025 & 190 (11) & 146 (11) & 124 (11) & 281 (11) & 1.53 \\
6-31G** & water200 & 5000 & 95.8 (8) & 97.2 (10) & 75.1 (10) & 167 (10) & 1.28 \\
cc-pVDZ & gly110 & 8275 & 160 (11) & 204 (12) & 153 (11) & 351 (12) & 1.05 \\
cc-pVDZ & water200 & 5000 & 123 (9) & 209 (10) & 130 (10) & 275 (10) & 0.95 \\
cc-pVTZ & gly40 & 7465 & 330 (10) & 557 (11) & 224 (11) & 774 (11) & 1.47 \\
cc-pVTZ & water120 & 7800 & 537 (9) & -- & 683 (10) & 1324 (10) & 0.79 \\
\bottomrule
\end{tabular}

\begin{tablenotes}
\item Largest system run by both GPU4PySCF and ForSCF (mixed), with iterations in parentheses, on one V100. --: exceeds the host memory of the node.
\item[\emph{b}] Time of GPU4PySCF divided by that of the fastest ForSCF scheme.
\end{tablenotes}
\end{threeparttable}
\end{table}
Both programs need about the same number of iterations (median \itmedMX{} and \itmedGP{} per basis set), so the time to solution follows the build times, diluted by the work outside the builds: setup, guess, diagonalization and DIIS take \cvnfMXPaGly\% and \cvnfMXPaWat\% of ForSCF's run for the 6-31G* systems and \cvnfMXTZGly\% and \cvnfMXTZWat\% for cc-pVTZ.
For gly$_{120}$ with 6-31G*, the 11 Fock builds take \SI{50}{s} of the \SI{97}{s}; the 11 diagonalizations, \SI{23}{s}; the setup of the one-electron integrals, the orthogonalization and the screening and pair data, \SI{14}{s}; and DIIS, \SI{5}{s}. Diagonalization, which scales as $N^3$, grows faster than the builds and weighs most for the large systems with the smaller basis sets.
QUICK needs a median of \itmedQK{} iterations.
The converged energies of ForSCF lie within \cvmaxdeMX\Eh{} (mixed), \cvmaxdeUM\Eh{} (opt-UM) and \cvmaxdeBRC\Eh{} (opt-Brc) of the references, those of GPU4PySCF within \cvmaxdeGP\Eh{} and those of QUICK within \cvmaxdeQK\Eh. The opt-UM and opt-Brc values above $10^{-5}\Eh$ (\cvmissUM{} and \cvmissBRC{} of about 140 calculations) are per-build errors just below the target plus the convergence tolerance.

\FloatBarrier
\subsection{Comparison to CPU Programs}
\label{sec:cpures}
\begin{figure}[!htb]
\centering
\begin{subfigure}[t]{0.49\textwidth}\centering
\includegraphics[width=\textwidth]{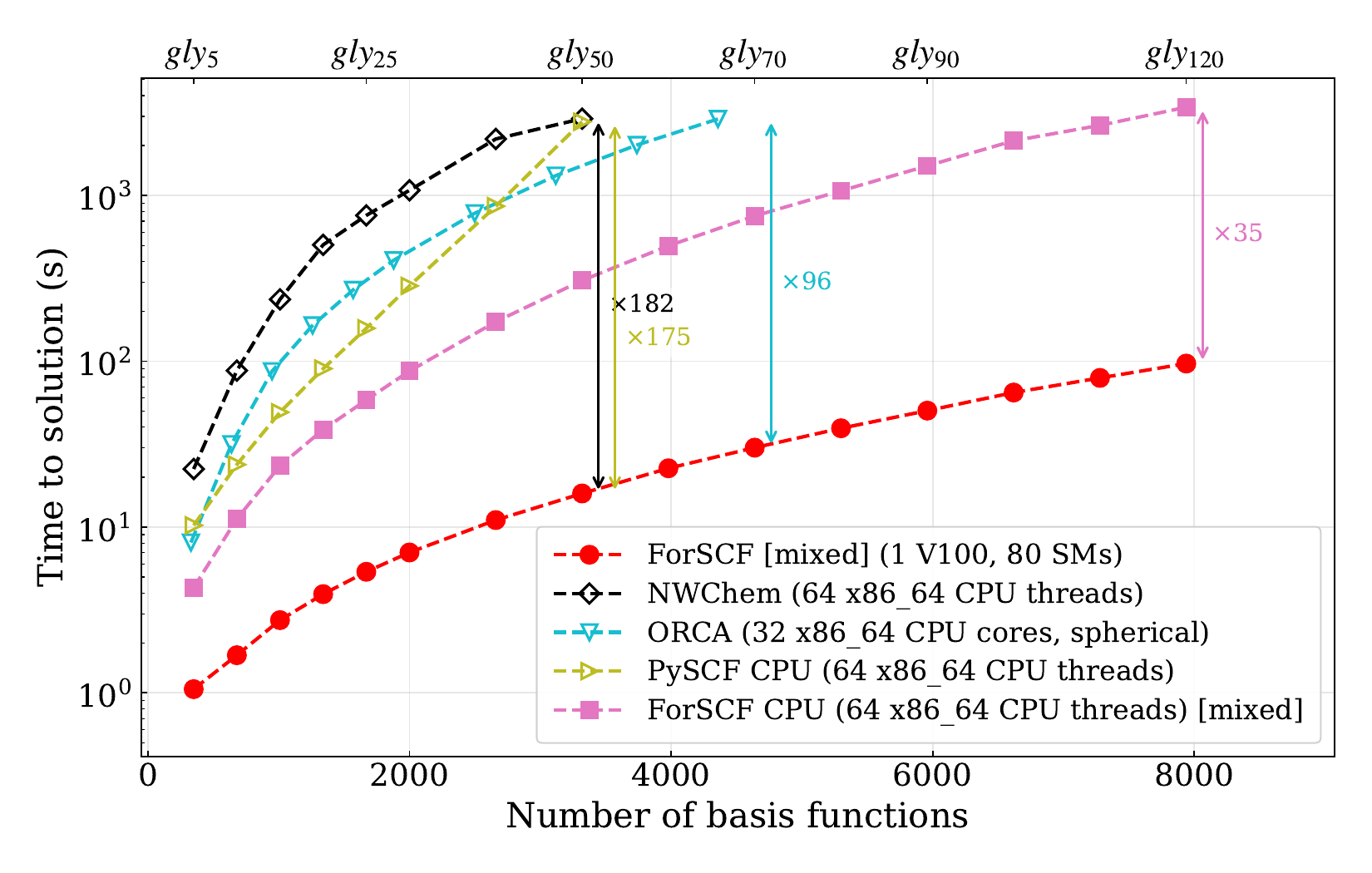}\caption{polyglycine, 6-31G*}
\end{subfigure}\hfill
\begin{subfigure}[t]{0.49\textwidth}\centering
\includegraphics[width=\textwidth]{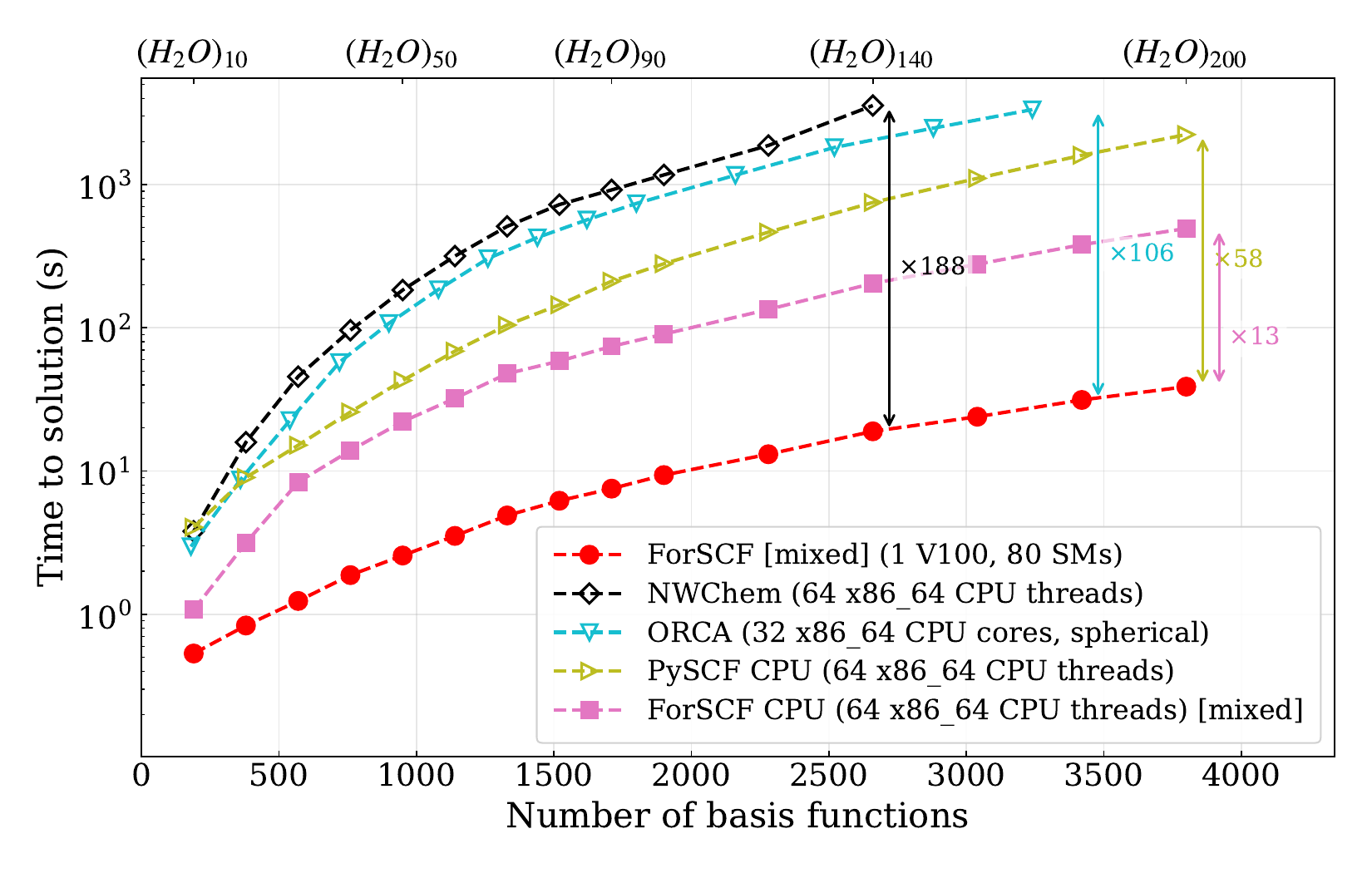}\caption{water clusters, 6-31G*}
\end{subfigure}
\caption{Time to solution of ForSCF on one V100 and on 64 CPU threads, and of NWChem (64 MPI ranks), PySCF (64 threads) and ORCA (32 MPI processes, drawn at its own number of spherical functions) on the same node, with arrows giving the ratio of each program to ForSCF on the GPU. 6-31G** and STO-3G are given in the Supporting Information.}
\label{fig:cpu}
\end{figure}

Figure~\ref{fig:cpu} and Table~\ref{tab:cpu} compare ForSCF on one V100 with ForSCF's CPU build and with NWChem, PySCF and ORCA on the whole node.
At the largest systems native NWChem completed, ForSCF on one V100 converges RHF calculations \cpuNwVsGpuMin--\cpuNwVsGpuMax{} times as fast as NWChem, \cpuPyVsGpuMin--\cpuPyVsGpuMax{} times as fast as PySCF and \cpuOrVsGpuMin--\cpuOrVsGpuMax{} times as fast as ORCA.
The CPU build of ForSCF, compiled from the same source, is \cpuRatioMin--\cpuRatioMax{} times slower than one V100 and \cpuVsNwMin--\cpuVsNwMax{} times faster than NWChem's direct SCF on the same node.
The two factors separate the sources of the speedup over NWChem, which is their product: the CPU build runs the same integral schemes, screening and SCF as the GPU build on the same processor as NWChem, so that its lead over NWChem comes from the algorithms, including the fewer Fock builds of DIIS against NWChem's Newton--Raphson solver (Section~\ref{sec:nwres}), and the ratio of the CPU and GPU builds of ForSCF is that of the hardware for the same algorithm.
PySCF and ORCA lie between NWChem and the CPU build of ForSCF on every system of Table~\ref{tab:cpu}.
\begin{table}[!htb]
\caption{Time to solution (s) of ForSCF on one V100 and of CPU programs on one node}
\label{tab:cpu}
\centering\small
\begin{threeparttable}
% Generated by scripts/make_results.py; do not edit.
\begin{tabular}{llrrrrrr}
\toprule
basis & system & $N_\mathrm{bf}$ & ForSCF GPU & ForSCF CPU & NWChem & PySCF & ORCA \\
\midrule
STO-3G & gly100 & 2307 & 8.80 & 126 & 1029 & 547 & 488 (2307) \\
STO-3G & water200 & 1400 & 10.7 & 85.2 & 703 & 312 & 460 (1400) \\
6-31G* & gly50 & 3319 & 15.9 & 307 & 2907 & 2786 & 1318 (3118) \\
6-31G* & water140 & 2660 & 19.0 & 204 & 3563 & 750 & 1816 (2520) \\
6-31G** & gly40 & 3025 & 13.3 & 249 & 2758 & 1572 & 1089 (2864) \\
6-31G** & water120 & 3000 & 21.1 & 263 & 3492 & 831 & 2061 (2880) \\
cc-pVDZ & gly30 & 2275 & 13.7 & 159 & 2330 & 456 & 1361 (2154) \\
cc-pVDZ & water100 & 2500 & 24.1 & 232 & 3078 & 540 & 2758 (2400) \\
\bottomrule
\end{tabular}

\begin{tablenotes}
\item Largest system completed by native NWChem. ForSCF: mixed scheme. ForSCF CPU and PySCF: 64 OpenMP threads. NWChem: 64 MPI ranks, direct SCF. ORCA: 32 MPI processes and spherical functions, with its own $N_\mathrm{bf}$ in parentheses.
\end{tablenotes}
\end{threeparttable}
\end{table}
As in Ref.~\citenum{Palethorpe2024}, the power efficiency is estimated from the nominal thermal design power (TDP) of the processors that do the work: \SI{\tdpGPU}{W} for one V100-SXM2 and \SI{\tdpCPU}{W} for the Threadripper PRO 7975WX, so that the CPU draws \tdpRatio{} times the power of the GPU.
At the largest systems ORCA completed, ForSCF on one V100 is \orSpMin--\orSpMax{} times as fast as ORCA on the whole node, which corresponds to a \orEffMin--\orEffMax{} times higher power efficiency (largest for \orEffMaxSys). The host CPU, which is mostly idle while ForSCF builds on the GPU, is not counted, as in Ref.~\citenum{Palethorpe2024}.

\FloatBarrier
\subsection{ForSCF as the Fock-Build Engine of NWChem}
\label{sec:nwres}
\begin{figure}[!htb]
\centering
\begin{subfigure}[t]{0.49\textwidth}\centering
\includegraphics[width=\textwidth]{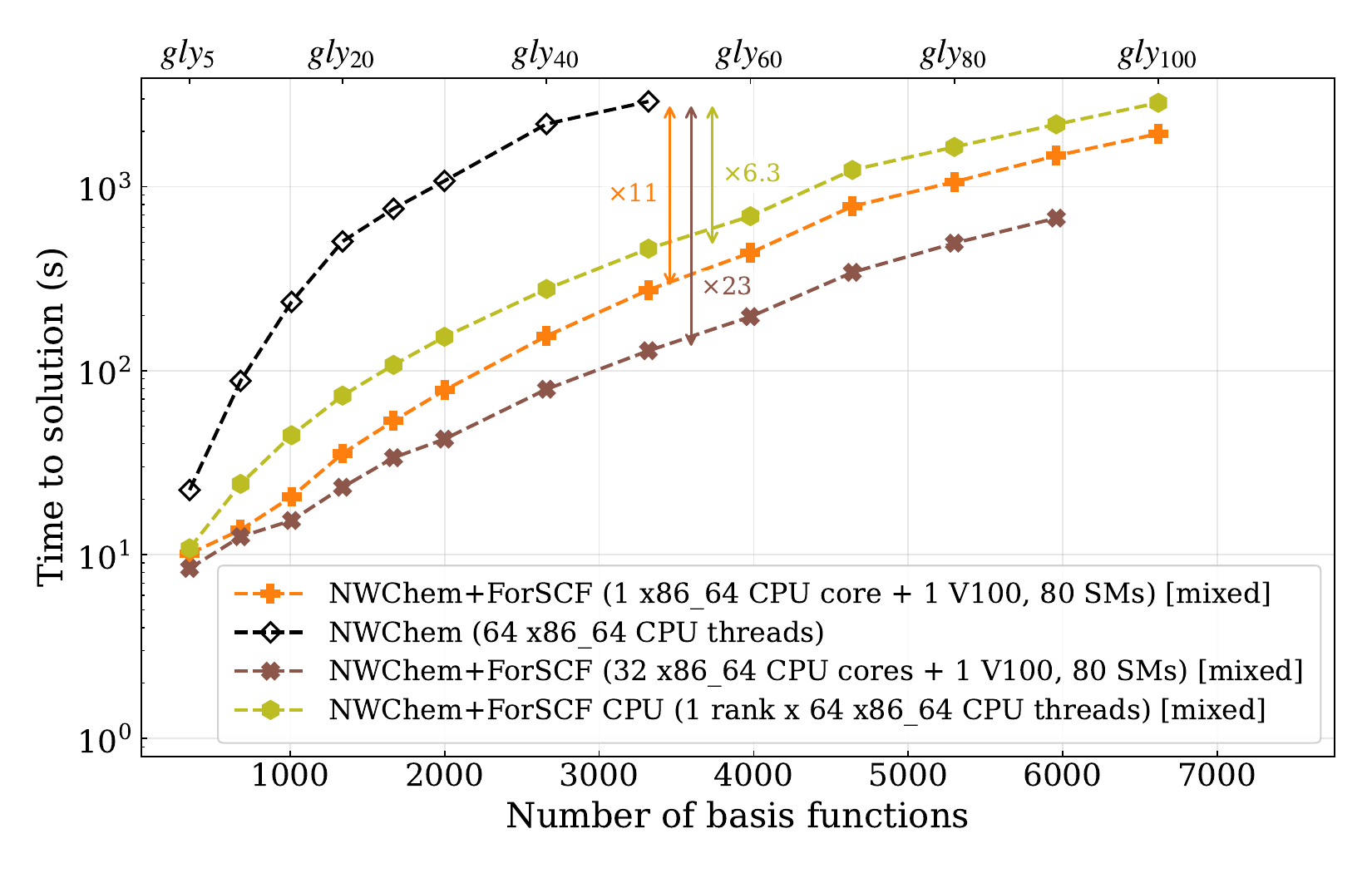}\caption{polyglycine, 6-31G*}
\end{subfigure}\hfill
\begin{subfigure}[t]{0.49\textwidth}\centering
\includegraphics[width=\textwidth]{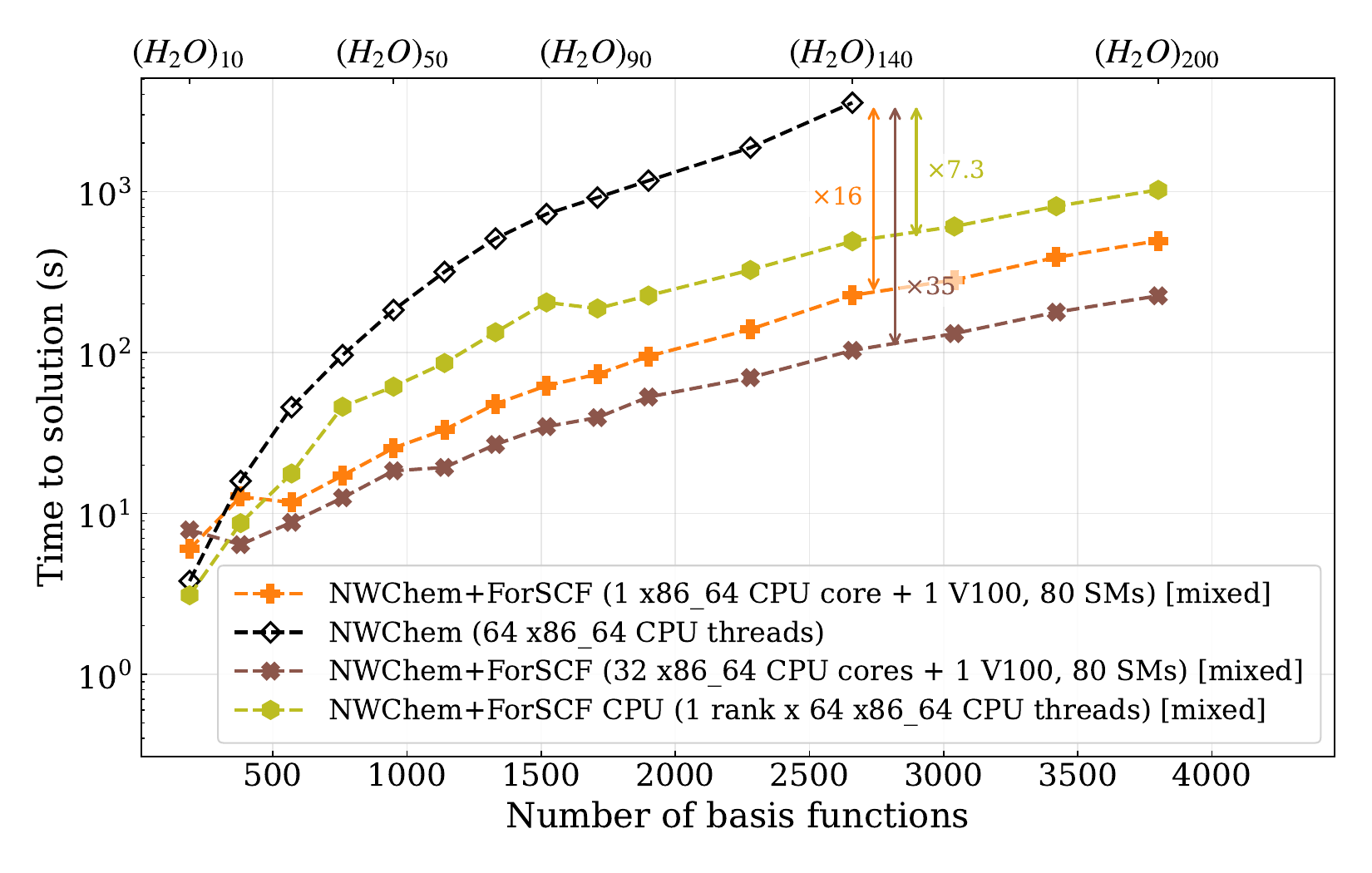}\caption{water clusters, 6-31G*}
\end{subfigure}\\[0.5em]
\begin{subfigure}[t]{0.49\textwidth}\centering
\includegraphics[width=\textwidth]{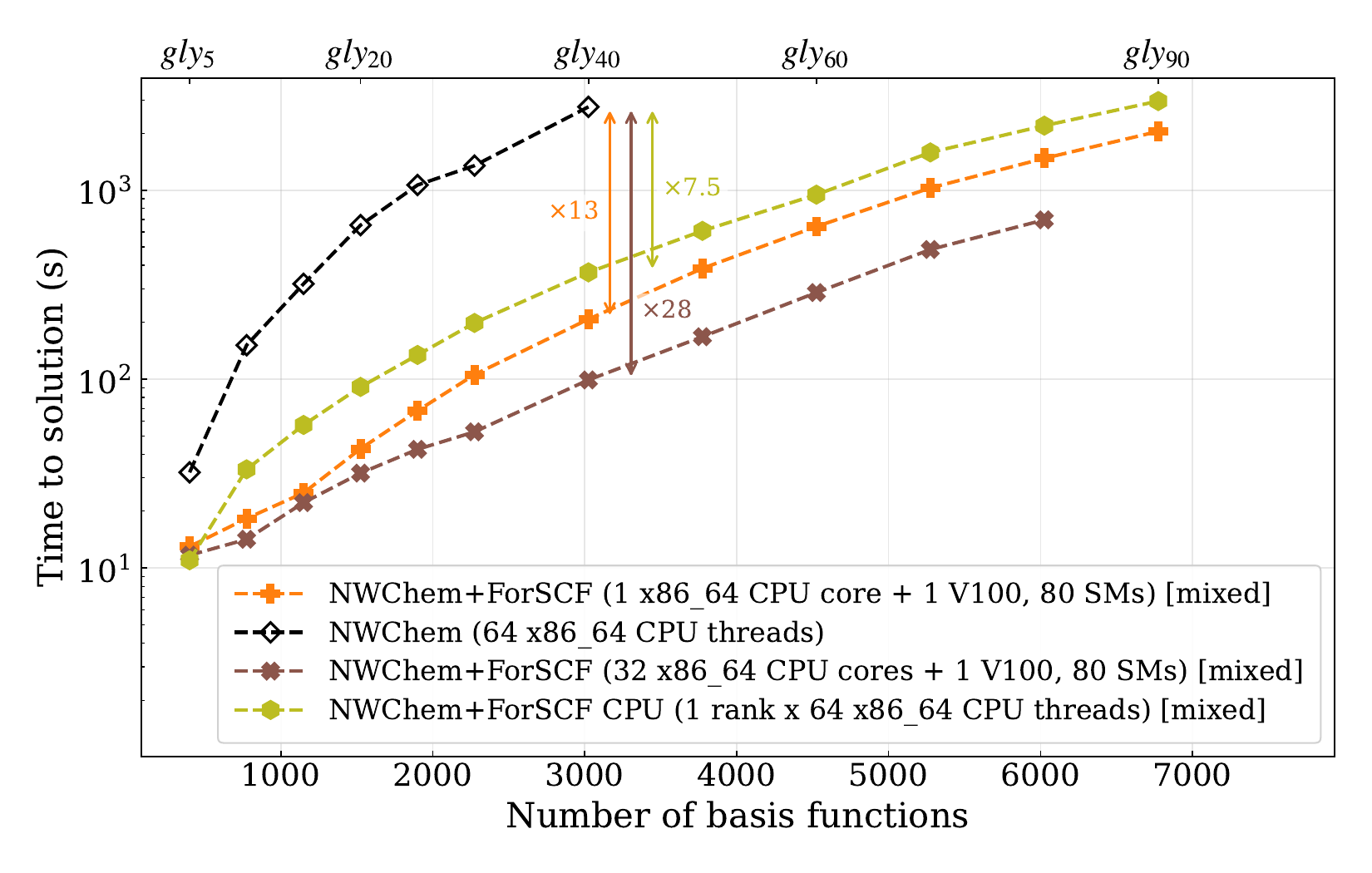}\caption{polyglycine, 6-31G**}
\end{subfigure}\hfill
\begin{subfigure}[t]{0.49\textwidth}\centering
\includegraphics[width=\textwidth]{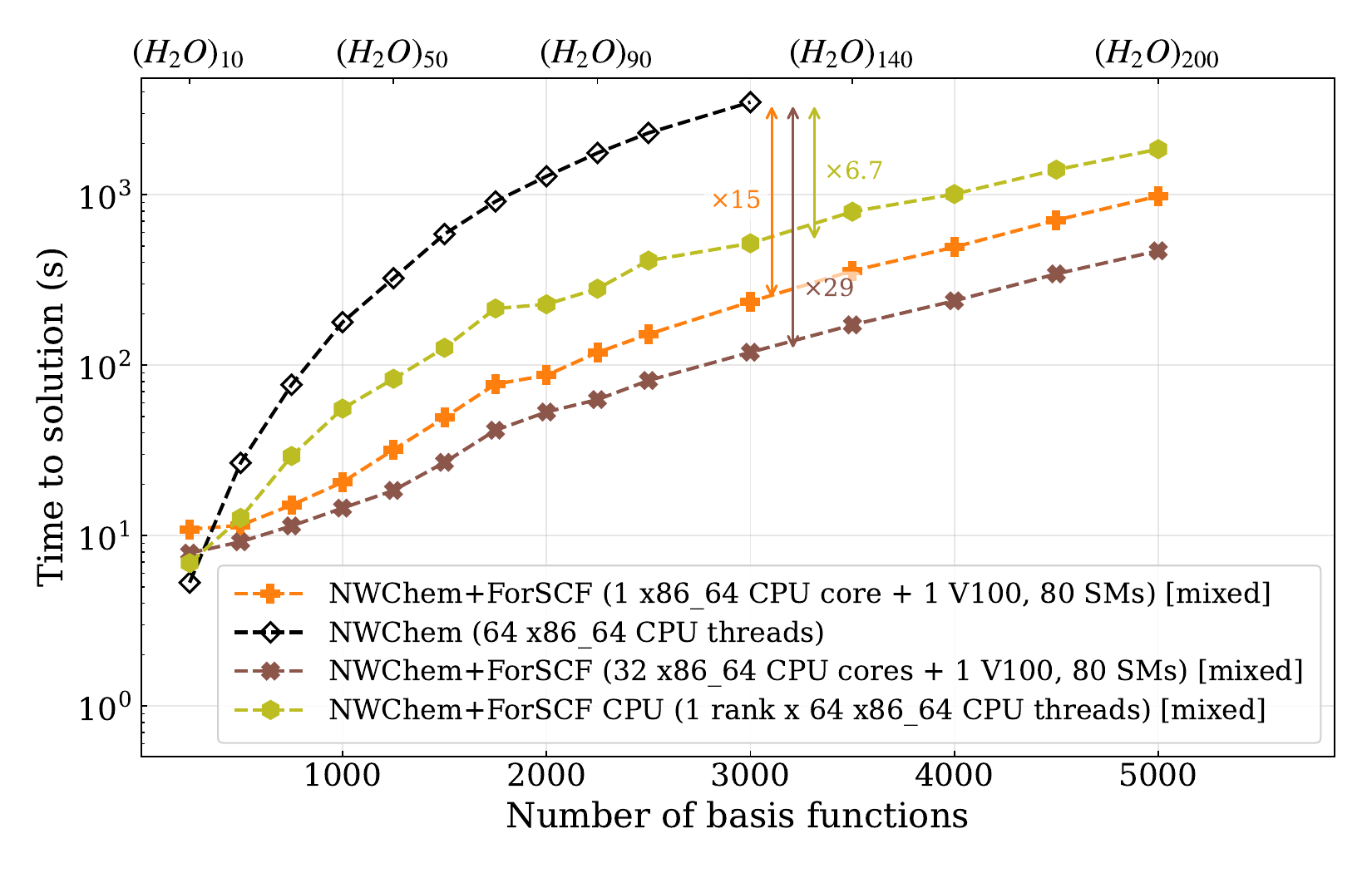}\caption{water clusters, 6-31G**}
\end{subfigure}
\caption{Time to solution of NWChem's direct RHF on 64 MPI ranks, alone and with ForSCF as its Fock builder on one V100 (with 32 ranks or one rank) or on 64 CPU threads, with arrows giving the speedup over native NWChem at the largest system native NWChem completed.}
\label{fig:nwchem}
\end{figure}

NWChem is the example of a host program (Section~\ref{sec:hosts}). Figure~\ref{fig:nwchem} and Table~\ref{tab:nwchem} compare NWChem's direct SCF on 64 MPI ranks with the same calculation using ForSCF as its Fock builder, with \NWSource.
With one V100, NWChem's SCF is \nwspMin--\nwspMax{} times faster at the largest systems that native NWChem completed within the one-hour limit and the memory of the node, and it continues to polyglycine chains of 80 to 120 units. With the CPU build of ForSCF on the same node, it is \nwCpuSpMin--\nwCpuSpMax{} times faster.
\begin{table}[!htb]
\caption{Time to solution (s) of NWChem with and without ForSCF}
\label{tab:nwchem}
\centering\small
\begin{threeparttable}
% Generated by scripts/make_results.py; do not edit.
\begin{tabular}{llrrrr}
\toprule
basis & system & $N_\mathrm{bf}$ & NWChem & + ForSCF GPU & + ForSCF CPU \\
\midrule
STO-3G & gly100 & 2307 & 1029 & 78.9 (13.0) & 317 (3.2) \\
STO-3G & water200 & 1400 & 703 & 27.6 (25.5) & 157 (4.5) \\
6-31G* & gly50 & 3319 & 2907 & 128 (22.7) & 460 (6.3) \\
6-31G* & water140 & 2660 & 3563 & 103 (34.6) & 491 (7.3) \\
6-31G** & gly40 & 3025 & 2758 & 99.0 (27.9) & 367 (7.5) \\
6-31G** & water120 & 3000 & 3492 & 119 (29.3) & 518 (6.7) \\
cc-pVDZ & gly30 & 2275 & 2330 & 71.6 (32.5) & 281 (8.3) \\
cc-pVDZ & water100 & 2500 & 3078 & 110 (27.9) & 539 (5.7) \\
\bottomrule
\end{tabular}

\begin{tablenotes}
\item Largest system completed by native NWChem, with the speedup over native NWChem in parentheses. NWChem: 64 MPI ranks, direct SCF. + ForSCF GPU: 32 MPI ranks, with ForSCF building on one V100. + ForSCF CPU: one rank with ForSCF on 64 OpenMP threads.
\end{tablenotes}
\end{threeparttable}
\end{table}
The times include all of NWChem's own work: its Newton--Raphson solver performs \nwBuildsLo--\nwBuildsHi{} Fock builds per calculation, up to about twice as many as the iterations of ForSCF's own SCF, and its other steps run on the CPU ranks.
All \cvnNWF{} calculations with ForSCF lie within \cvmaxdeNWF\Eh{} of the references.

\FloatBarrier
\subsection{Additional Benchmarks}
\label{sec:additional}
\subsubsection{Multi-GPU Scaling}
\label{sec:multigpures}
\begin{figure}[!htb]
\centering
\includegraphics[width=0.55\textwidth]{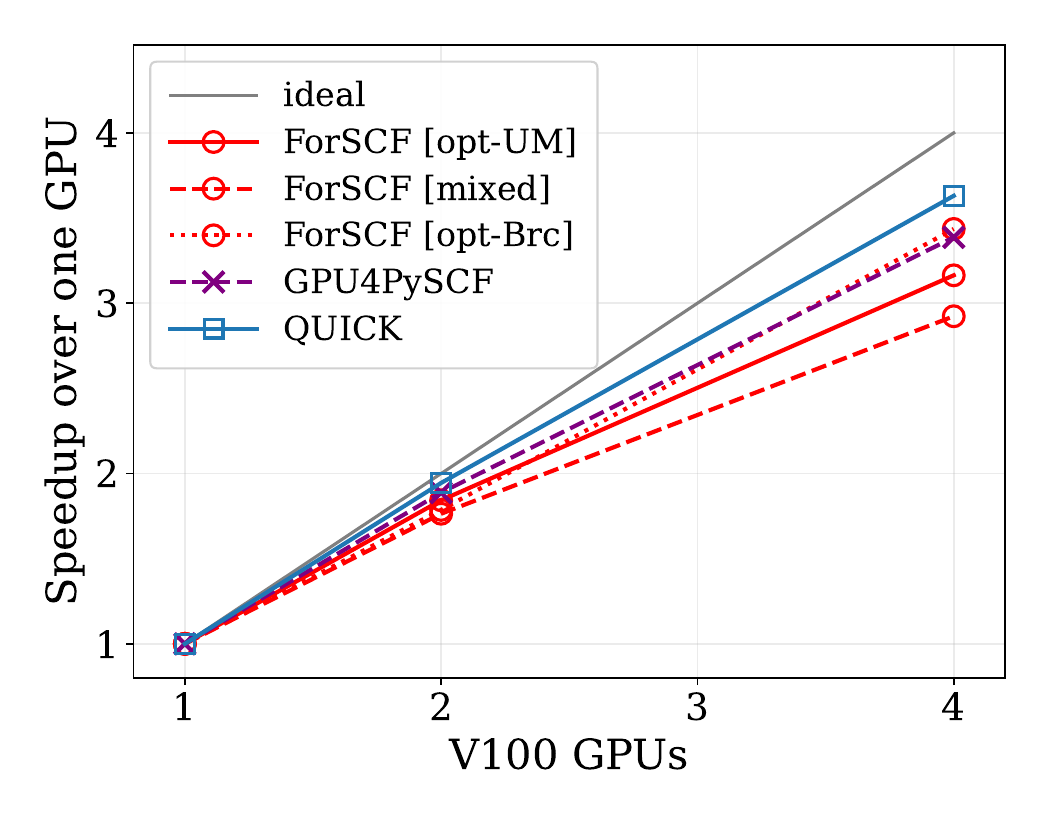}
\caption{Speedup of the Fock build of \mgSys{} over one V100 on one, two and four NVLink-connected V100s, each program at its selected setting (Table~\ref{tab:thresholds}), with the ideal speedup in gray.}
\label{fig:multigpu}
\end{figure}

Table~\ref{tab:multigpu} and Figure~\ref{fig:multigpu} give the Fock-build times and speedups of \mgSys{} (\mgNbf{} basis functions) on one, two and four V100s at the settings of Table~\ref{tab:thresholds}. \mgSys{} is the largest polyglycine chain at which four processes of every program fit in the host memory of the node.
QUICK runs its CUDA and MPI build with one MPI process per GPU.
\begin{table}[!htb]
\caption{Fock-build time (s) on one to four V100 GPUs}
\label{tab:multigpu}
\centering\small
\begin{threeparttable}
% Generated by scripts/make_results.py; do not edit.
\begin{tabular}{lrrrrrr}
\toprule
program & 1 GPU & 2 GPUs & 4 GPUs & $S_2$ & $S_4$ & $E_4$ \\
\midrule
ForSCF (opt-UM) & 2.54 & 1.38 & 0.802 & 1.84 & 3.16 & 79\% \\
ForSCF (mixed) & 1.61 & 0.913 & 0.551 & 1.76 & 2.92 & 73\% \\
ForSCF (opt-Brc) & 3.18 & 1.78 & 0.926 & 1.79 & 3.44 & 86\% \\
GPU4PySCF & 1.91 & 1.01 & 0.564 & 1.89 & 3.39 & 85\% \\
QUICK & 34.9 & 17.9 & 9.61 & 1.94 & 3.63 & 91\% \\
\bottomrule
\end{tabular}

\begin{tablenotes}
\item \mgSys, settings of Table~\ref{tab:thresholds}, NVLink-connected V100s. $S_n$ is the speedup over one GPU, and $E_4=S_4/4$. ForSCF was built without $f$ functions here, which lowers its host memory. QUICK: CUDA and MPI build, one MPI process per GPU.
\end{tablenotes}
\end{threeparttable}
\end{table}
opt-Brc scales best among the ForSCF schemes, with a parallel efficiency of \mgEffBRC\% on four GPUs against \mgEffUM\% for opt-UM and \mgEffMX\% for the mixed scheme, and GPU4PySCF reaches \mgEffGP\%.
The mixed scheme gives the shortest build on one, two and four GPUs, \mgTMX\,s on four against \mgTGP\,s for GPU4PySCF.
QUICK scales best of all programs, with \mgEffQK\%, but its single-GPU build is \mgOneQKvsMX{} times slower than that of the mixed scheme, so the work that every process repeats is small next to the integral work they share; on four GPUs it takes \mgTQK\,s.
As for opt-UM in Ref.~\citenum{Palethorpe2024}, the losses come from the parts of each build that every process repeats (mainly the screening setup and the transfer of the density and of the result between host and GPU); from imbalance between the processes, to which the slower host transfers of two of the four GPUs contribute; and from the reduction through host memory. These losses weigh most on the mixed scheme, whose single-GPU build is the fastest.

\FloatBarrier
\subsubsection{Coulomb and Exchange Builds}
\label{sec:jkres}\label{sec:jksplit}\label{sec:jkscaling}
Table~\ref{tab:jksplit} separates the Coulomb and exchange builds at the largest systems, and Table~\ref{tab:jkscaling} gives the effective exponents $t\propto N^{x}$ over the polyglycine chains.
At the largest systems, the opt-UM Coulomb build is faster than the J engine of GPU4PySCF for 6-31G* and cc-pVTZ and slower for cc-pVDZ, and opt-Brc exchange is faster in every case except the cc-pVTZ water clusters (ratio \jkKrTZWat).
The trends of Section~\ref{sec:throughput} follow from how the two ratios and the shares of the two builds change with size.
\begin{table}[!htb]
\caption{Coulomb and exchange build times (s) at matched accuracy}
\label{tab:jksplit}
\centering\small
\begin{threeparttable}
% Generated by scripts/make_results.py; do not edit.
\begin{tabular}{llrrrrrrr}
\toprule
 & & & \multicolumn{2}{c}{ForSCF (mixed)} & \multicolumn{2}{c}{GPU4PySCF} & \multicolumn{2}{c}{ratio\textsuperscript{\emph{a}}} \\
\cmidrule(lr){4-5}\cmidrule(lr){6-7}\cmidrule(lr){8-9}
basis & system & $N_\mathrm{bf}$ & $\J$ & $\K$ & $\J$ & $\K$ & $\J$ & $\K$ \\
\midrule
6-31G* & $\mathrm{gly}_{120}$ & 7939 & 3.93 & 1.73 & 5.12 & 4.09 & 1.30 & 2.36 \\
6-31G* & $(\mathrm{H_2O})_{200}$ & 3800 & 1.35 & 2.49 & 1.75 & 4.70 & 1.30 & 1.89 \\
cc-pVDZ & $\mathrm{gly}_{110}$ & 8275 & 7.65 & 2.38 & 5.06 & 4.91 & 0.66 & 2.06 \\
cc-pVDZ & $(\mathrm{H_2O})_{200}$ & 5000 & 3.93 & 8.64 & 3.19 & 8.92 & 0.81 & 1.03 \\
cc-pVTZ & $\mathrm{gly}_{50}$ & 9315 & 10.2 & 16.5 & 21.8 & 19.3 & 2.14 & 1.17 \\
cc-pVTZ & $(\mathrm{H_2O})_{140}$ & 9100 & 12.7 & 97.3 & 22.2 & 57.1 & 1.74 & 0.59 \\
\bottomrule
\end{tabular}

\begin{tablenotes}
\item Separate Coulomb-only and exchange-only builds on one V100 on the fixed densities, at the settings of Table~\ref{tab:thresholds}, for the largest systems both programs ran.
\item[\emph{a}] Time of GPU4PySCF divided by that of ForSCF.
\end{tablenotes}
\end{threeparttable}
\end{table}
\begin{table}[!htb]
\caption{Effective scaling exponents of the Coulomb and exchange builds}
\label{tab:jkscaling}
\centering\small
\begin{threeparttable}
% Generated by scripts/make_results.py; do not edit.
\begin{tabular}{lcccccc}
\toprule
 & \multicolumn{2}{c}{ForSCF (mixed)} & \multicolumn{2}{c}{ForSCF (opt-UM)} & \multicolumn{2}{c}{GPU4PySCF} \\
\cmidrule(lr){2-3}\cmidrule(lr){4-5}\cmidrule(lr){6-7}
basis & $\J$ & $\K$ & $\J$ & $\K$ & $\J$ & $\K$ \\
\midrule
6-31G* & 2.05 & 1.40 & 2.03 & 1.14 & 1.82 & 1.13 \\
cc-pVDZ & 2.04 & 1.20 & 2.04 & 1.14 & 1.81 & 1.45 \\
cc-pVTZ & 2.03 & 1.25 & -- & -- & 1.85 & 1.22 \\
\bottomrule
\end{tabular}

\begin{tablenotes}
\item $t\propto N^{x}$ fitted over the polyglycine chains with at least \num{2000} basis functions (gly$_{40}$--gly$_{120}$ for 6-31G*, gly$_{30}$--gly$_{110}$ for cc-pVDZ, gly$_{15}$--gly$_{50}$ for cc-pVTZ), on one V100. --: not measured.
\end{tablenotes}
\end{threeparttable}
\end{table}
The Coulomb ratio falls along every polyglycine series, from \jkcFirstJrPaGly{} to \jkcLastJrPaGly{} between \jkcFirstSysPaGly{} and \jkcLastSysPaGly{} with 6-31G*, from \jkcFirstJrDZGly{} to \jkcLastJrDZGly{} with cc-pVDZ and from \jkcFirstJrTZGly{} to \jkcLastJrTZGly{} with cc-pVTZ, since the opt-UM Coulomb build scales as $N^{\xJMXDZ}$ against $N^{\xJGPDZ}$. Its primitive screening keeps more far-field quartets than the contracted screening of GPU4PySCF, and their number grows with the size of the system (Section~\ref{sec:sweep}).
The exchange ratio falls less or not at all over the chains: it stays near three with 6-31G*, falling only to \jkcLastKrPaGly{} at \jkcLastSysPaGly; it rises from \jkcFirstKrDZGly{} to \jkcLastKrDZGly{} with cc-pVDZ, as opt-Brc exchange scales as $N^{\xKMXDZ}$ against $N^{\xKGPDZ}$; and it stays near \jkcLastKrTZGly{} with cc-pVTZ.
Since the exchange build of ForSCF also grows more slowly than its Coulomb build, the Coulomb share of its build rises along the chains, from \jkcFirstJsharePaGly\% to \jkcLastJsharePaGly\% with 6-31G* and from \jkcFirstJshareDZGly\% to \jkcLastJshareDZGly\% with cc-pVDZ, and the ratio of the full build approaches the Coulomb ratio.
With 6-31G* this lowers the ratio steadily, whereas with cc-pVDZ the rising exchange ratio offsets the falling Coulomb ratio over the shorter chains, which keeps the full ratio near \spPeakDZGly{} up to gly$_{40}$.
With cc-pVTZ, exchange dominates the build over the whole series, as the Coulomb share rises only from \jkcFirstJshareTZGly\% to \jkcLastJshareTZGly\%, and the Coulomb ratio stays above two, so that the full ratio stays nearly constant. The series ends at gly$_{50}$, at the memory of the node, before the Coulomb build could dominate as it does from gly$_{30}$ on with cc-pVDZ.
For the water clusters, exchange is the larger part of ForSCF's build at the largest sizes of every basis set (Table~\ref{tab:jksplit}) and at every size with cc-pVTZ, whose exchange ratio falls below one beyond \jkcKcrossTZWat{} and reaches \jkcLastKrTZWat{} at \jkcLastSysTZWat, as the exchange build of ForSCF grows as $N^{\jkcxKMXTZWat}$ against $N^{\jkcxKGPTZWat}$ over the clusters with at least \num{2000} basis functions, while over the chains both grow as $N^{\xKMXTZ}$ and $N^{\xKGPTZ}$.
In the compact clusters the number of exchange quartets that survive screening keeps growing with size, and most of them fall in the highly contracted $d$ and $f$ classes of cc-pVTZ. The faster growth is attributed to these classes, whose opt-Brc kernels are split to limit register spills and run with one bra pair per thread block at low occupancy (Section~\ref{sec:brcgpu}), whereas GPU4PySCF stages their data in shared memory for larger thread blocks.
Along the chains the surviving quartets per unit of length stop growing early, so that both exchange builds grow nearly linearly and their ratio stays constant.
opt-UM exchange scales nearly linearly, as $N^{\xKUMPa}$ for 6-31G*, as in Ref.~\citenum{Palethorpe2024}, but over the sizes studied it is slower than opt-Brc exchange, whose cost grows as $N^{\xKMXDZ}$ to $N^{\xKMXPa}$.

\section{Conclusions}
\label{sec:conclusions}
ForSCF implements the opt-Brc and opt-UM schemes of Palethorpe et al.\ with kernels generated from Head-Gordon--Pople and McMurchie--Davidson recurrence graphs.
It differs from the original CUDA implementation mainly in its programming model. Both schemes are written in Fortran with compiler directives, and one source builds with OpenACC for NVIDIA and AMD GPUs and with OpenMP for multicore CPUs, so that the same algorithms run on both without a second implementation.
The other adaptations are of three kinds. The first concerns the smaller L2 cache and register file of the V100, chiefly the splitting of opt-Brc classes over several kernels. The second concerns the directive model, where per-thread exits and shared-memory reductions become group-wide cut-offs and partial sums combined across the group. The third concerns details that the published description leaves open, such as the density-decay fit, $N^*$ and the Boys function.
The block bound of Luehr et al.\cite{Luehr2016} and a budgeted tail exit extend opt-UM exchange, and a mixed scheme builds the Coulomb matrix with opt-UM and the exchange matrix with opt-Brc, each with its own threshold, which makes it the fastest ForSCF scheme from medium system sizes on, whereas opt-Brc is the fastest for most small systems.

Programs with equal nominal screening thresholds do not compute Fock matrices of equal accuracy.
At the published thresholds, the primitive Coulomb screening of opt-UM lets converged energies deviate by up to \pubMaxDE\Eh{} on the largest systems, and calibration on fixed converged densities shows that it needs $\tau_J$ of $10^{-11}$ to $3\times10^{-11}$, whereas opt-Brc and GPU4PySCF meet the $10^{-5}\Eh$ target at looser values than the published ones.
At matched accuracy on one V100, ForSCF builds the Fock matrix on average \spAvgPopleMin--\spAvgPopleMax{} times as fast as GPU4PySCF with STO-3G and the Pople basis sets and \spAvgCcMin--\spAvgCcMax{} times as fast with the correlation-consistent ones, and converges complete RHF calculations up to \cvspMax{} times as fast.
The lead shrinks with system size: the opt-UM Coulomb build scales as $N^{\xJMXDZ}$ against $N^{\xJGPDZ}$ for the J engine of GPU4PySCF, while the opt-Brc exchange of the mixed scheme remains up to \jkKrPaGly{} times as fast as GPU4PySCF's.
Against CPU programs on all 32 cores of the same node, ForSCF on one V100 converges RHF calculations \cpuNwVsGpuMin--\cpuNwVsGpuMax{} times as fast as NWChem at the largest systems that NWChem completed, a factor that is the product of the lead of ForSCF's algorithms, which its CPU build shows on the same processor (\cpuVsNwMin--\cpuVsNwMax{} times), and of the GPU over the CPU for the same algorithms (\cpuRatioMin--\cpuRatioMax{} times).
Against ORCA, the speedup of \orSpMin--\orSpMax{} times corresponds to a power efficiency up to \orEffMax{} times higher, estimated from the nominal power of the processors.
As the Fock-build engine of another program, shown here for NWChem, ForSCF makes direct SCF \nwspMin--\nwspMax{} times faster with one GPU than NWChem alone on the same node, and \nwCpuSpMin--\nwCpuSpMax{} times faster with its CPU build on the same cores, without changing NWChem's algorithms.

Four limitations remain.
For the larger water clusters with the correlation-consistent basis sets, GPU4PySCF is faster, most clearly with cc-pVTZ: the opt-Brc exchange kernels of ForSCF are latency-bound at low occupancy with one warp per thread block, whereas GPU4PySCF stages its data in shared memory for larger thread blocks.
The Coulomb build scales less favorably than that of GPU4PySCF.
On four GPUs, opt-Brc reaches \mgEffBRC\% parallel efficiency, but the mixed scheme only \mgEffMX\%, since every process repeats parts of each build and holds a full copy of the screening and pair data, and the largest systems were limited by host memory.
The AMD GPU build has not been validated.
The next steps are exchange kernels with several warps per thread block for the highly contracted classes; a Coulomb build with lower scaling; distributed screening and pair data, so that more GPUs also extend the reachable system size; validation of the AMD GPU build; analytic gradients; and the use of ForSCF as the Fock-build engine of further quantum chemistry programs.

\section*{Data and Software Availability}
The ForSCF source code, including the kernel generators, will be made available at \url{https://github.com/mqsdk/forscf} with the final journal publication.
The molecular geometries and the basis sets are given in the Supporting Information.

\begin{suppinfo}
The Supporting Information is available via Zenodo (\url{https://doi.org/10.5281/zenodo.24560500}).
\begin{itemize}
\item Fock-build, time-to-solution, CPU and NWChem plots for STO-3G. Threshold-sweep plots and full threshold grids of the opt-UM and mixed schemes for every basis set. Fock-build times of all six programs at common system sizes. The differences of ForSCF from the published schemes, element by element. Settings and patches of the comparison programs. Example inputs of every program (PDF).
\item Geometries of all polyglycine chains and water clusters, taken from Ref.~\citenum{Palethorpe2024} and the basis sets in Gaussian94 format (ZIP).
\end{itemize}
\end{suppinfo}

\bibliography{references}

@article{Palethorpe2024,
  author = {Palethorpe, Elise and Stocks, Ryan and Barca, Giuseppe M. J.},
  title = {Advanced Techniques for High-Performance {Fock} Matrix Construction on {GPU} Clusters},
  journal = {Journal of Chemical Theory and Computation},
  year = {2024},
  volume = {20},
  pages = {10424--10442},
  doi = {10.1021/acs.jctc.4c00994},
  number = {23},
  publisher = {American Chemical Society (ACS)},
}

@article{Palethorpe2025,
  author = {Palethorpe, Elise and Barca, Giuseppe M. J.},
  title = {High-Performance, High-Angular-Momentum {J} Engine on Graphics Processing Units},
  journal = {Journal of Chemical Theory and Computation},
  year = {2025},
  volume = {21},
  pages = {9388--9403},
  doi = {10.1021/acs.jctc.5c00775},
  number = {19},
  publisher = {American Chemical Society (ACS)},
}

@article{Barca2021,
  author = {Barca, Giuseppe M. J. and Alkan, Melisa and Vallejo, Jorge L. Galvez and Poole, David L. and Rendell, Alistair P. and Gordon, M.},
  title = {Faster Self-Consistent Field ({SCF}) Calculations on {GPU} Clusters},
  journal = {Journal of Chemical Theory and Computation},
  year = {2021},
  volume = {17},
  pages = {7486--7503},
  doi = {10.1021/acs.jctc.1c00720},
  number = {12},
  publisher = {American Chemical Society (ACS)},
}

@article{Barca2020,
  author = {Barca, Giuseppe M. J. and Vallejo, Jorge L. Galvez and Poole, David L. and Rendell, Alistair P. and Gordon, M.},
  title = {High-Performance, Graphics Processing Unit-Accelerated {Fock} Build Algorithm},
  journal = {Journal of Chemical Theory and Computation},
  year = {2020},
  volume = {16},
  pages = {7232--7238},
  number = {12},
  doi = {10.1021/acs.jctc.0c00768},
  publisher = {American Chemical Society (ACS)},
}

@article{Yasuda2008,
  author = {Yasuda, Koji},
  title = {Two-Electron Integral Evaluation on the Graphics Processor Unit},
  journal = {Journal of Computational Chemistry},
  year = {2008},
  volume = {29},
  pages = {334--342},
  doi = {10.1002/jcc.20779},
  number = {3},
  publisher = {Wiley},
}

@article{Ufimtsev2008,
  author = {Ufimtsev, Ivan S. and Martínez, T.},
  title = {Quantum Chemistry on Graphical Processing Units. 1. Strategies for Two-Electron Integral Evaluation},
  journal = {Journal of Chemical Theory and Computation},
  year = {2008},
  volume = {4},
  pages = {222--231},
  number = {2},
  doi = {10.1021/ct700268q},
  publisher = {American Chemical Society (ACS)},
}

@article{Ufimtsev2009,
  author = {Ufimtsev, Ivan S. and Martínez, T.},
  title = {Quantum Chemistry on Graphical Processing Units. 2. Direct Self-Consistent-Field Implementation},
  journal = {Journal of Chemical Theory and Computation},
  year = {2009},
  volume = {5},
  pages = {1004--1015},
  doi = {10.1021/ct800526s},
  number = {4},
  publisher = {American Chemical Society (ACS)},
}

@article{Asadchev2012,
  author = {Asadchev, Andrey and Gordon, Mark S.},
  title = {New Multithreaded Hybrid {CPU/GPU} Approach to {Hartree--Fock}},
  journal = {Journal of Chemical Theory and Computation},
  year = {2012},
  volume = {8},
  pages = {4166--4176},
  doi = {10.1021/ct300526w},
  number = {11},
  publisher = {American Chemical Society (ACS)},
}

@article{Manathunga2021,
  author = {Manathunga, Madushanka and Jin, Chi and Cruzeiro, Vinícius Wilian D. and Miao, Yipu and Mu, Dawei and Arumugam, Kamesh and Keipert, Kristopher and Aktulga, Hasan Metin and Merz, Jr., Kenneth M. and G{\"o}tz, Andreas W.},
  title = {Harnessing the Power of Multi-{GPU} Acceleration into the Quantum Interaction Computational Kernel Program},
  journal = {J. Chem. Theory Comput.},
  year = {2021},
  volume = {17},
  pages = {3955--3966},
  doi = {10.1021/acs.jctc.1c00145.s002},
  publisher = {American Chemical Society (ACS)},
}

@article{Manathunga2023,
  author = {Manathunga, M. and Aktulga, H. and Götz, A. and Merz, K.},
  title = {Quantum Mechanics/Molecular Mechanics Simulations on {NVIDIA} and {AMD} Graphics Processing Units},
  journal = {Journal of Chemical Information and Modeling},
  year = {2023},
  volume = {63},
  pages = {711--717},
  doi = {10.1021/acs.jcim.2c01505},
  publisher = {American Chemical Society (ACS)},
}

@article{Li2025,
  author = {Li, Rui and Sun, Qiming and Zhang, Xing and Chan, Garnet Kin-Lic},
  title = {Introducing {GPU} Acceleration into the {Python}-Based Simulations of Chemistry Framework},
  journal = {Journal of Physical Chemistry A},
  year = {2025},
  volume = {129},
  pages = {1459--1468},
  number = {5},
  doi = {10.1021/acs.jpca.4c05876},
  publisher = {American Chemical Society (ACS)},
}

@article{Wu2025,
  author = {Wu, Xiao-Jie and Sun, Qi-Ming and Pu, Zhi-Chen and Zheng, Tian-Ze and Ma, Wenzhi and Yan, Wen and Xia, Yu and Wu, Zhen and Huo, Mian and Li, Xiang and others},
  title = {Enhancing {GPU}-Acceleration in the {Python}-Based Simulations of Chemistry Frameworks},
  journal = {WIREs Computational Molecular Science},
  year = {2025},
  volume = {15},
  pages = {e70008},
  number = {2},
  doi = {10.1002/wcms.70008},
  publisher = {Wiley},
}

@article{Sun2020,
  author = {Sun, Qi-Ming and Zhang, Xing and Banerjee, S. and Bao, Peng and Barbry, M. and Blunt, N. S. and Bogdanov, N. and Booth, G. H. and Chen, Jia and Cui, Z. and others},
  title = {Recent Developments in the {PySCF} Program Package},
  journal = {Journal of Chemical Physics},
  year = {2020},
  volume = {153},
  pages = {024109},
  doi = {10.1063/5.0006074},
}

@article{Seritan2021,
  author = {Seritan, Stefan and Bannwarth, C. and Fales, Bryan S. and Hohenstein, E. and Isborn, C. and Kokkila-Schumacher, S. and Li, Xin and Liu, F. and Luehr, Nathan and Snyder, James W. and others},
  title = {{TeraChem}: A Graphical Processing Unit-Accelerated Electronic Structure Package for Large-Scale Ab Initio Molecular Dynamics},
  journal = {WIREs Computational Molecular Science},
  year = {2021},
  volume = {11},
  pages = {e1494},
  number = {2},
  doi = {10.1002/wcms.1494},
  publisher = {Wiley},
}

@article{Wang2024,
  author = {Wang, Yuan-Heng and Hait, D. and Johnson, K. G. and Fajen, O. J. and Zhang, J. and Guerrero, R. D. and Martínez, Todd J.},
  title = {Extending {GPU}-Accelerated {Gaussian} Integrals in the {TeraChem} Software Package to f Type Orbitals: Implementation and Applications},
  journal = {Journal of Chemical Physics},
  year = {2024},
  volume = {161},
  pages = {174118},
  number = {17},
  doi = {10.1063/5.0233523},
  publisher = {AIP Publishing},
}

@article{Asadchev2023,
  author = {Asadchev, Andrey and Valeev, Edward F.},
  title = {High-Performance Evaluation of High Angular Momentum 4-Center {Gaussian} Integrals on Modern Accelerated Processors},
  journal = {Journal of Physical Chemistry A},
  year = {2023},
  volume = {127},
  pages = {10889--10895},
  number = {51},
  doi = {10.1021/acs.jpca.3c04574},
  publisher = {American Chemical Society (ACS)},
}

@article{Asadchev2024,
  author = {Asadchev, Andrey and Valeev, Edward F.},
  title = {3-Center and 4-Center 2-Particle {Gaussian} {AO} Integrals on Modern Accelerated Processors},
  journal = {Journal of Chemical Physics},
  year = {2024},
  volume = {160},
  pages = {244109},
  number = {24},
  doi = {10.1063/5.0217001},
  publisher = {AIP Publishing},
}

@incollection{Luehr2016,
  author = {Luehr, Nathan and Sisto, Aaron and Martínez, T.},
  title = {Gaussian Basis Set {Hartree--Fock}, Density Functional Theory, and Beyond on {GPUs}},
  booktitle = {Electronic Structure Calculations on Graphics Processing Units},
  editor = {Walker, Ross C. and G{\"o}tz, Andreas W.},
  publisher = {Wiley},
  address = {Chichester},
  year = {2016},
  pages = {67--100},
  doi = {10.1002/9781118670712.CH4},
}

@article{Apra2020,
  author = {Aprà, E. and Bylaska, E. and Jong, De W. A. and Govind, N. and Kowalski, K. and Straatsma, T. and Valiev, M. and Dam, H. V. van and Alexeev, Y. and Anchell, J. and others},
  title = {{NWChem}: Past, Present, and Future},
  journal = {Journal of Chemical Physics},
  year = {2020},
  volume = {152},
  pages = {184102},
  doi = {10.1063/5.0004997},
  number = {18},
  publisher = {AIP Publishing},
}

@article{GANSU,
  author = {Ito, Yasuaki and Tsuji, Satoki and Nakano, Koji and Kasagi, Akihiko},
  title = {{GANSU}: A {GPU}-Native Quantum Chemistry Framework for Efficient {Hartree--Fock} and Post-{HF} Calculations},
  journal = {Engineer},
  year = {2026},
  volume = {7},
  number = {5},
  pages = {205},
  doi = {10.3390/eng7050205},
  publisher = {MDPI AG},
}

@misc{OpenACC33,
  author = {{OpenACC-Standard.org}},
  title = {The {OpenACC} Application Programming Interface, Version 3.3},
  howpublished = {\url{https://www.openacc.org/specification}},
  year = {2022},
}

@misc{OpenMP52,
  author = {{OpenMP Architecture Review Board}},
  title = {{OpenMP} Application Programming Interface, Version 5.2},
  howpublished = {\url{https://www.openmp.org/specifications/}},
  year = {2021},
}

@book{Szabo1996,
  author = {Szabo, Attila and Ostlund, Neil S.},
  title = {Modern Quantum Chemistry: Introduction to Advanced Electronic Structure Theory},
  publisher = {Dover Publications},
  address = {Mineola, NY},
  year = {1982},
}

@article{Whitten1973,
  author = {Whitten, J. L.},
  title = {Coulombic Potential Energy Integrals and Approximations},
  journal = {The Journal of Chemical Physics},
  year = {1973},
  volume = {58},
  pages = {4496--4501},
  number = {10},
  doi = {10.1063/1.1679012},
  publisher = {AIP Publishing},
}

@article{Haser1989,
  author = {Häser, Marco and Ahlrichs, R.},
  title = {Improvements on the Direct {SCF} Method},
  journal = {Journal of Computational Chemistry},
  year = {1989},
  volume = {10},
  pages = {104--111},
  number = {1},
  doi = {10.1002/jcc.540100111},
  publisher = {Wiley},
}

@article{Almlof1982,
  author = {Almlöf, J. and Fægri, K. and Korsell, K.},
  title = {Principles for a Direct {SCF} Approach to {LCAO-MO} Ab-Initio Calculations},
  journal = {J. Comput. Chem.},
  year = {1982},
  volume = {3},
  pages = {385--399},
  doi = {10.1002/jcc.540030314},
}

@article{Density_decay_1,
  title = {Decay properties of the finite-temperature density matrix in metals},
  author = {Goedecker, S.},
  journal = {Physical Review B},
  volume = {58},
  issue = {7},
  pages = {3501--3502},
  numpages = {0},
  year = {1998},
  month = {Aug},
  publisher = {American Physical Society (APS)},
  doi = {10.1103/PhysRevB.58.3501},
  url = {https://link.aps.org/doi/10.1103/PhysRevB.58.3501},
  number = {7},
}

@article{Ochsenfeld1998,
  author = {Ochsenfeld, Christian and White, Christopher A. and Head-Gordon, Martin},
  title = {Linear and Sublinear Scaling Formation of {Hartree--Fock}-Type Exchange Matrices},
  journal = {The Journal of Chemical Physics},
  year = {1998},
  volume = {109},
  pages = {1663--1669},
  number = {5},
  doi = {10.1063/1.476741},
  publisher = {AIP Publishing},
}

@article{Maslen1998,
  author = {Maslen, P. E. and Ochsenfeld, C. and White, C. A. and Lee, M. S. and Head-Gordon, M.},
  title = {Locality and Sparsity of Ab Initio One-Particle Density Matrices and Localized Orbitals},
  journal = {The Journal of Physical Chemistry A},
  year = {1998},
  volume = {102},
  pages = {2215--2222},
  doi = {10.1021/jp972919j},
  number = {12},
  publisher = {American Chemical Society (ACS)},
}

@article{IsmailBeigi1999,
  author = {Ismail-Beigi, Sohrab and Arias, T. A.},
  title = {Locality of the Density Matrix in Metals, Semiconductors, and Insulators},
  journal = {Physical Review Letters},
  year = {1999},
  volume = {82},
  pages = {2127--2130},
  doi = {10.1103/PhysRevLett.82.2127},
  number = {10},
  publisher = {American Physical Society (APS)},
}

@article{Kohn1959,
  author = {Kohn, W.},
  title = {Analytic Properties of {Bloch} Waves and {Wannier} Functions},
  journal = {Physical Review},
  year = {1959},
  volume = {115},
  pages = {809--821},
  number = {4},
  doi = {10.1103/PHYSREV.115.809},
  publisher = {American Physical Society (APS)},
}

@article{McMurchie1978,
  author = {McMurchie, Larry E. and Davidson, Ernest R.},
  title = {One- and Two-Electron Integrals over {Cartesian} {Gaussian} Functions},
  journal = {Journal of Computational Physics},
  year = {1978},
  volume = {26},
  pages = {218--231},
  doi = {10.1016/0021-9991(78)90092-X},
  number = {2},
  publisher = {Elsevier BV},
}

@article{HeadGordon1988,
  author = {Head-Gordon, Martin and Pople, John A.},
  title = {A Method for Two-Electron {Gaussian} Integral and Integral Derivative Evaluation Using Recurrence Relations},
  journal = {The Journal of Chemical Physics},
  year = {1988},
  volume = {89},
  pages = {5777--5786},
  number = {9},
  doi = {10.1063/1.455553},
  publisher = {AIP Publishing},
}

@article{Obara1986,
  author = {Obara, S. and Saika, A.},
  title = {Efficient Recursive Computation of Molecular Integrals over {Cartesian} {Gaussian} Functions},
  journal = {The Journal of Chemical Physics},
  year = {1986},
  volume = {84},
  pages = {3963--3974},
  number = {7},
  doi = {10.1063/1.450106},
  publisher = {AIP Publishing},
}

@article{Johnson1991,
  author = {Johnson, Benny G. and Gill, Peter M. W. and Pople, John A.},
  title = {Exact and Approximate Solutions to the One-Center {McMurchie--Davidson} Tree-Search Problem},
  journal = {International Journal of Quantum Chemistry},
  year = {1991},
  volume = {40},
  pages = {809--827},
  doi = {10.1002/qua.560400610},
  number = {6},
  publisher = {Wiley},
}

@article{Ryu1993,
  author = {Ryu, Ungsik and Kim, Myeongcheol and Lee, Yoon Sup},
  title = {Optimal Use of the Recurrence Relations for the Evaluation of Molecular Integrals over {Cartesian} {Gaussian} Basis Functions},
  journal = {Journal of Computational Chemistry},
  year = {1993},
  volume = {14},
  pages = {30--36},
  doi = {10.1002/jcc.540140107},
  number = {1},
  publisher = {Wiley},
}

@article{Gill1989,
  author = {Gill, Peter M. W. and Head-Gordon, Martin and Pople, John A.},
  title = {An Efficient Algorithm for the Generation of Two-Electron Repulsion Integrals over {Gaussian} Basis Functions},
  journal = {Int. J. Quantum Chem.},
  year = {1989},
  volume = {36},
  number = {S23},
  pages = {269--280},
  doi = {10.1002/qua.560360831},
}

@article{Davidson1996,
  author = {Davidson, Ernest R.},
  title = {Comment on ``{Comment} on {Dunning's} Correlation-Consistent Basis Sets''},
  journal = {Chemical Physics Letters},
  year = {1996},
  volume = {260},
  pages = {514--518},
  number = {3-4},
  doi = {10.1016/0009-2614(96)00917-7},
  publisher = {Elsevier BV},
}

@article{Hehre1969,
  author = {Hehre, W. J. and Stewart, R. F. and Pople, J. A.},
  title = {Self-Consistent Molecular-Orbital Methods. {I}. Use of {Gaussian} Expansions of {Slater}-Type Atomic Orbitals},
  journal = {The Journal of Chemical Physics},
  year = {1969},
  volume = {51},
  pages = {2657--2664},
  doi = {10.1063/1.1672392},
  number = {6},
  publisher = {AIP Publishing},
}

@article{Hehre1972,
  author = {Hehre, W. J. and Ditchfield, R. and Pople, J. A.},
  title = {Self-Consistent Molecular Orbital Methods. {XII}. Further Extensions of {Gaussian}-Type Basis Sets for Use in Molecular Orbital Studies of Organic Molecules},
  journal = {The Journal of Chemical Physics},
  year = {1972},
  volume = {56},
  pages = {2257--2261},
  number = {5},
  doi = {10.1063/1.1677527},
  publisher = {AIP Publishing},
}

@article{Hariharan1973,
  author = {Hariharan, P. C. and Pople, J. A.},
  title = {The Influence of Polarization Functions on Molecular Orbital Hydrogenation Energies},
  journal = {Theoretica Chimica Acta},
  year = {1973},
  volume = {28},
  pages = {213--222},
  doi = {10.1007/BF00533485},
  number = {3},
  publisher = {Springer Science and Business Media LLC},
}

@article{Dunning1989,
  author = {Dunning, Jr., Thom H.},
  title = {Gaussian Basis Sets for Use in Correlated Molecular Calculations. {I}. {The} Atoms Boron through Neon and Hydrogen},
  journal = {The Journal of Chemical Physics},
  year = {1989},
  volume = {90},
  pages = {1007--1023},
  doi = {10.1063/1.456153},
  number = {2},
  publisher = {AIP Publishing},
}

@article{Pulay1980,
  author = {Pulay, P{\'e}ter},
  title = {Convergence Acceleration of Iterative Sequences. The Case of {SCF} Iteration},
  journal = {Chemical Physics Letters},
  year = {1980},
  volume = {73},
  pages = {393--398},
  doi = {10.1016/0009-2614(80)80396-4},
  number = {2},
  publisher = {Elsevier BV},
}

@article{Pulay1982,
  author = {Pulay, P.},
  title = {Improved {SCF} Convergence Acceleration},
  journal = {Journal of Computational Chemistry},
  year = {1982},
  volume = {3},
  pages = {556--560},
  doi = {10.1002/jcc.540030413},
  number = {4},
  publisher = {Wiley},
}

@article{VanLenthe2006,
  author = {Lenthe, J. van and Zwaans, R. and Dam, H. V. Van and Guest, M. F.},
  title = {Starting {SCF} Calculations by Superposition of Atomic Densities},
  journal = {Journal of Computational Chemistry},
  year = {2006},
  volume = {27},
  pages = {926--932},
  doi = {10.1002/jcc.20393},
  number = {8},
  publisher = {Wiley},
}

@article{Lehtola2019,
  author = {Lehtola, Susi},
  title = {Assessment of Initial Guesses for Self-Consistent Field Calculations. Superposition of Atomic Potentials: Simple yet Efficient},
  journal = {Journal of Chemical Theory and Computation},
  year = {2019},
  volume = {15},
  pages = {1593--1604},
  doi = {10.1021/acs.jctc.8b01089},
  number = {3},
  publisher = {American Chemical Society (ACS)},
}

@article{Roothaan1951,
  author = {Roothaan, C. C. J.},
  title = {New Developments in Molecular Orbital Theory},
  journal = {Reviews of Modern Physics},
  year = {1951},
  volume = {23},
  pages = {69--89},
  doi = {10.1103/RevModPhys.23.69},
  number = {2},
  publisher = {American Physical Society (APS)},
}

@article{Hall1951,
  author = {Hall, G. G.},
  title = {The Molecular Orbital Theory of Chemical Valency. {VIII}. {A} Method of Calculating Ionization Potentials},
  journal = {Proceedings of the Royal Society of London. Series A, Mathematical and physical sciences},
  year = {1951},
  volume = {205},
  pages = {541--552},
  doi = {10.1098/rspa.1951.0048},
  number = {1083},
  publisher = {The Royal Society},
}

@article{Kohn1965,
  author = {Kohn, W. and Sham, L. J.},
  title = {Self-Consistent Equations Including Exchange and Correlation Effects},
  journal = {Physical Review},
  year = {1965},
  volume = {140},
  pages = {A1133--A1138},
  doi = {10.1103/PhysRev.140.A1133},
  number = {4A},
  publisher = {American Physical Society (APS)},
}

@article{Moller1934,
  author = {Møller, C. and Plesset, M.},
  title = {Note on an Approximation Treatment for Many-Electron Systems},
  journal = {Physical Review},
  year = {1934},
  volume = {46},
  pages = {618--622},
  doi = {10.1103/PhysRev.46.618},
  number = {7},
  publisher = {American Physical Society (APS)},
}

@article{Bartlett2007,
  author = {Bartlett, R. and Musiał, M.},
  title = {Coupled-Cluster Theory in Quantum Chemistry},
  journal = {Reviews of Modern Physics},
  year = {2007},
  volume = {79},
  pages = {291--352},
  doi = {10.1103/RevModPhys.79.291},
  number = {1},
  publisher = {American Physical Society (APS)},
}

@misc{CUDA,
  author = {{NVIDIA Corporation}},
  title = {{CUDA} {C++} Programming Guide},
  howpublished = {\url{https://docs.nvidia.com/cuda/cuda-c-programming-guide/}},
  note = {Accessed 2026-10-22},
}

@misc{HIP,
  author = {{Advanced Micro Devices}},
  title = {{HIP} Documentation},
  howpublished = {\url{https://rocm.docs.amd.com/projects/HIP/}},
  note = {Accessed 2026-10-22},
}

@article{Neese2020,
  author = {Neese, F. and Wennmohs, F. and Becker, U. and Riplinger, C.},
  title = {The {ORCA} Quantum Chemistry Program Package},
  journal = {Journal of Chemical Physics},
  year = {2020},
  volume = {152},
  pages = {224108},
  doi = {10.1063/5.0004608},
  number = {22},
  publisher = {AIP Publishing},
}

\end{document}